\documentclass{jfm}
\usepackage{rotating, graphicx}
\usepackage{xcolor}
\usepackage{epstopdf, epsfig}
\usepackage{epsfig}

\usepackage{lscape}
\usepackage{float}

\shorttitle{Marangoni stresses in drop-interface coalescence}
\shortauthor{Constante-Amores et al.}

\title{Role of surfactant-induced Marangoni stresses in drop-interface coalescence} 

\author{C. R. Constante-Amores\aff{1},
A. Batchvarov\aff{1},
L. Kahouadji\aff{1},
S. Shin\aff{2},
J. Chergui\aff{3},
D. Juric\aff{3}
 \and O. K. Matar\aff{1}\corresp{\email{o.matar@imperial.ac.uk}}}
\affiliation{\aff{1}Department of Chemical Engineering, Imperial College London, South Kensington Campus, London SW7 2AZ, United Kingdom
\aff{2}Department of Mechanical and System Design Engineering, Hongik University, Seoul 04066, Republic of Korea.
\aff{3}Universit\'e Paris Saclay, Centre National de la Recherche Scientifique (CNRS), Laboratoire Interdisciplinaire des Sciences du Num\'erique (LISN), 91400 Orsay, France.}
\begin{document}
\maketitle

\begin{abstract}
We study the effect of surfactants on the dynamics  of a drop-interface coalescence using %; this represents a canonical example of interfacial singularity as a result of topological change. We use 
full three-dimensional direct numerical simulations. We employ a hybrid interface-tracking/level-set method, which takes into account Marangoni stresses that arise from surface tension gradients, interfacial and bulk diffusion, and sorption kinetic effects.
We validate our predictions against the experimental data %A physical configuration for which the outcome involves a partial-coalescence type is validated against 
of Blanchette and Bigioni [\textit{Nat. Phys.} \textbf{2}(4):254-257 (2006)] %, in terms of its temporal evolution for the surfactant-free case. A 
and perform a parametric study 
that demonstrates the delicate interplay between the flow fields and those associated with the surfactant bulk and interfacial concentrations. 
%considering the strength of surfactant, and sorption and desorption kinetics is performed.
The results of this work unravel %and shed light  
the crucial role of the %tangential 
Marangoni stresses in the flow physics of %the 
coalescence %phenomena 
with particular attention paid to their influence on neck reopening dynamics %. This study demonstrates the delicate interplay of the different flow fields and provides remarkable insights into the reopening of the neck 
in terms of stagnation-point inhibition, and near-neck vorticity generation. %from the neck. Additionally, 
We demonstrate that surfactant-laden cases feature a  rigidifying effect on the interface compared to the surfactant-free case, a mechanism that underpins the observed surfactant-induced phenomena. %with clean interfaces. 
\end{abstract}

%%%%%%%%%%%%%%%%%%%%%%%%%%%%%%%%%%%%%%%%%
%%%%%%%%%%%%%%%%%%%%%%%%%%%%%%%%%%%%%%%%%
%%%%%%%%%%%%%%%%%%%%%%%%%%%%%%%%%%%%%%%%%
\section{Introduction}
%%%%%%%%%%%%%%%%%%%%%%%%%%%%%%%%%%%%%%%%%
%%%%%%%%%%%%%%%%%%%%%%%%%%%%%%%%%%%%%%%%%
%%%%%%%%%%%%%%%%%%%%%%%%%%%%%%%%%%%%%%%%%

The occurrence of drop-interface coalescence has been observed in a wide range of natural phenomena and industrial applications, such as  rain/cloud formation \citep{Raes_ae_2000}, atomisation \citep{Villermaux_arfm_2007}, and also emulsification or de-emulsification processes \citep{Ziegler_pol_2005}. Over half a century ago, \citet{Charles_jcs_1960} observed coalescence in their ground-breaking experiments, and, ever since, researchers have been in constant pursuit of a better physical understanding of this phenomenon. However, it was not until the advent of high-speed imaging %important advancements in experimental technology, such as the existence of high spatio-temporal resolutions cameras, made it
that it became possible for \cite{Thoroddsen_pof_2000} to observe the self-similar coalescence cascade phenomena of a drop before its total coalescence. Since then, the significant interest in the field has led to a recent comprehensive review on the topic by \cite{Kavehpour_arfm_2015} who concluded that further work is needed to understand the Marangoni-effect during the drop-interface coalescence dynamics.

The %drop-interface coalescence 
dynamics commence with the drainage of the fluid between the drop and an interface separating this fluid from another bulk phase whose extent is typically much larger than the drop diameter. This drainage leads to the formation of a thin fluid layer between the drop and the interface. As the layer thickness decreases, van der Waals forces trigger its rupture with the generation of a hole which expands driven by capillarity. The hole expansion has been widely studied by \citet{Eggers_jfm_1999,Paulsen_prl_2011,Anthony_prf_2017} among others, concluding that the interfacial dynamics are solely governed by a balance between viscous and capillary forces, and subsequently the Ohnesorge number $Oh$ (e.g., ratio of viscous to capillary forces) is the most appropriate control parameter for this phenomenon. Different coalescence regimes have been identified depending on the order of magnitude of $Oh$: (i) %\textit{(i)} 
the inertial/capillary coalescence regime ($Oh \ll1$), which is characterised by a nearly  inviscid liquid, and the dynamics are surface-tension-driven; (ii) %\textit{(ii)} 
the viscous/capillary regime ($Oh >1$) where the viscous forces play a major role in the interfacial dynamics; finally,  (iii) %\textit{(iii)} 
an intermediate regime which bridges the inertial/capillary and viscous/capillary regimes (e.g., no dominance by either viscosity or surface tension).

As pointed out earlier for intermediate values of $Oh$, \citet{Thoroddsen_pof_2000} observed the  so-called `coalescence-cascade of a drop' in which the drop coalescence leads to the generation of a smaller daughter droplet which results in a cascade of self-similar events until this successive coalescence process is completed. The process of formation of a  daughter droplet is known as a `partial coalescence' phenomenon, and its physical understanding came from the insightful %unprecedented set of 
experimental and numerical results of  \citet{Blanchette_np_2006} who suggested that the occurrence of pinch-off depends solely on the competition between the vertical (inertia-viscous) and horizontal (capillary) pulls (i.e., the former aids the total coalescence and the latter the capillary breakup), rather than the mechanism of Rayleigh-Plateau instability. Additionally \citet{Blanchette_np_2006} have provided  an extensive $Bo$-$Oh$ phase diagram delineating the boundaries between partial and total coalescence. Here, $Bo$ is the %dimensionless 
Bond number which compares the importance of gravitational to surface tension forces. 

Importantly, \citet{Blanchette_pof_2009} and \citet{Sun_prf_2018} have also considered situations in which there is a surface tension mismatch between the drop and the interface triggering the generation of tangential Marangoni stresses in the plane of the  common interface formed post coalescence. In their experimental and numerical investigation of the coalescence of a water drop with an ethanol reservoir, they reported that Marangoni-induced flow leads to the  ejection of an additional drop from its summit during its vertical  stretching. Similarly, the  generation of gradients of surface tension can also be triggered by the use of %surface-active-agents (i.e., 
surfactants %) 
\citep{Manikantan_jfm_2020}.

Current understanding of the coalescence dynamics in such surfactant-laden systems came from \citet{Dong_pof_2019} who experimentally showed for the first time surfactant concentration profiles for systems characterised by high Bond numbers.  Additionally, they suggested that surfactants have a strong effect  on the interfacial dynamics inducing interfacial rupture  (i.e., hole formation) in an off-axis location. Their interfacial concentration profiles agree qualitatively with the previous numerical work performed by  \citet{Martin_pof_2015}. Finally, \citet{Shim_prf_2017} suggested that the presence of surfactants decreases the air-drainage-time between the drop and interface (so-called,  `damped-coalescence-cascade mechanism').

\citet{Craster_pof_2002} have shown that the presence of surfactants does not affect the dynamics of interfacial singularities, as they are swept away from the pinch-off point; thus, the  scaling laws predicted by \citet{Eggers_prl_1993} are remain unaltered.  Nonetheless, recently, \citet{Constante-Amores_prf_2020} showed the  deleterious effect of Marangoni stresses on interfacial singularities in the context of capillary retraction of a liquid thread. They demonstrated that Marangoni stresses drive reopening of the neck during the capillary retraction (i.e., escape from the `end-pinching' mechanism) as a result of the suppression of stagnation point formation by flow-reversal in the vicinity of the neck, and the higher generation of vorticity from the neck. 

Although, \citet{Martin_pof_2015} have shown that the presence of surfactants is responsible for the inhibition of the partial coalescence event, there is still a lack of explanation of what causes this pinch-off inhibition. Moreover, the appreciation of Marangoni stresses profiles, and other significant insights into the flow fields close to the pinch-off are also missing.  The present study aims to clarify and answer these questions and to overcome all numerical difficulties presented in \citet{Martin_pof_2015} by taking into account the nonlinear relation between the surfactant concentration and surface tension, which is of central importance in non-dilute systems. Additionally, we are able to explore parameter ranges corresponding to large density and viscosity contrasts, corresponding to air-water systems, %use real physical properties 
without %any 
numerical difficulties, %such as water-air density and viscosity ratios, 
and to go beyond the surfactant elasticity range studied by \citet{Martin_pof_2015}. 

The rest of this article is organised as follows: Section \ref{problem_formulation} presents the governing equations, numerical setup, and the validation of the surfactant-free case against the experimental observations of \citet{Blanchette_np_2006}. Section \ref{results} provides a discussion of the results which are focused on the origin of the inhibition of the interfacial singularity, and a parametric study accounting for the strength of the Marangoni stress, and sorption dynamics. Finally, concluding remarks are summarised in Section \ref{conclusion}.

%%%%%%%%%%%%%%%%%%%%%%%%%%%%%%%%%%%%%%%%%
%%%%%%%%%%%%%%%%%%%%%%%%%%%%%%%%%%%%%%%%%
%%%%%%%%%%%%%%%%%%%%%%%%%%%%%%%%%%%%%%%%%
\section{Problem formulation and numerical method \label{problem_formulation}}
%%%%%%%%%%%%%%%%%%%%%%%%%%%%%%%%%%%%%%%%%
%%%%%%%%%%%%%%%%%%%%%%%%%%%%%%%%%%%%%%%%%
%%%%%%%%%%%%%%%%%%%%%%%%%%%%%%%%%%%%%%%%%

With the purpose of studying the dynamics of  interfacial coalescence in the presence of surfactants, we perform direct numerical simulations of the two-phase Navier-Stokes equations in a three-dimensional Cartesian domain $\mathbf{x} = \left(x, y, z \right)$ (see Figure \ref{configuration}a). The treatment of  the interface and its surface tension forces  is handled using a hybrid front-tracking/level-set technique, also known as the Level Contour Reconstruction Method (LCRM) (\cite{Shin_ijnmf_2009}, \cite{Shin_jmst_2017}), with surfactant transport being  resolved both in the bulk and on the interface. More information on the numerical technique applied to surfactant transport can be found in the work of \cite{Shin_jcp_2018}. Moreover, the dependence of the surface tension  on the interfacial surfactant concentration is described by a nonlinear Langmuir equation of state \citep{Muradoglu_jcp_2014,Shin_jcp_2018}. 

\subsection{Scaling}
In what follows, all variables will be made dimensionless (represented by tildes) using
\begin{equation} \label{scales}
\quad \tilde{\mathbf{x}}=\frac{\mathbf{x}}{R_o},
\quad \tilde{t}=\frac{t}{T}, 
\quad \tilde{\textbf{u}}=\frac{\textbf{u}} {U},
\quad \tilde{p}=\frac{p}{\rho_l U^2}, 
\quad \tilde{\sigma}=\frac{\sigma}{\sigma_s},
\quad \tilde{\Gamma}=\frac{\Gamma}{\Gamma_\infty},
\quad \tilde{C}=\frac{C}{C_\infty},
\quad \tilde{C_s}=\frac{C_s}{C_\infty},
\end{equation}
\noindent	
where, $t$, $\textbf{u}$, and $p$ stand for time, velocity, and pressure, respectively. The physical parameters correspond to the liquid density $\rho_l$, viscosity, $\mu_l$, surface tension, $\sigma$, surfactant-free surface tension, $\sigma_s$, and  gravitational acceleration, $g$; $T=\sqrt{\rho_l R_o^3/\sigma_s}$ is the capillary time scale and $R_o$ is the initial drop radius; hence the velocity scale is $U=R_o/T= \sqrt{\sigma_s/(\rho_l R_o)}$. The interfacial surfactant concentration, $\Gamma$, is scaled on the saturation interfacial concentration, $\Gamma_{\infty}$, whereas the the bulk and bulk sub-phase (the region immediately adjacent to the interface) surfactant concentrations given by $C$ and $C_s$, respectively, are scaled on the initial bulk surfactant concentration, $C_{\infty}$. As a result of the scaling in equation (\ref{scales}), the dimensionless forms of the governing equations for the flow and the surfactant transport are respectively expressed as
\begin{equation}\label{div}
 \nabla \cdot \tilde{\textbf{u}}=0,
\end{equation}
\begin{equation}\label{NS_Eq}
\tilde{\rho} (\frac{\partial \tilde{\textbf{u}}}{\partial \tilde{t}}+\tilde{\textbf{u}} \cdot\nabla \tilde{\textbf{u}}) + \nabla \tilde{p}  = -Bo \textbf{i}_z + Oh ~ \nabla\cdot  \left [ \tilde{\mu} (\nabla \tilde{\textbf{u}} +\nabla \tilde{\textbf{u}}^T) \right ] +
\int \limits_{\tilde{A}\tilde{(t)}} \left(\tilde{\sigma} \tilde{\kappa} \textbf{n}  +  \nabla_s  \tilde{\sigma} \right) \delta \left(\tilde{\textbf{x}} - \tilde{\textbf{x}}_{_f}  \right)\mbox{d}\tilde{A},
\end{equation}
\begin{equation} 
\frac{\partial \tilde{C}} {\partial \tilde{t}}+\tilde{\textbf{u}}\cdot \nabla \tilde{C}= \frac{1}{Pe_b} \nabla^2 \tilde{C},
\end{equation}
 \begin{equation} 
 \label{equation_surfactant}
 \frac{\partial \tilde{\Gamma}}{\partial \tilde{t}}+\nabla_s \cdot (\tilde{\Gamma} \tilde{\textbf{u}} _{\rm t}) =\frac{1}{Pe_s} \nabla^2_s \tilde{\Gamma}+ Bi \left ( k  \tilde{C_s} (1-\tilde{\Gamma})- \tilde{\Gamma}  \right ),
 \end{equation}
\begin{equation} 
\tilde{\sigma}=1 + \beta_s \ln{\left(1 -\tilde{\Gamma}\right)},
\label{marangoni_eq}
 \end{equation}
\noindent
which correspond to the equations of mass and momentum conservation, the convective-diffusion equations for the surfactant bulk and interfacial concentrations, and the nonlinear surfactant equation of state, respectively. 
Here, the density $\tilde{\rho}$ and viscosity $\tilde{\mu}$ are expressed by $\tilde{\rho}=\rho_g/\rho_l + \left(1 -\rho_g/\rho_l\right) \mathcal{H}\left(\tilde{\textbf{x}},\tilde{t}\right)$ and $\tilde{\mu}=\mu_g/\mu_l+ \left(1 -\mu_g/\mu_l\right) \mathcal{H}\left( \tilde{\textbf{x}},\tilde{t}\right)$ wherein $\mathcal{H}\left( \tilde{\textbf{x}},\tilde{t}\right)$ represents a smoothed Heaviside function, which is zero in the gas phase and unity in the liquid phase, where the subscript $g$ designates the gas phase; $\tilde{\textbf{u}}_{\rm{t}}= \left ( \tilde{\textbf{u}}_{\rm{s}} \cdot \textbf{t} \right ) \textbf{t}$ represents the velocity vector tangential to the interface in which $\tilde{\textbf{u}}_{\rm{s}}$ corresponds to the interfacial velocity; 
$\kappa$ is twice  the mean interface curvature calculated from the Lagrangian interface grid;
$\nabla_s=\left({\mathbf{I}}-\mathbf{n}\mathbf{n}\right)\cdot \nabla$ stands for the surface gradient operator wherein $\mathbf{I}$ is the identity tensor and $\mathbf{n}$ is the outward-pointing unit normal to the interface; $\tilde{\mathbf{x}}_f$ is the parameterisation of the interface $\tilde{A} (\tilde{t})$; finally, 
$\delta$ represents  a Dirac delta function that is non-zero when $\tilde{\mathbf{x}}=\tilde{\mathbf{x}}_f$ only. 
The numerical method used to solve the above equations %within the remits of this work 
is described in detail by \citet{Shin_jcp_2018}.

The dimensionless groups that appear in equations (\ref{div})-(\ref{marangoni_eq}) are defined as 
\begin{equation}
Bo=\frac{\rho_l g R_o^2}{\sigma_s }, ~~~
Oh=\frac{\mu_l}{\sqrt{\rho_l \sigma_s R}} ,     ~~~
\end{equation}
\begin{equation}\label{surf_param}
Bi=\frac{k_d R_o}{U},        ~~~
k=\frac{k_a C_\infty }{k_d}, ~~~
Pe_s=\frac{ U R_o}{D_s},     ~~~ 
Pe_b=\frac{ U R_o}{D_b},     ~~~ 
\beta_s= \frac{\Re T \Gamma_\infty }{\sigma_s },
\end{equation}
\noindent
where $Bo$ and $Oh$ are the Bond number (ratio of gravitational to capillary forces) and Ohnesorge number (ratio of viscous to surface tension forces), respectively. The surfactant elasticity parameter, $\beta_s$,  measures the sensitivity of the surface tension to the surfactant concentration in which the parameter $\Re$ represents  the thermodynamic ideal gas constant value $8.314$ J K$^{-1}$ mol$^{-1}$, and $T$ denotes temperature. The parameters $Pe_s$ and $Pe_b$ are the interfacial and bulk Peclet numbers that represent the ratio of convective to diffusive time-scales in the plane of the interface and the bulk, respectively. The Biot number, $Bi$, stands for the ratio of characteristic desorptive to convective time-scales. Finally, $k$ is the ratio of adsorption to desorption time scales where $k_a$ and $k_d$ refer to the surfactant adsorption and desorption coefficients, respectively. 

At equilibrium, there is no surfactant exchange between the interface and the bulk, and the last term on the right-hand-side of equation (\ref{equation_surfactant}) %, also know as a source term, 
reduces to the Langmuir adsorption isotherm  
 \begin{equation}
\chi=\frac{\Gamma_{eq}}{\Gamma_{\infty}}=\frac{k}{(1+k)},
\end{equation}
\noindent
where $\chi$ stands for the fraction of the interface covered by adsorbed surfactant. Furthermore, the Marangoni stress, $\tilde{\tau}$, which appears in the third term on the right-hand-side of equation (\ref{NS_Eq}), is expressed as a function of $\tilde{\Gamma}$ as follows:
 \begin{equation}
 \label{langmuir}
 \tilde{\tau} \equiv  \nabla_s \tilde{\sigma} \cdot  \textbf{t} =-\frac{\beta_s}{1 -\tilde{\Gamma}} \nabla_s\tilde{\Gamma} \cdot \textbf{t},
\end{equation}
\noindent
where $\mathbf{t}$ is the unit tangent to the interface. 
In all cases considered in the present study, the Marangoni time-scale, $\mu R_o/ \Delta \sigma =O(10^{-4})$ s, as compared to the  capillary and sorptive/desorptive time-scales, which are of $O(10^{-3})$ s and $O(10^{-3})-O(10^{-4})$ s, respectively;  thus Marangoni stresses will play a crucial role in the coalescence phenomenon. Finally, the tildes are dropped henceforth with the understanding that hereafter all variables discussed are dimensionless unless stated otherwise.  

\subsection{Numerical setup, validation, and parameters}

\begin{figure}
\begin{center} 
\begin{tabular}{ccc}
\includegraphics[width=0.33\linewidth]{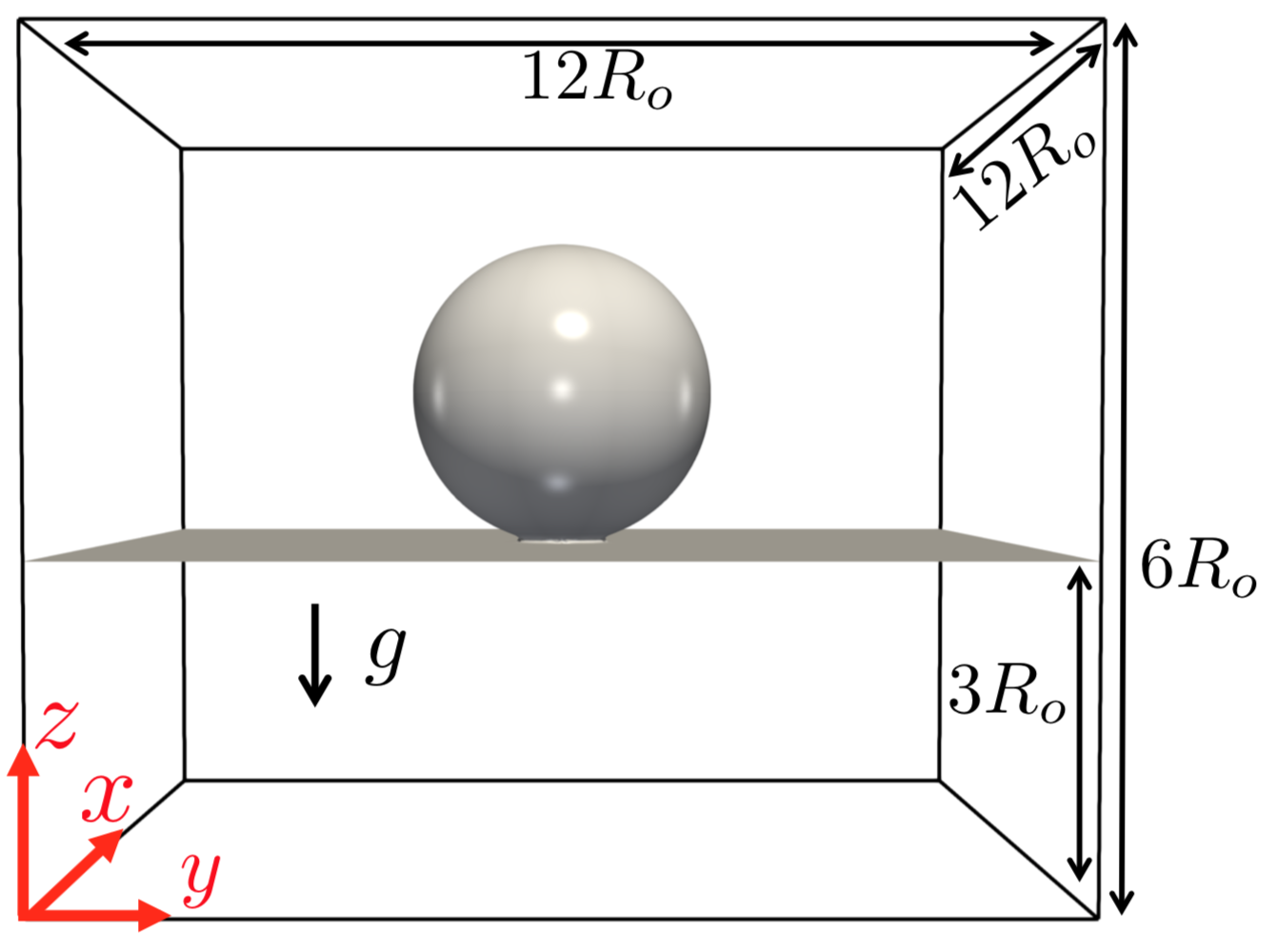}& \includegraphics[width=0.35\linewidth]{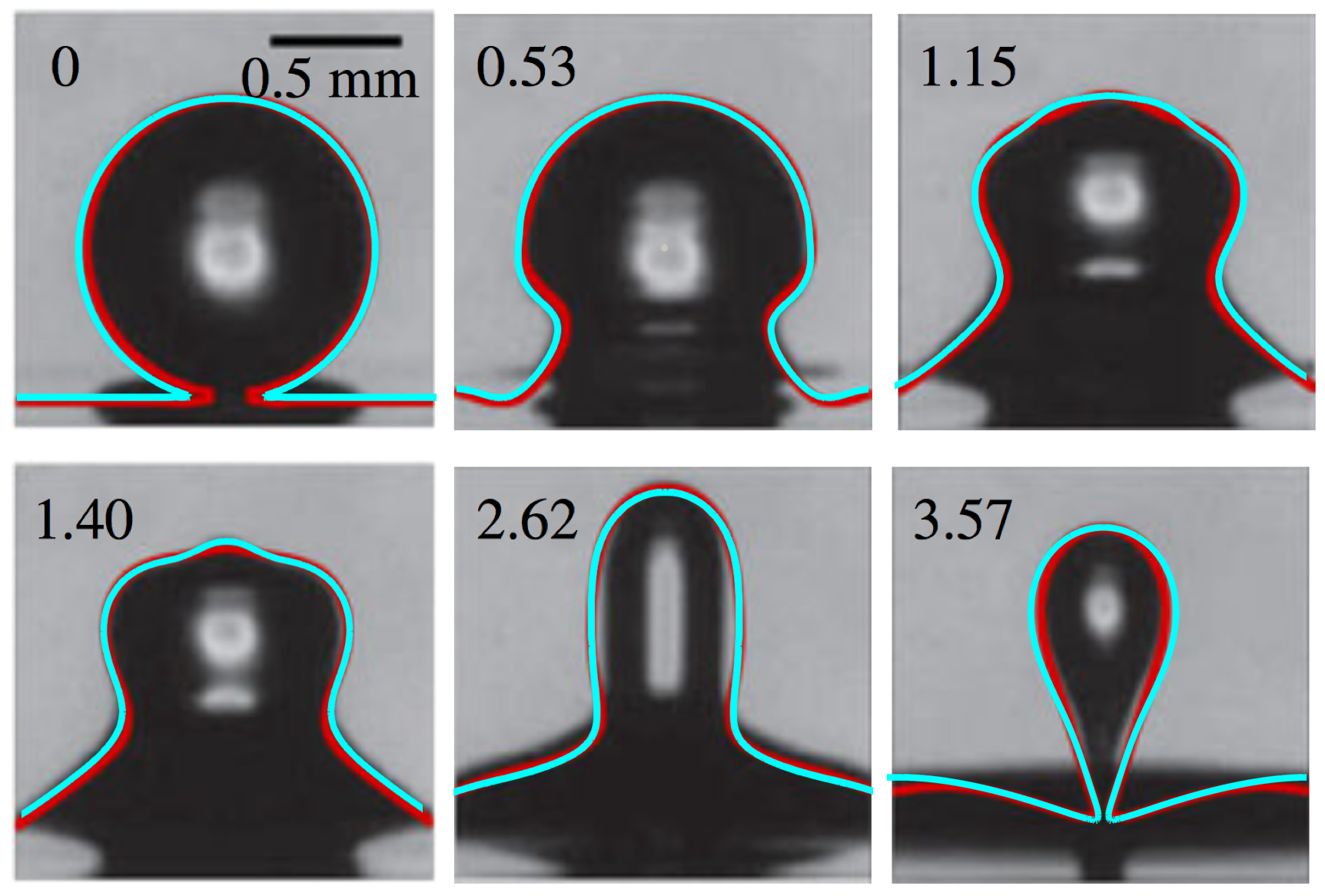}& \includegraphics[height=0.23\linewidth]{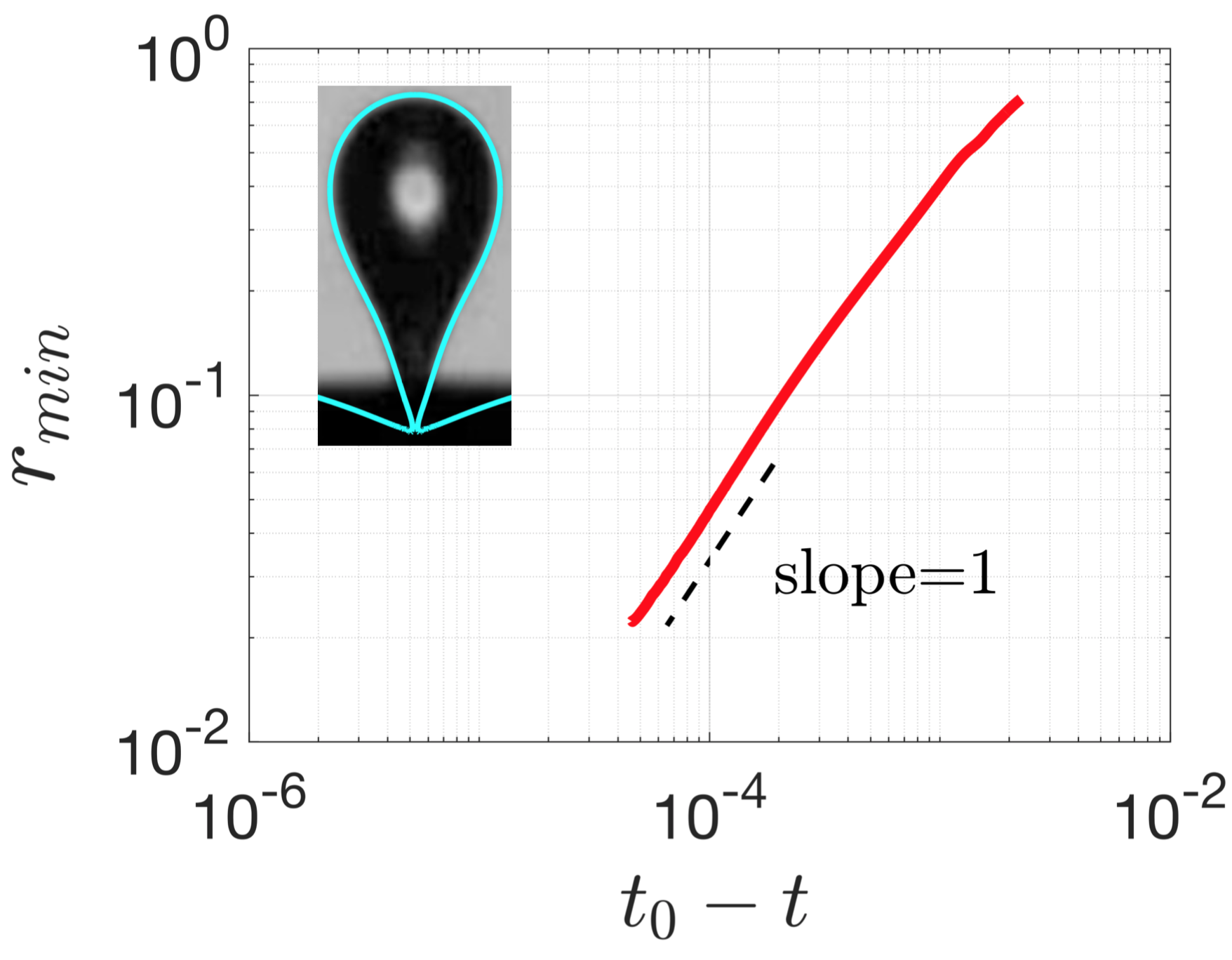}\\
(a) & (b) & (c) 
\end{tabular}
\end{center} 
\caption{\label{configuration} Schematic representation of the flow configuration, and validation of the numerical procedure: (a) initial shape of the drop resting close to the interface, highlighting the computational domain of size $12R_o\times12R_o\times 6R_o$ (not-to-scale) in a three-dimensional Cartesian domain, $\mathbf{x} = (x, y, z)$, with a resolution of $386^3$; (b) direct comparisons of our numerical predictions for a surfactant-free case (blue line) with experimental results reported by \citet{Blanchette_np_2006} for the post-coalescence dynamics of an ethanol drop in air prior to  interfacial singularity formation with  $Oh=0.011$ and $Bo=0.09$; also shown in red lines are the numerical solutions by \citet{Deka_jfm_2019} for the same case; (c) evolution of the neck radius as a function of the time to break-up, $t_o$, which agrees with the well-known inertial-viscous scaling theory of \citep{Eggers_prl_1993} shown as a dashed line with unity slope.} 
\end{figure}

The numerical setup closely follows the work done by \citet{Sun_prf_2018}, and \citet{Martin_pof_2015}. Thus, the size of the dimensionless computational domain is chosen as $12R_o\times12R_o\times 6R_o$, which is found to be sufficiently large to avoid the effect of artificial reflections from the boundaries.  We define a radial component as $r=\sqrt{\left(x-x_o\right)^2 + \left(y-y_o\right)^2}$ where $x_o$ and $y_o$ are the abscissa and ordinate drop position, respectively.  Solutions are sought subject to Neumann boundary conditions on all variables at the lateral boundaries, $p=0$ at the top boundary $z=6R_o$, and no-slip at  the bottom $z=0$. The %numerical 
initialisation of the interface corresponds to a spherical drop resting immediately above a horizontal flat interface before its interfacial rupture (e.g., all the velocities set to zero) where both drop and liquid pool are made up of the same liquid. Importantly, the drop is connected to the flat-interface by a neck of radius $0.25R_o$ for the initialisation of the dynamics; a similar approach has been previously used by \cite{Blanchette_np_2006,Blanchette_jfm_2009} and \cite{Martin_pof_2015}. The assumption is based on the time scale associated to the retraction of the neck $t_{CR}=R_o/\sqrt{2\sigma_s/\rho \delta}$, which is too short to have an influence on the phenomenon. 

Figures \ref{configuration}b-c highlight qualitative and quantitative validation of our numerical framework with results from the literature \citep{Eggers_prl_1993,Blanchette_np_2006,Deka_jfm_2019}. The numerical simulations have been benchmarked against the surfactant-free experimental results of \cite{Blanchette_np_2006} in terms of the temporal interfacial dynamics of the coalescence of an ethanol drop surrounded by air (displayed in figure \ref{configuration}b). Our numerical results are provided as snapshots of the interface location at times corresponding to those given by \citet{Blanchette_np_2006}. We have also included the numerical predictions from \citet{Deka_jfm_2019}.  Figure \ref{configuration}b demonstrates that our numerical framework is capable of predicting accurately the interfacial dynamics of the coalescence phenomenon for `clean' interfaces. Additionally, the temporal evolution of the neck towards its pinch-off is in good agreement with the well-known inertial-viscous scaling theory of \cite{Eggers_prl_1993} (see figure \ref{configuration}c). 

The dimensionless quantities for the studied phenomenon are consistent with experimentally-realisable systems. %have been chosen carefully to describe the system in an experimental-friendly framework. 
The Ohnesorge number was set to $Oh=2\times10^{-2}$ because it allows the observation of the interplay between the full range of dynamics as there is a competition between inertial, viscous, and capillary  forces.  The chosen density and viscosity ratios,  $\rho_g/ \rho_l =1.2 \times 10^{-3}$ and $\mu_g/\mu_l= 0.018$, respectively, are representative of an air-water system. The elasticity number $\beta_s$ depends on the interfacial concentration at saturation, $\Gamma_\infty$, which, in turn, is related to the critical micelle concentration (CMC) that is of $O(10^{-6})~$mol/m$^{2}$. We have explored the range of $0.1<\beta_s<0.5$ which corresponds to $2.9 \times 10^{-6}<$ CMC $<1.4 \times 10 ^{-5}$ mol/m$^{2}$. Typical values for the interfacial diffusion coefficient for surfactants such as %DS, 
Sodium Dodecyl Sulphate (SDS),  %SDS, 
N-Dodecyl-N,N-dimethylammonio-3-propane sulfonate (SB12), %SB12 
and similar %other 
monomers in aqueous solution are within the range of $10^{-12}<D_s<10^{-8}$ m$^2$/s  when $\Gamma$ is below the CMC %critical-micelle-concentration 
\citep{Joos_jcp_1982,Siderius_jsc_2002}; this range also covers phospholipid-based pulmonary surfactants, such as  N-(7-Nitrobenz-2-Oxa-1,3-Diazol-4-yl)-1,2-Dihexadecanoyl-sn-Glycero-3-
phosphocholine (NBD-PC), which are considered effectively insoluble \citep{Fallest_njp_2010,Strickland_jfm_2015}. 
Therefore, the interfacial Peclet number $Pe_s$ lies in the range  $10^{3}<Pe_s<10^{6}$. Recently, \cite{Bachvarov_prf_2020} and \cite{Constante-Amores_prf_2020} suggested that the investigated interfacial dynamics reach saturation above $Pe_s=100$; thus, the selected interfacial Peclet is set to $Pe_s=100$. % to ease the numerical solution.  
In terms of the chosen bulk Peclet number, \cite{Agrawal_jcis_1988} suggested that the interfacial and bulk Peclet numbers are of the same order of magnitude; on this basis, hereafter we set $Pe_b=Pe_s$.  
In summary, we have chosen the values of the surfactant-related parameters to ensure that all of the relevant physical processes associated with surfactant transport such as Marangoni stresses, surface/bulk diffusion, and sorption kinetics are represented in the present study. 

In terms of mesh resolution studies, we have ensured that our numerical simulations are mesh-independent, and subsequently  for a resolution of $(386)^3$, the results do not change with decreasing cell size. We have also ensured that %assessed and monitored 
the liquid volume and surfactant mass conservation are satisfied with an error under $10 ^{-3}~\%$ (see Appendix for more information).  Extensive mesh studies for surface tension-driven phenomena using the same computational method have been  published previously (\cite{Bachvarov_prf_2020,Constante-Amores_prf_2020}). 
A discussion of the results is presented next.

%%%%%%%%%%%%%%%%%%%%%%%%%%%%%%%%%%%%%%%%%
%%%%%%%%%%%%%%%%%%%%%%%%%%%%%%%%%%%%%%%%%
%%%%%%%%%%%%%%%%%%%%%%%%%%%%%%%%%%%%%%%%%
\section{Results\label{results}}
%%%%%%%%%%%%%%%%%%%%%%%%%%%%%%%%%%%%%%%%%
%%%%%%%%%%%%%%%%%%%%%%%%%%%%%%%%%%%%%%%%%
%%%%%%%%%%%%%%%%%%%%%%%%%%%%%%%%%%%%%%%%%

\begin{figure}   
\begin{tabular}{cccc}
\hline
$t=0.17$ & $t=0.69$ & $t=1.20$ & $t=1.69$ \\
\hline
\includegraphics[width=0.245\linewidth]{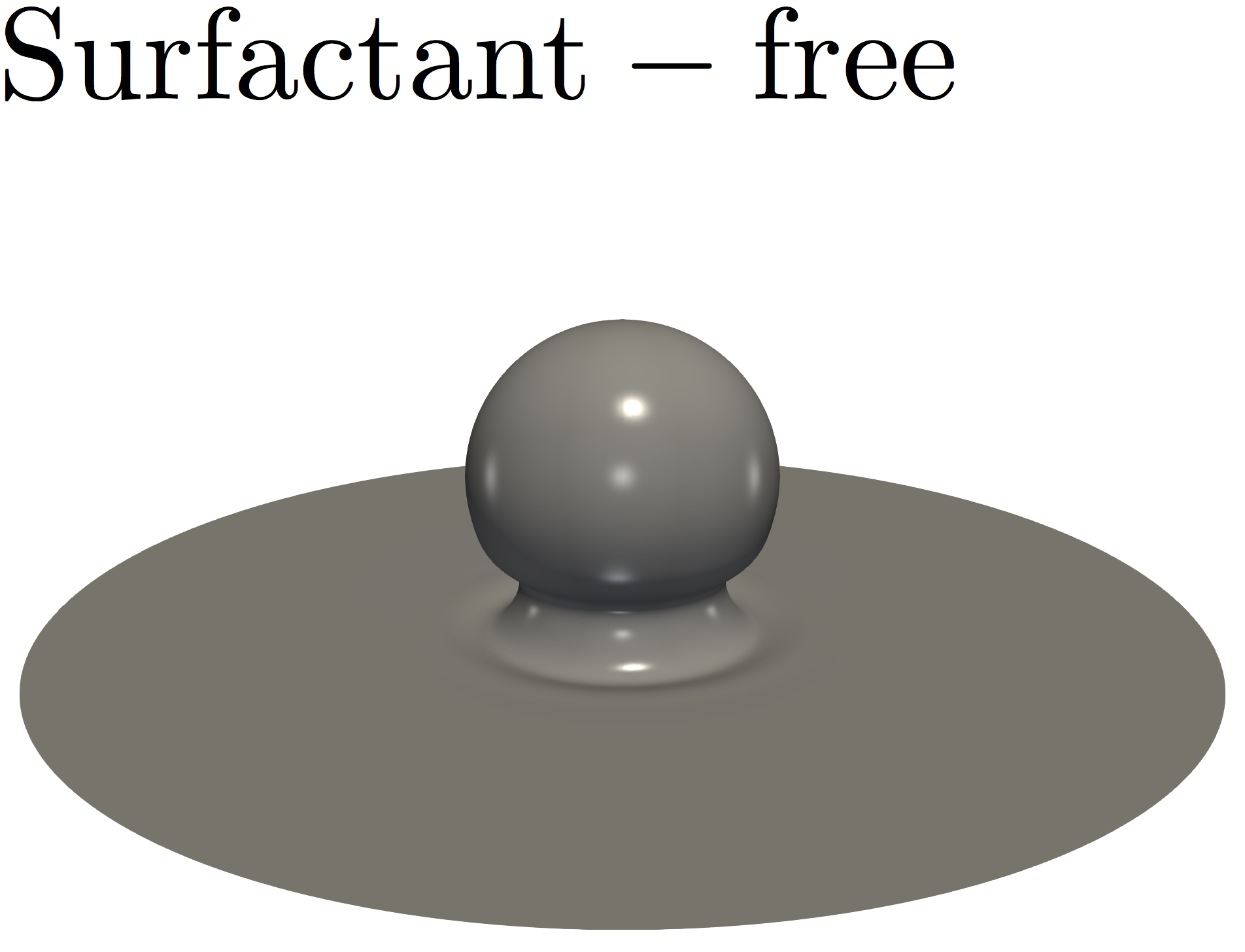}&
\includegraphics[width=0.245\linewidth]{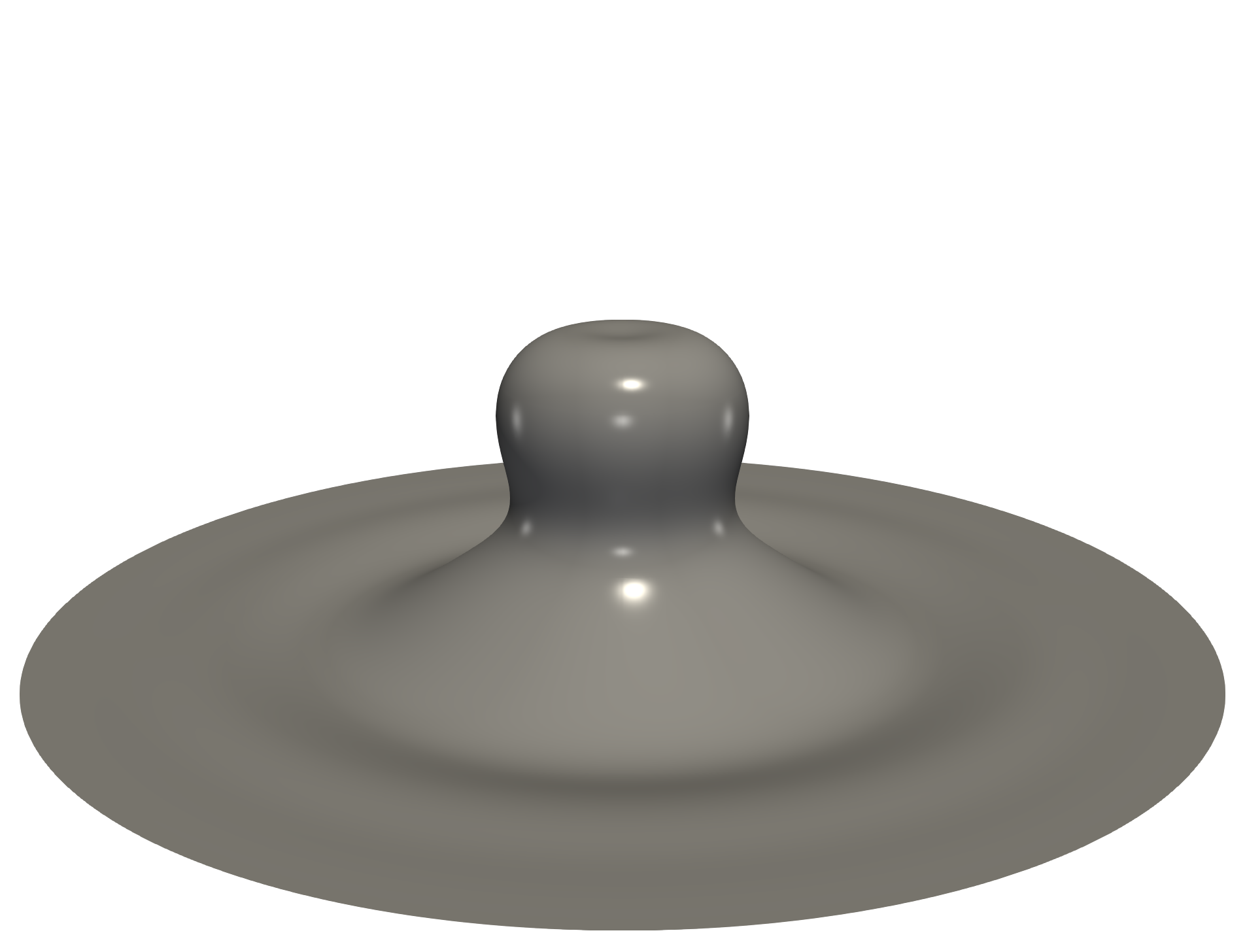}&
\includegraphics[width=0.245\linewidth]{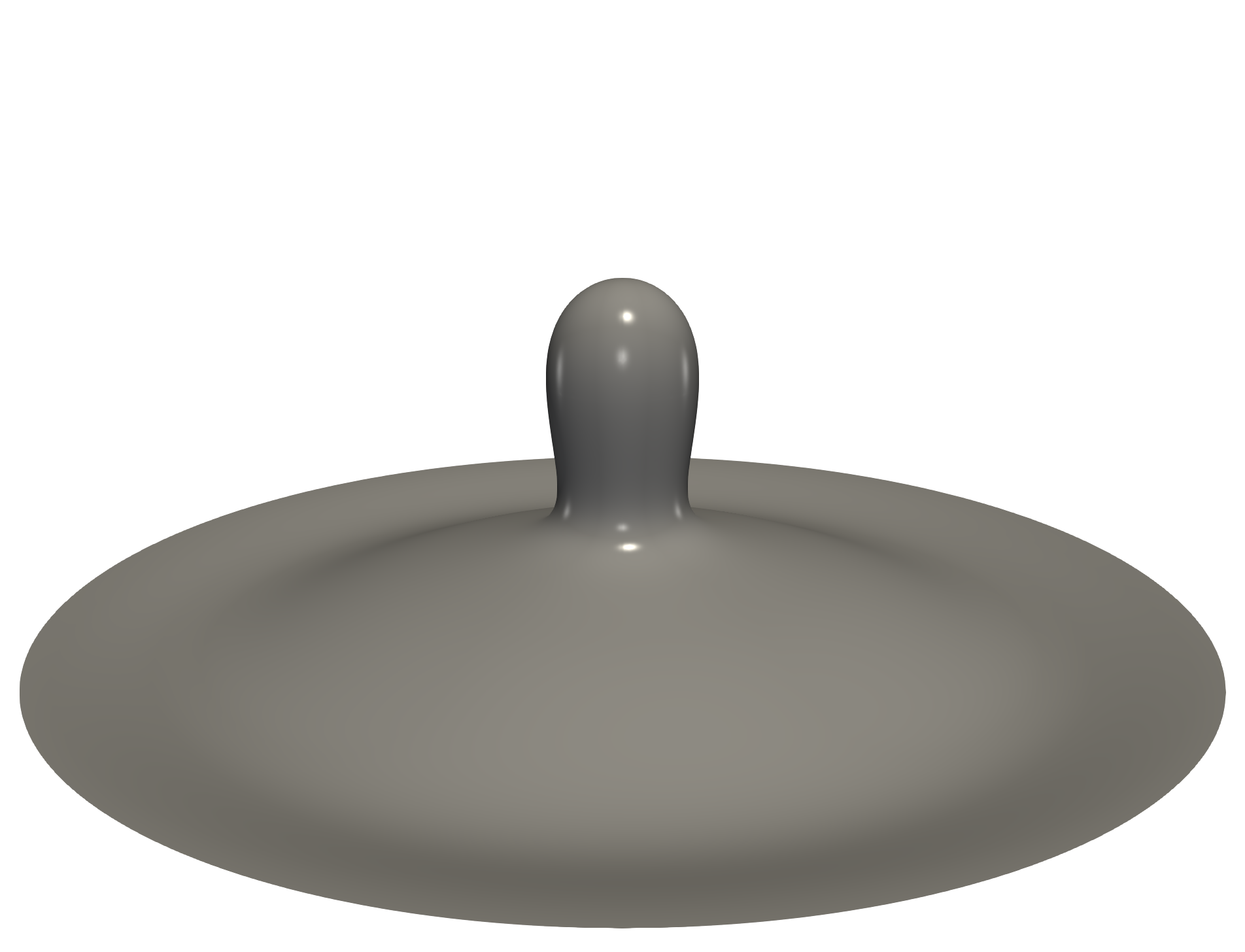}&
\includegraphics[width=0.245\linewidth]{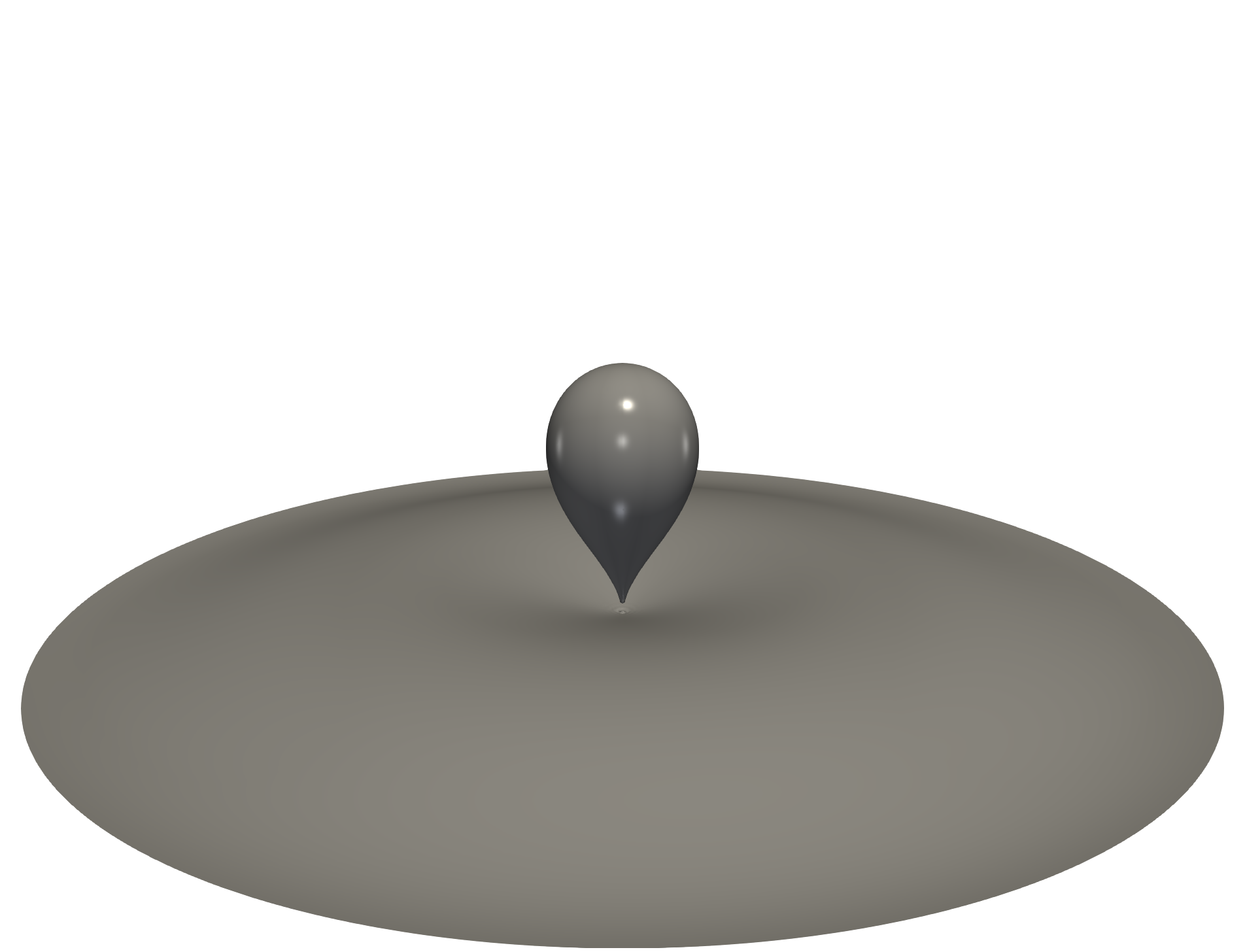}\\
(a) & (b) & (c) & (d)\\
\hline
\includegraphics[width=0.245\linewidth]{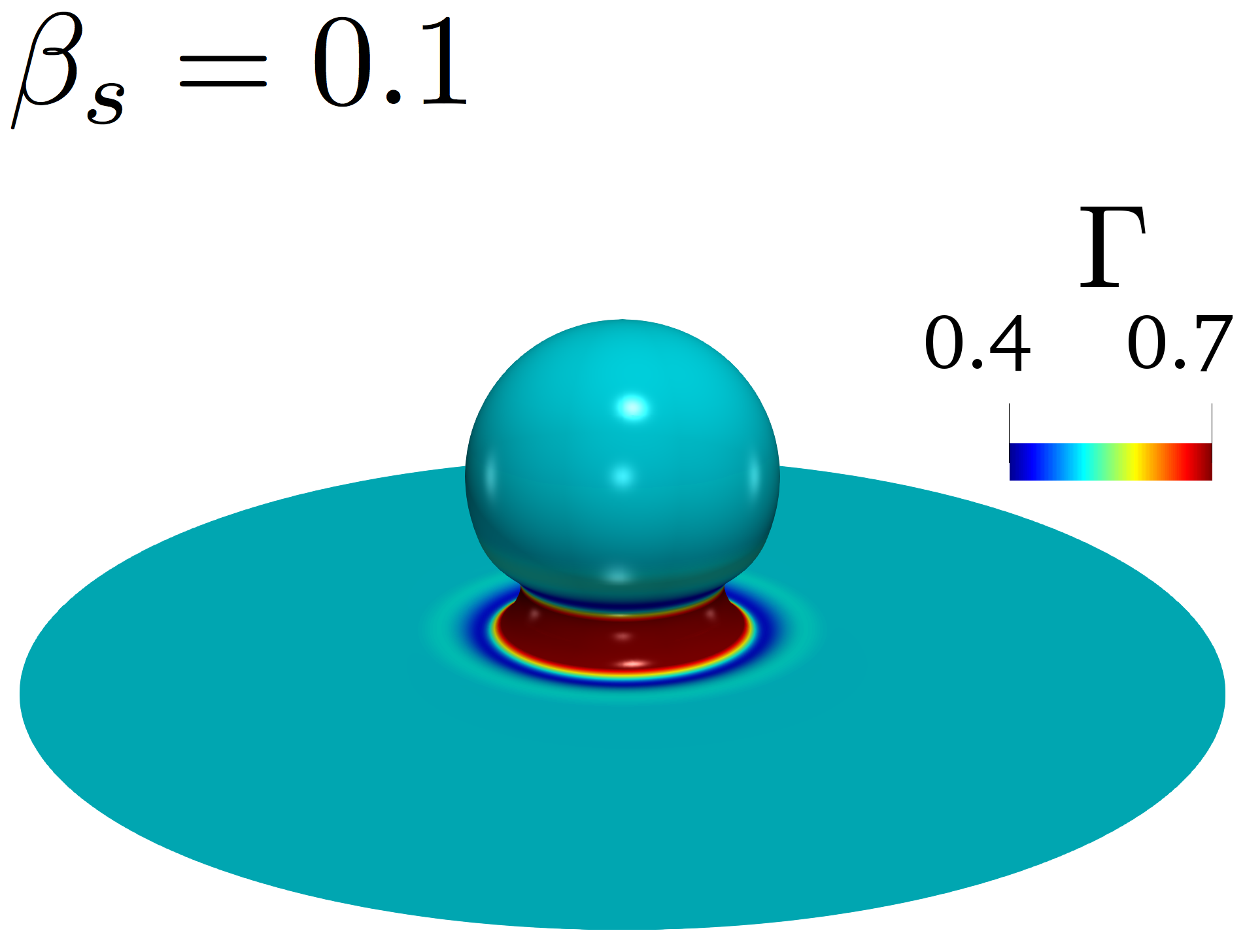}&
\includegraphics[width=0.245\linewidth]{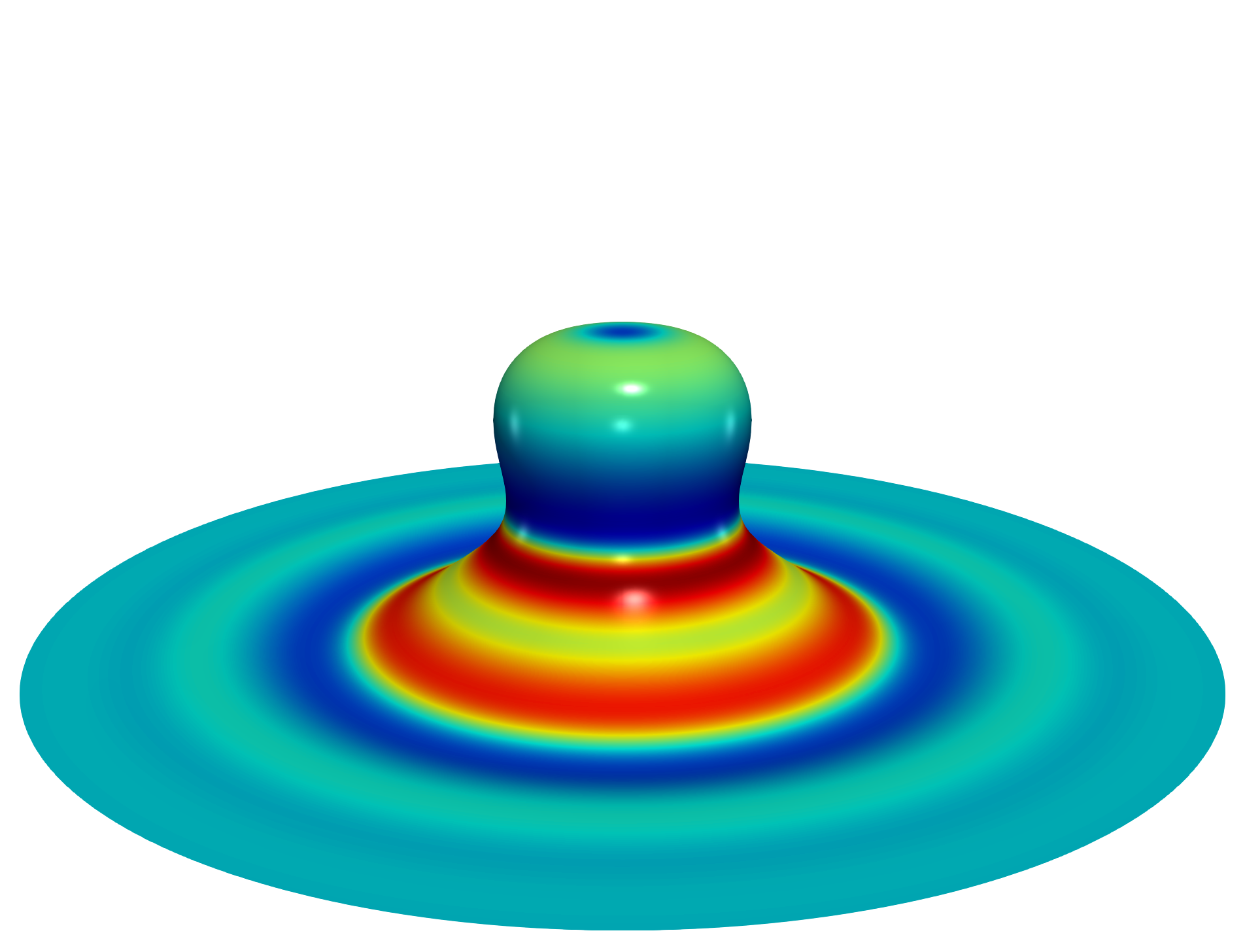}&
\includegraphics[width=0.245\linewidth]{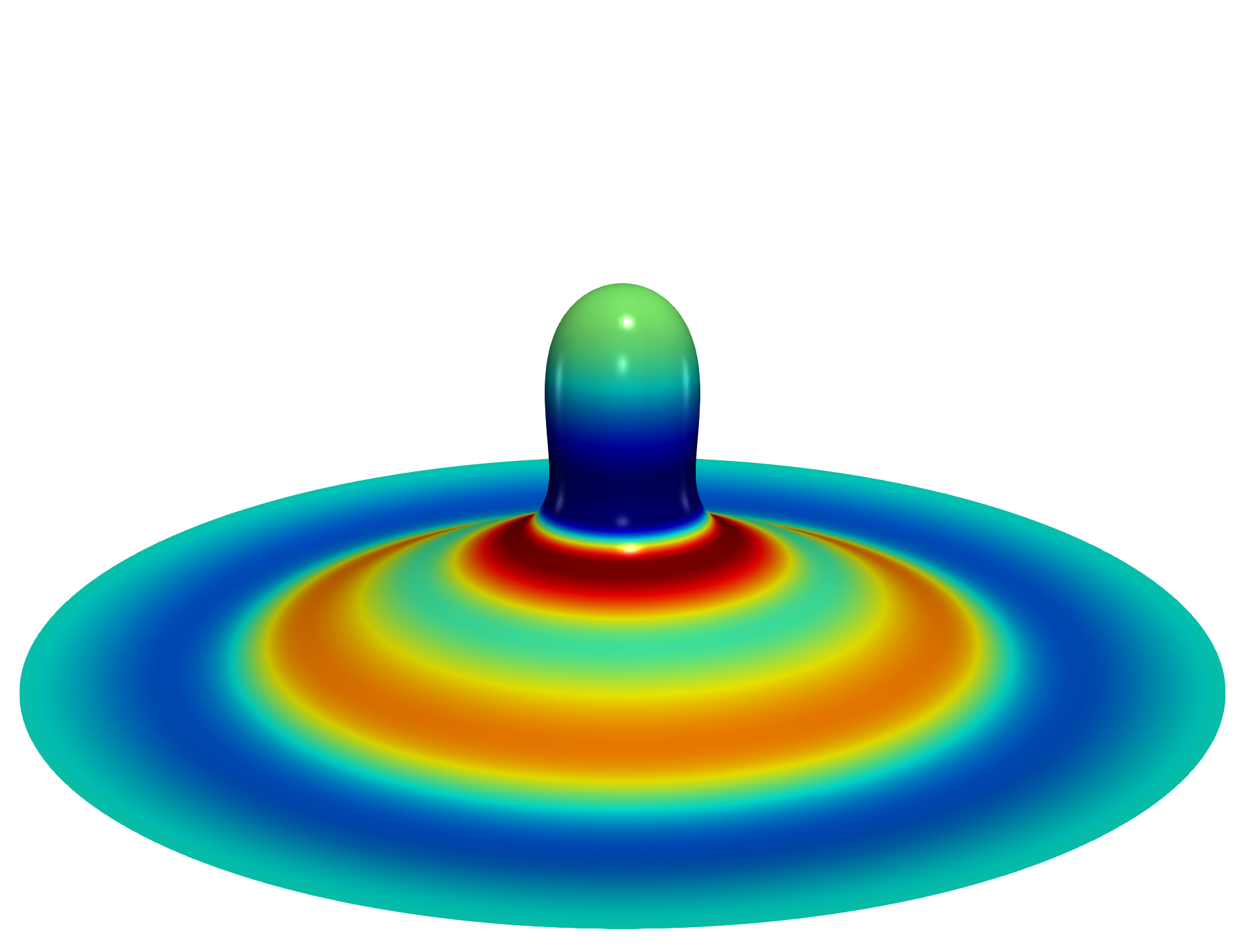}&
\includegraphics[width=0.245\linewidth]{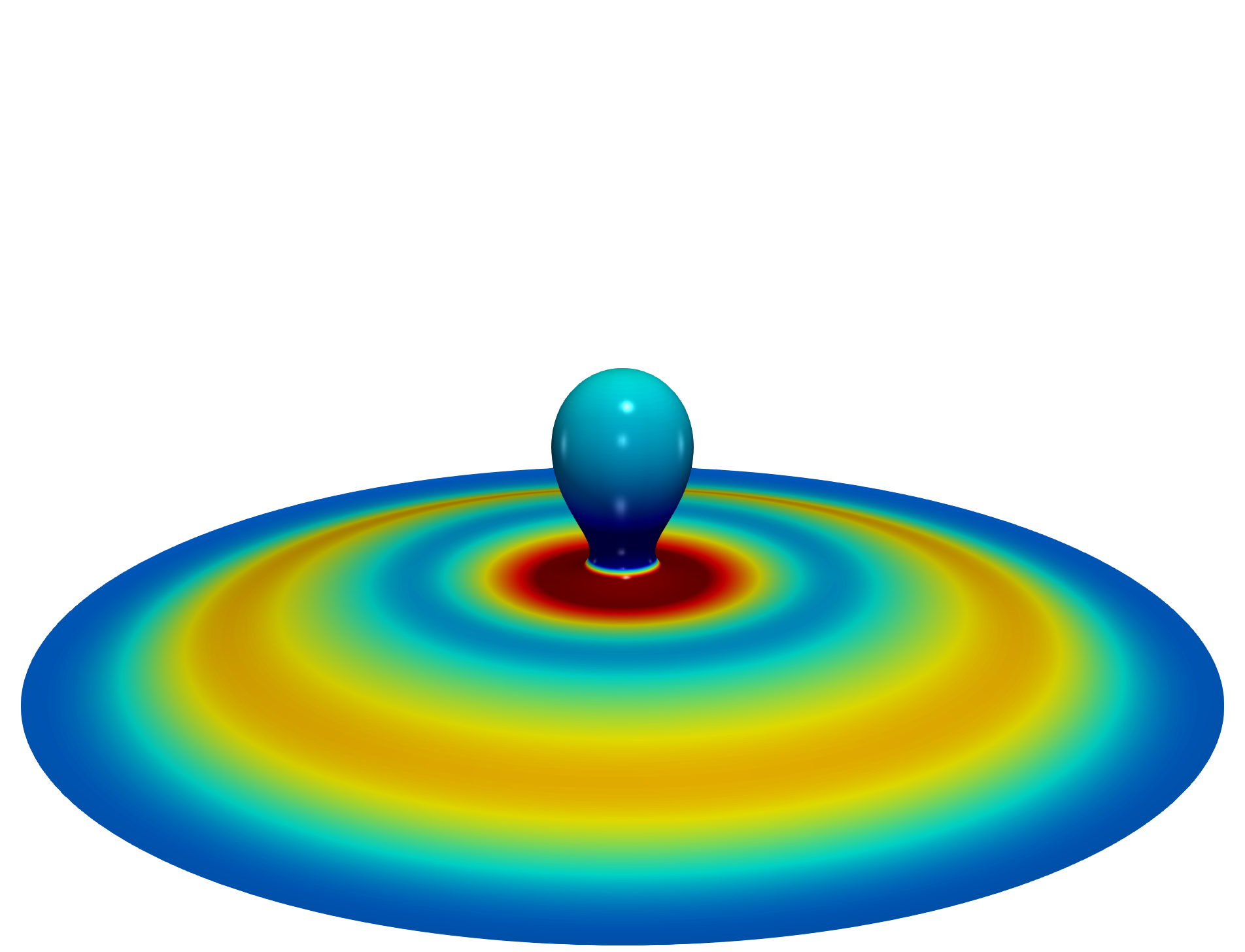}\\
(e) & (f) & (g) & (h)\\
\hline
\includegraphics[width=0.245\linewidth]{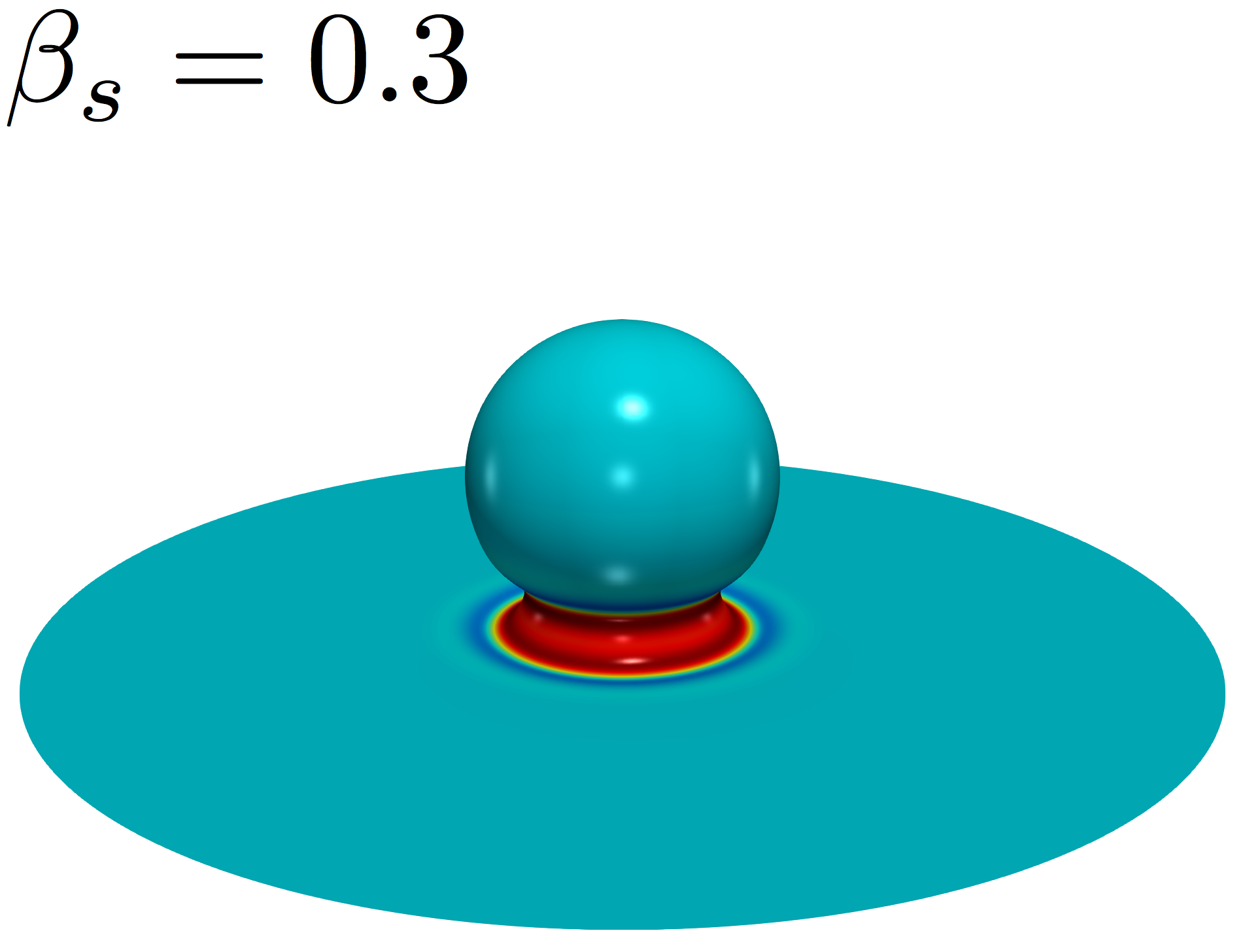}&
\includegraphics[width=0.245\linewidth]{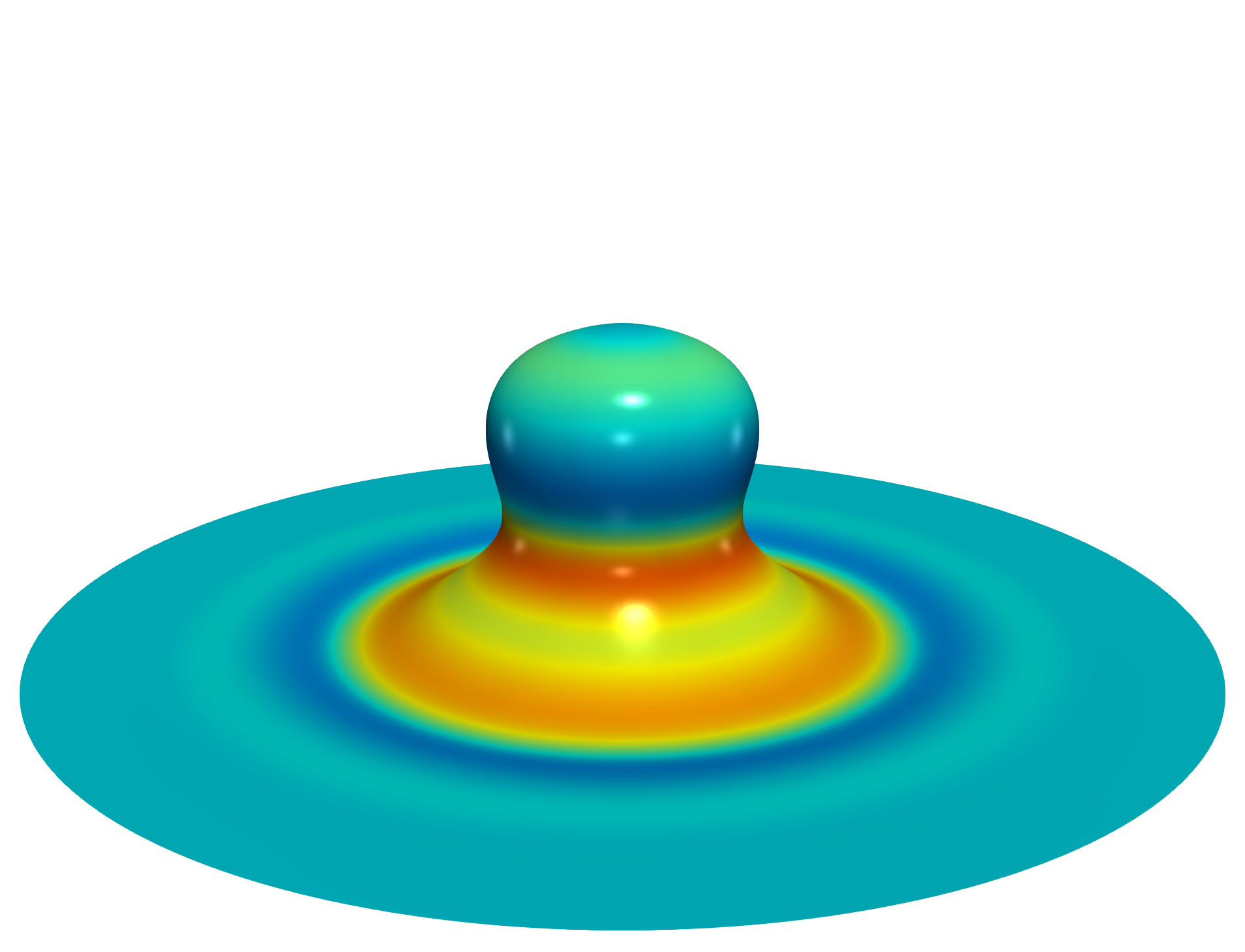}&
\includegraphics[width=0.245\linewidth]{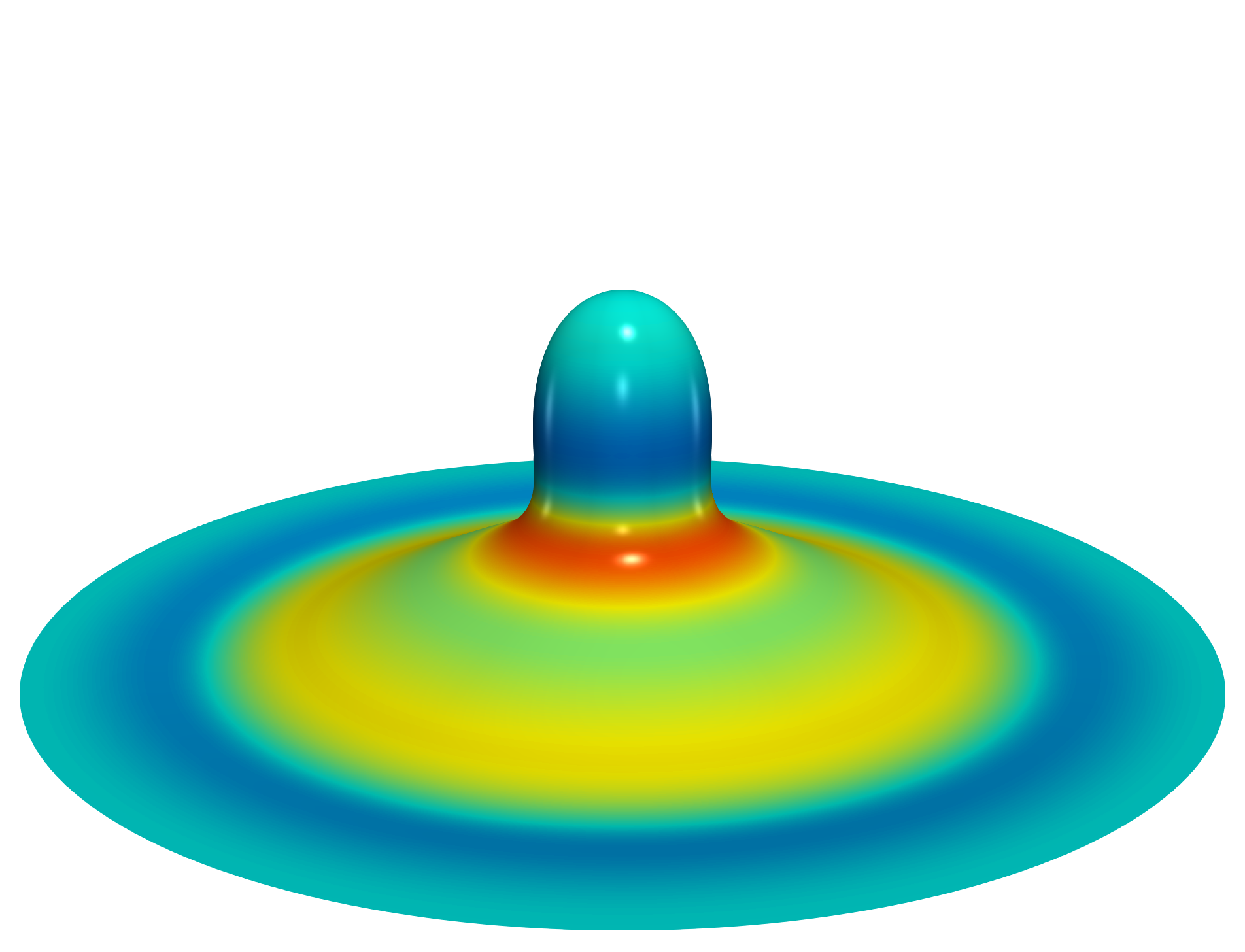}&
\includegraphics[width=0.245\linewidth]{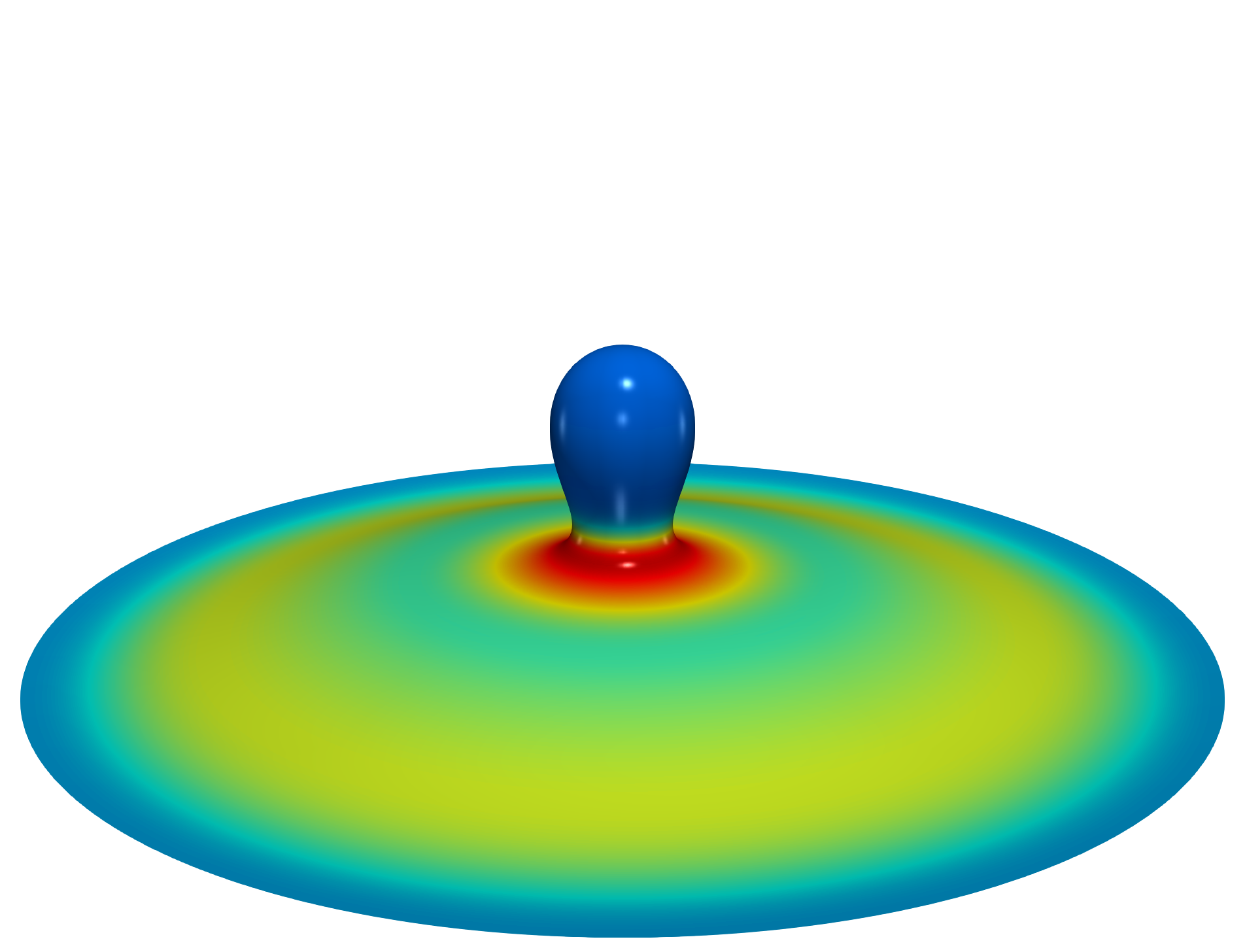}\\
(i) & (j) & (k) & (l)\\
\hline
\includegraphics[width=0.245\linewidth]{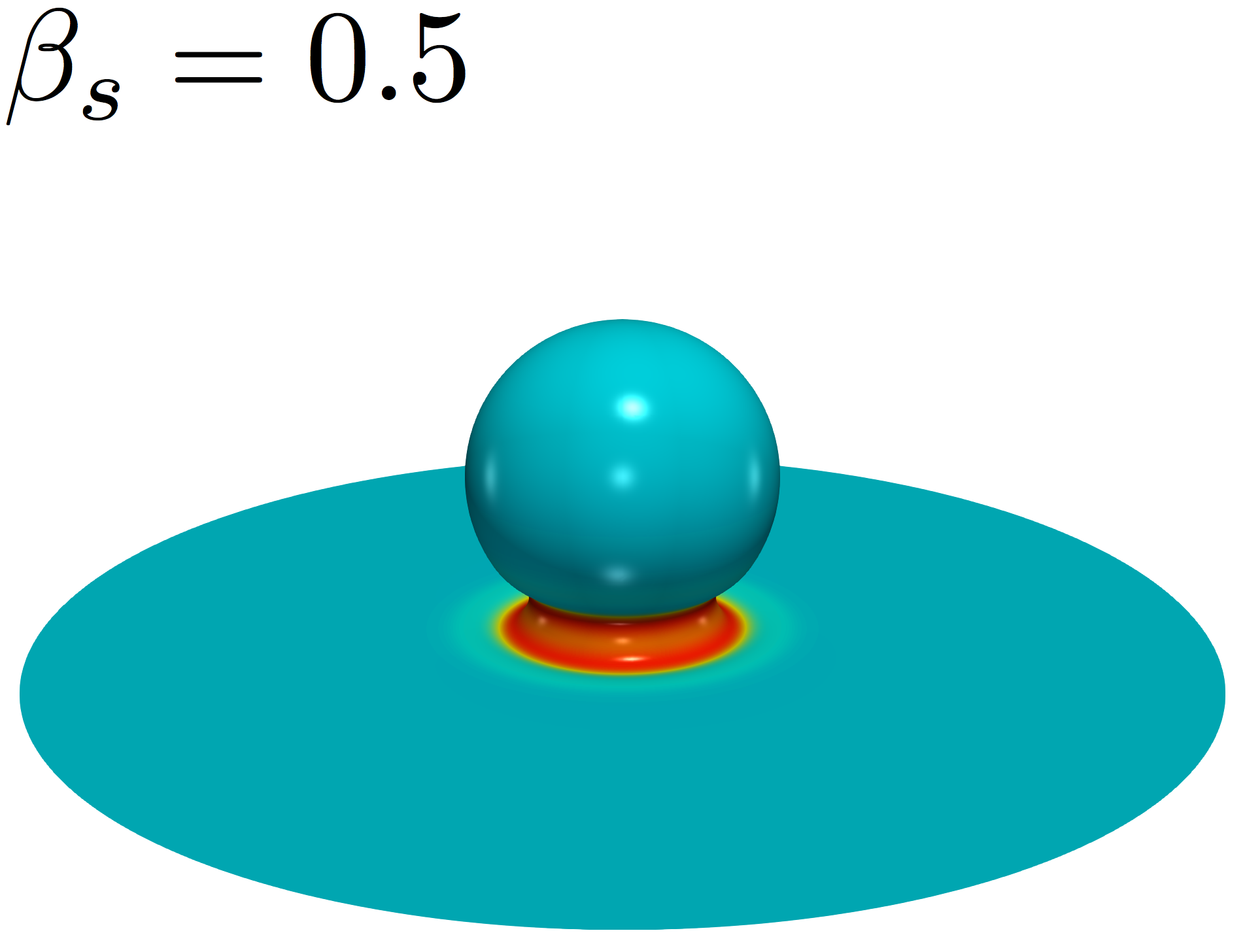}&
\includegraphics[width=0.245\linewidth]{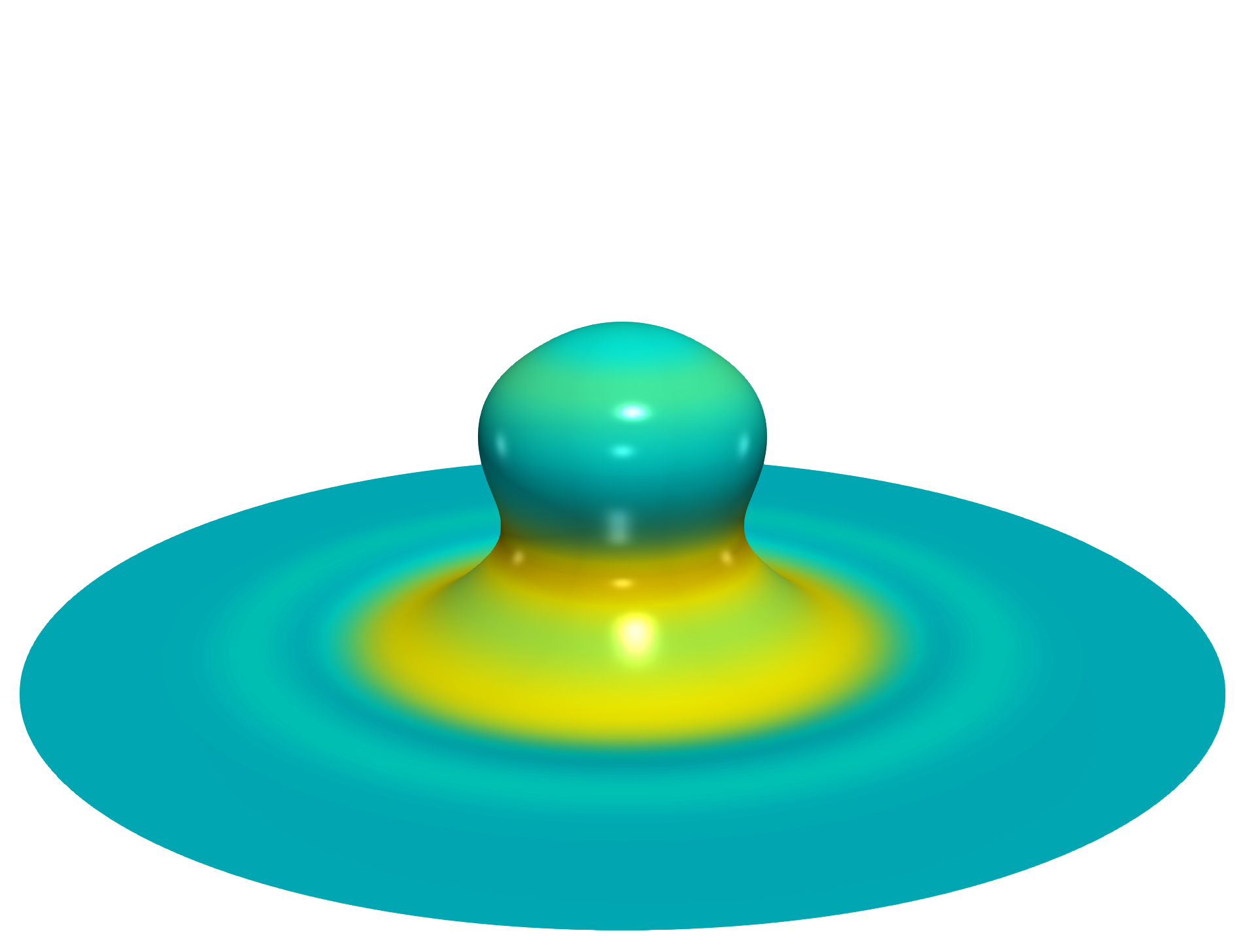}&
\includegraphics[width=0.245\linewidth]{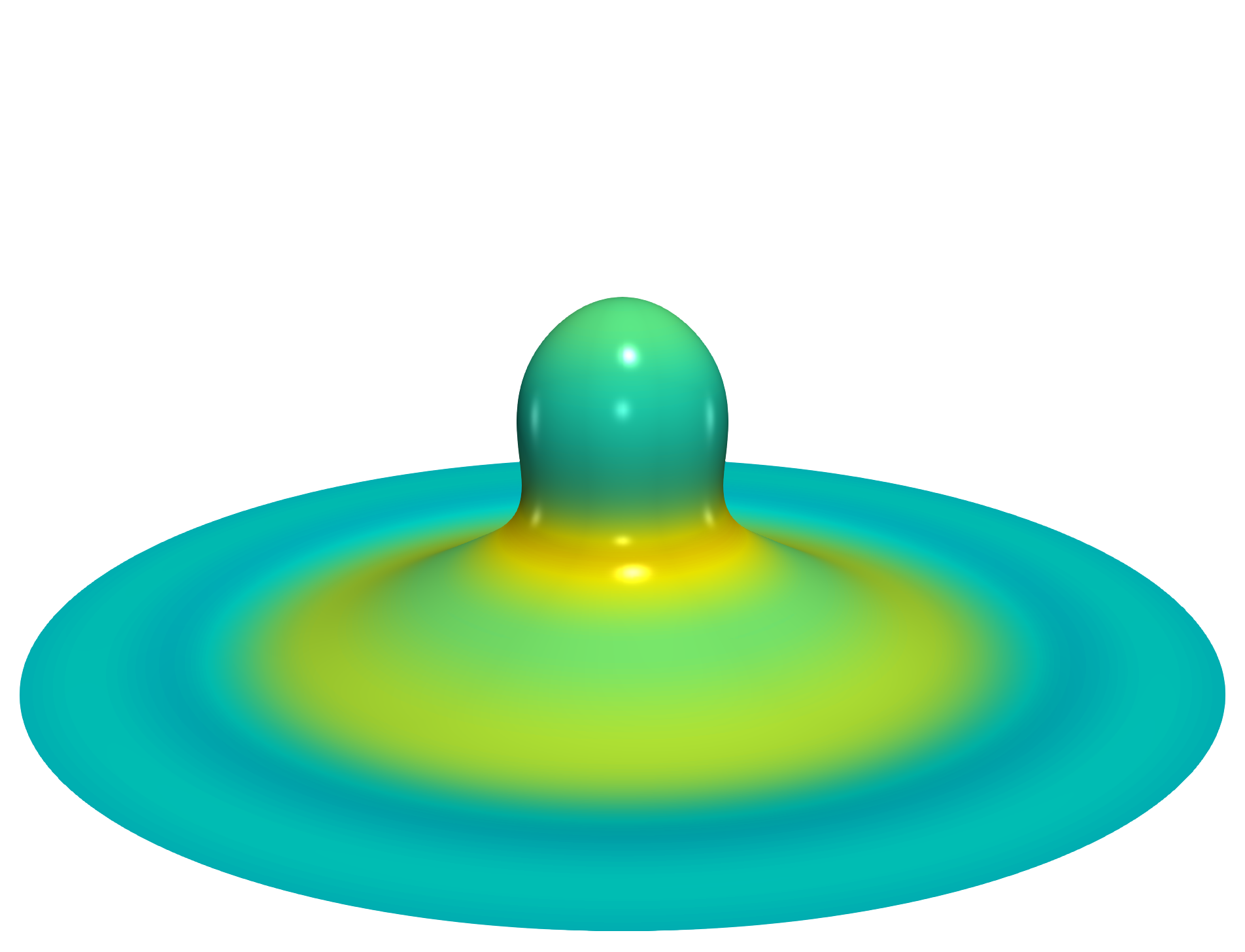}&
\includegraphics[width=0.245\linewidth]{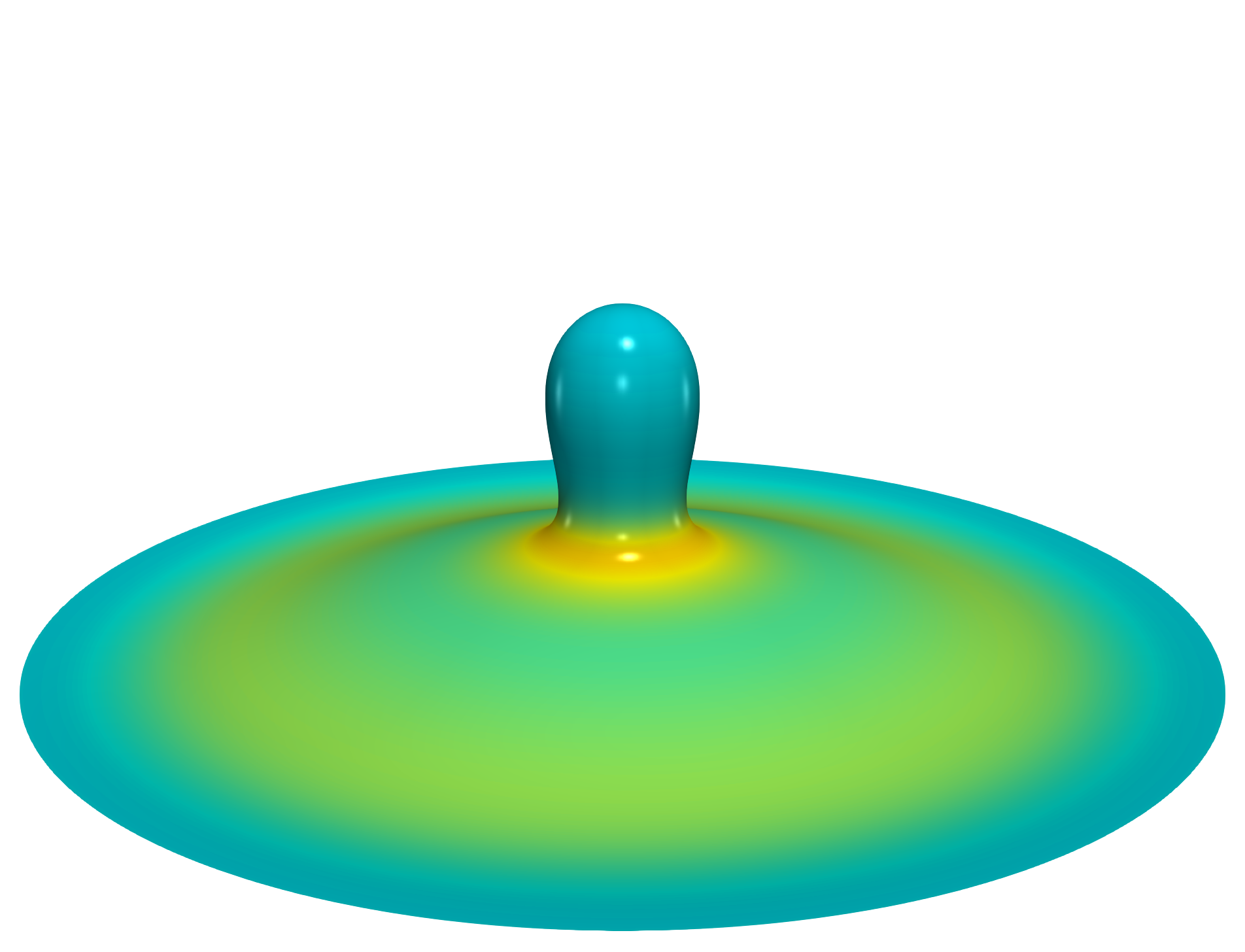}\\
(m) & (n) & (o) & (p)\\
\hline
\end{tabular}
\caption{\label{temporal_evolution} Effect of $\beta_s$ on the drop-interface coalescence dynamics for insoluble surfactants. Spatio-temporal evolution of the three-dimensional interface shape for surfactant-free, (a)-(d), and surfactant-laden coalescence for $\beta_s=0.1$, (e)-(h), $\beta_s=0.3$, (i)-(l),  and $\beta_s=0.5$, (m)-(p). Here, the dimensionless parameters are $Oh=0.02$ and $Bo=10^{-3}$, and for the surfactant-laden cases, $Pe_s=100$ and $\Gamma_o=\Gamma_\infty/2$. The colour indicates the magnitude of $\Gamma$, and legend  is shown in (e).}
\end{figure}

Following the good agreement between the surfactant-free coalescence simulation and the  experimental results of \citet{Blanchette_np_2006}, key surfactant effects will be investigated in this section. We first display our results related to the effect of insoluble surfactant, e.g. NBD-PC \citep{Fallest_njp_2010,Strickland_jfm_2015},
showing ultimately the insights regarding the surfactant-driven-escape from a potential pinch-off singularity such as the one depicted in figure \ref{configuration}b. %  its interfacial singularity. 
We then present the effect of the sorptive kinetics on the phenomenon throughout the use of soluble surfactants, e.g. SDS \citep{Siderius_jsc_2002}. It is also worth mentioning that all surfactant simulations have been carried out until the neck has either pinched off or undergone reopening.
Additionally, we provide with conclusive evidence that the neck-reopening is driven by Marangoni stresses.

%there is a change of the neck dynamics (i.e., either reopening or pinchoff of the neck), 
%as this is the main objective of the present study. 

%%%%%%%%%%%%%%%%%%%%%%%%%%%%%%%%%%%%%%%%%
%%%%%%%%%%%%%%%%%%%%%%%%%%%%%%%%%%%%%%%%%
%%%%%%%%%%%%%%%%%%%%%%%%%%%%%%%%%%%%%%%%%
\subsection{Insoluble surfactants}
\label{sec:insoluble}
%%%%%%%%%%%%%%%%%%%%%%%%%%%%%%%%%%%%%%%%%
%%%%%%%%%%%%%%%%%%%%%%%%%%%%%%%%%%%%%%%%%
%%%%%%%%%%%%%%%%%%%%%%%%%%%%%%%%%%%%%%%%%
We start the discussion of the results by presenting the  effect of the surface-active agents through the analysis of the elasticity parameters $\beta_s$ with $Oh=0.02$, $Bo=10^{-3}$, $Pe_s=100$, and $\Gamma_o=\Gamma_\infty/2$. Figure \ref{temporal_evolution} shows the spatio-temporal interfacial dynamics for the surfactant-free and surfactant-laden cases as a function of the  elasticity parameter. At the early stages of the dynamics, the neck expands as a result of the capillary retraction of the liquid bridge which separates the drop from the underlying liquid pool. The capillary retraction gives rise to the formation of capillary waves that travel upwards towards the drop summit. As pointed out by  \citet{Blanchette_np_2006}, the oscillations caused by the travelling capillary waves yield vertical stretching forming a nearly-cylindrical drop, as shown in figure \ref{temporal_evolution}c, before capillarity acts to drive  the dynamics towards a more energy-favourable state by pulling on the sides of the drop. % (e.g., horizontal pulls). 
This capillary action leads to pinch-off of the liquid bridge via a singularity which culminates in the formation of a  secondary droplet; this, in turn,  follows a `cascade of coalescence events' until the coalescence process is completed as also shown by \cite{Thoroddsen_pof_2000}, \cite{Blanchette_np_2006}, and \cite{Blanchette_jfm_2009}.  

For all surfactant-laden cases, the generation of a secondary droplet is %successfully 
avoided even for the lower end of the elasticity parameter range. For $\beta_s=0.1$, the dynamics follow closely those of the surfactant-free case, where significant vertical stretching of the original droplet is observed (see figures \ref{temporal_evolution}e-h). At the point where surface tension is expected to dominate the narrowing of the neck, the presence of non-uniform surfactant concentration generates Marangoni stresses that change the outcome of the dynamics. By increasing $\beta_s$, surfactant redistribution along the interface is enhanced as displayed in figures \ref{temporal_evolution}i-l and figures \ref{temporal_evolution}m-p, for $\beta_s=0.3$ and $\beta_s=0.5$, respectively. The surfactant concentration gradients, and associated transport, is seen to suppress the capillary waves and to limit the vertical stretch of the drop.

\begin{figure}
\begin{center}
\begin{tabular}{ccc}
\includegraphics[ width=0.33\linewidth]{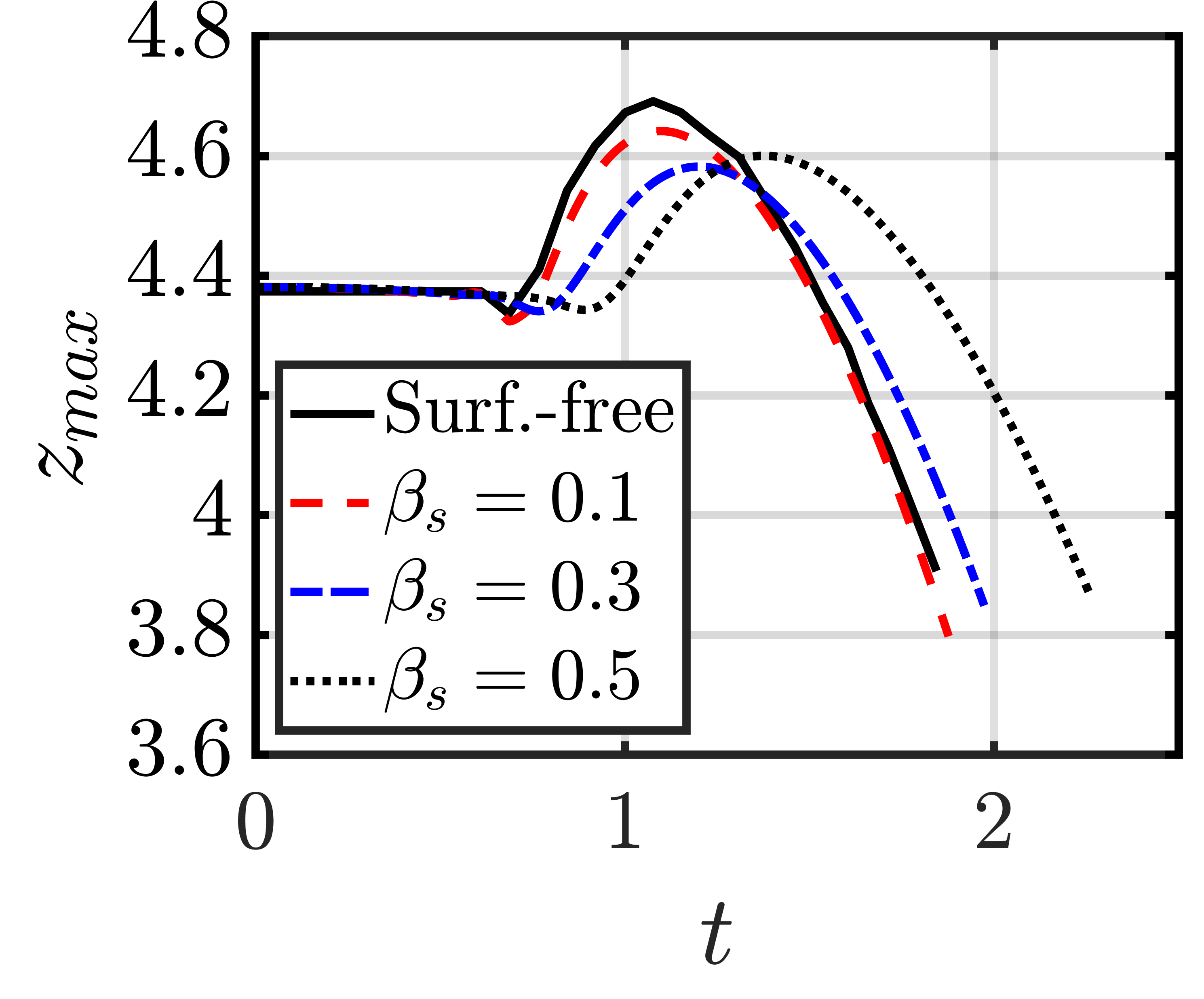}&
\includegraphics[ width=0.33\linewidth]{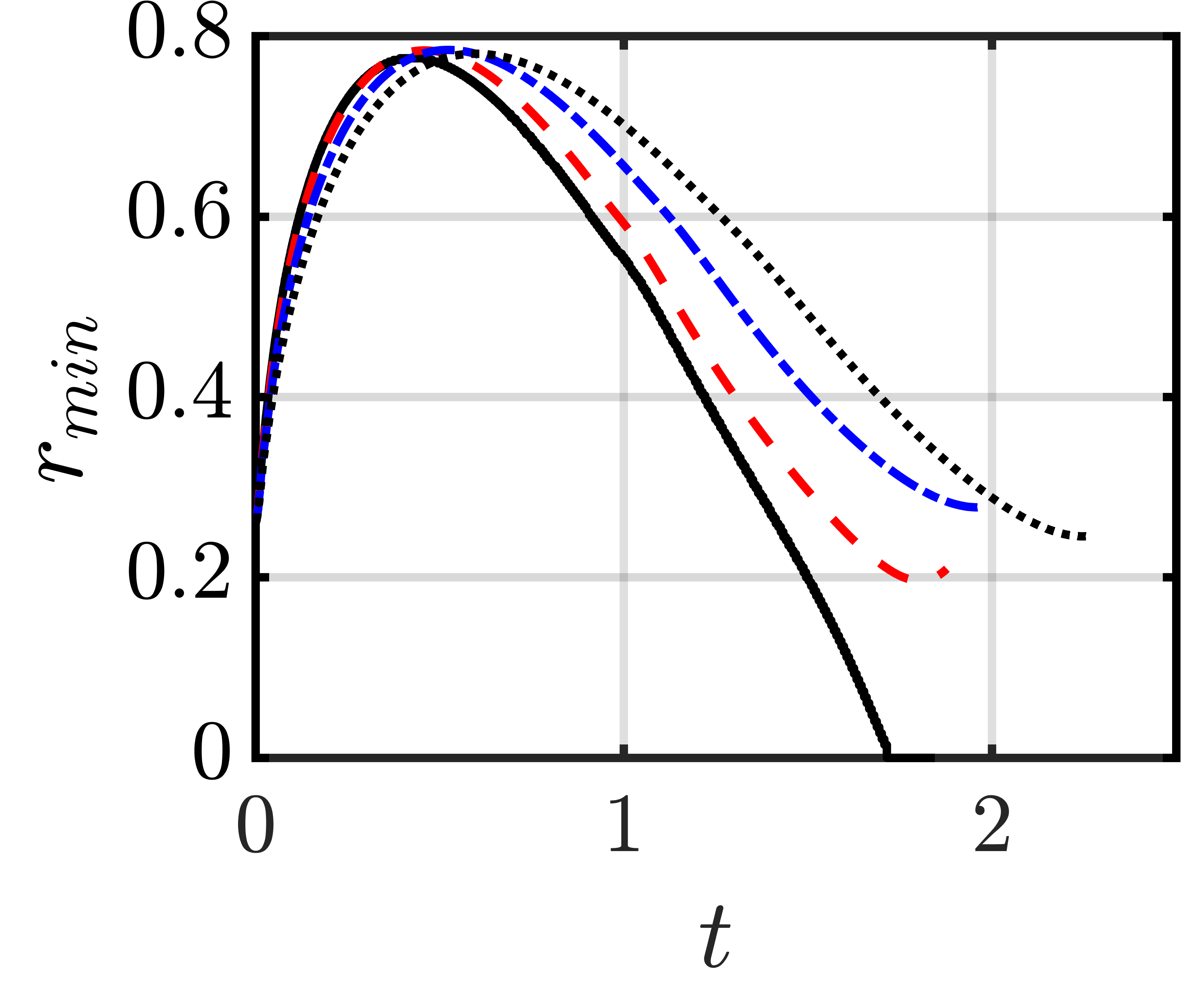}&
\includegraphics[ width=0.33\linewidth]{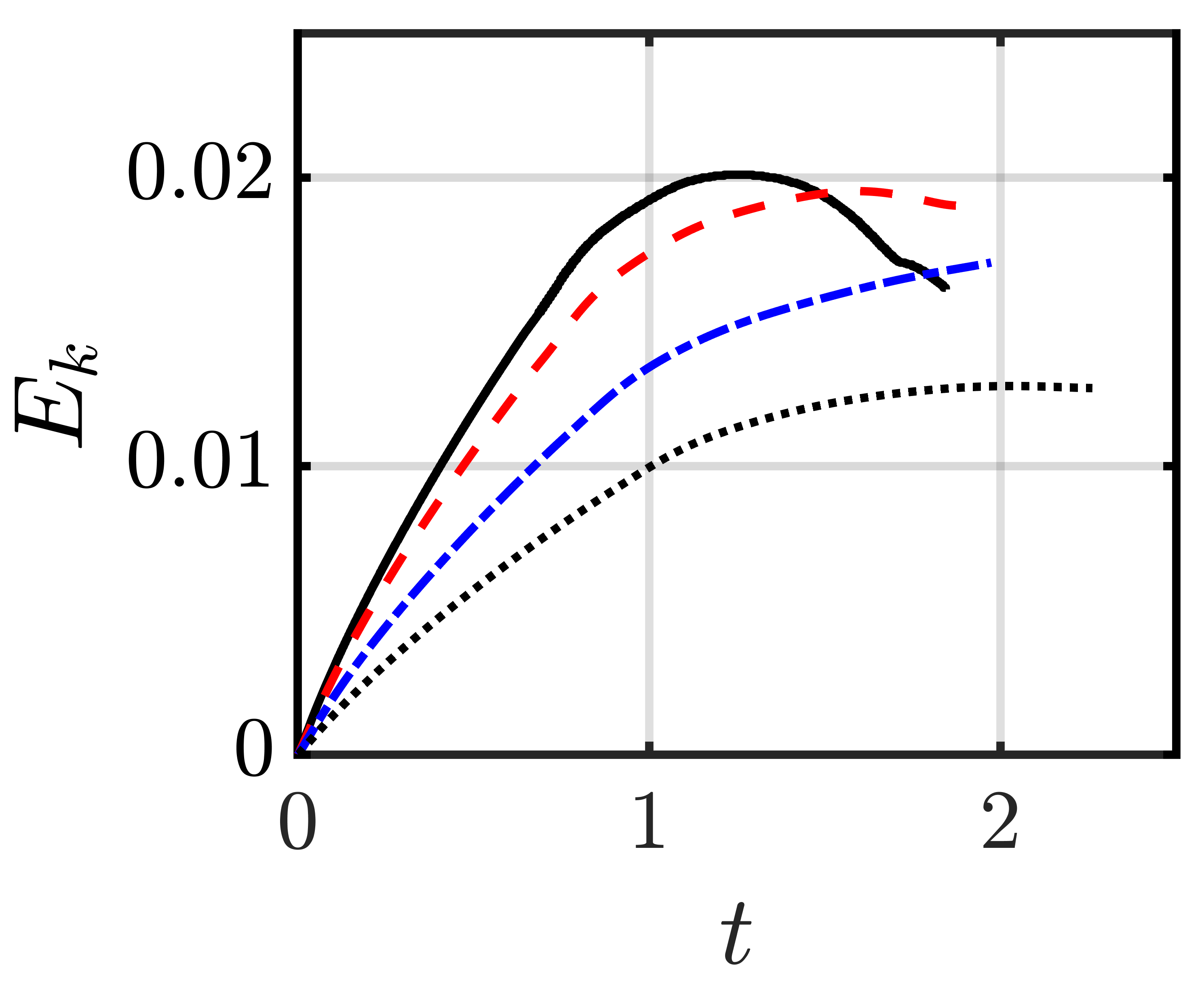}\\
 (a) & (b) & (c)
\end{tabular}
\end{center} 
\caption{\label{metrics_beta}Effect of $\beta_s$ on the temporal dynamics of the vertical extent of the drop, (a), its minimum neck radius, (b), and the system kinetic energy $E_k$, (c), for $Oh=0.02$, $Bo=10^{-3}$, $Pe_s=100$, and $\Gamma_o=\Gamma_\infty/2$.} 
\end{figure}

Figure \ref{metrics_beta} shows the immobilising effect brought about by the presence of surfactants through the analysis of the temporal dynamics of the maximum vertical stretch of the droplet, $z_{max}$, the neck radius, $r_{min}$, and the kinetic energy, $E_k=\int_V (\rho \textbf{u}^2/2) dV$. Here, the $E_k$ values have been normalised by the  surface energy $E_s=S \sigma_s$, where $S$ is the initial superficial area of the system. 
The evidence for damping of the upward drop oscillation can be seen in figure \ref{metrics_beta}a. Here, the increase in $\beta_s$ is seen to depress the maximum crest location of the drop. These observations confirm the expectations of \citet{Martin_pof_2015} of suppression of the axial oscillation with an increase in the surfactant strength (though these authors were only able to run simulations for $\beta_s \leq 0.2$). Interestingly, the temporal evolution of $z_{max}$ exhibits a non-monotonic dependence on  $\beta_s$, with the most suppressed crest being associated with the intermediate value of $\beta_s=0.3$. The physical explanation of this outcome will be provided in the discussion of figure \ref{beta_flow_fields} below. Furthermore, investigation of the temporal variation of the minimum neck radius, $r_{min}$ (see figure \ref{metrics_beta}b), confirms  neck reopening for all surfactant-laden cases, with a 50\% rise in the time associated with the onset of re-opening, $t_r$, corresponding to an increase in $\beta_s$ from 0.1 to 0.5. %where the difference between the time associated with $r_{min}$  reversal between lowest and highest $\beta_s$ parameters is nearly $50\%$. 
The non-monotonic dependence on $\beta_s$ is also exhibited by $t_r$: 
%Another non-monotonic response is observed 
%in relation to the neck reopening point. 
even though the longest delay in neck closure is observed for the highest $\beta_s$ studied, %Marangoni parameter, 
the $t_r$ value %the reopening deflection point 
for $\beta_s=0.3$ is associated with the largest $r_{min}$. 
Further quantification of these physical phenomena is provided below. 

Finally, % figure \ref{metrics_beta}c shows the kinetic energy, %which is defined as $E_k=\int_V (\rho \textbf{u}^2/2) dV$ %(e.g., calculated in the entire system as $\rho$ is defined by a Heaviside function). %and normalised with the initial surface energy $E_s=S\sigma_s$, where $S$ is the superficial area of the interface.  
inspection of the kinetic energy plots shows that the presence of surfactants induces a monotonically-decreasing overall value of $E_k$ with $\beta_s$ over the range in time encompassing the creation of the cylindrically-shaped drop (see figure \ref{metrics_beta}c). This reduction of the kinetic energy is a result of the rigidification of the interface brought on by the tangential  Marangoni stresses in agreement with \citet{Asaki_prl_1995}. It is evident, however, that for the surfactant-free case, $E_K$ decreases rapidly, as the drop breaks up via neck pinch-off, eventually dipping below those associated with  $\beta_s=0.1$ and $\beta_s=0.3$.

\begin{figure}
\begin{center} 
\begin{tabular}{cccc}
\includegraphics[ width=0.25\linewidth]{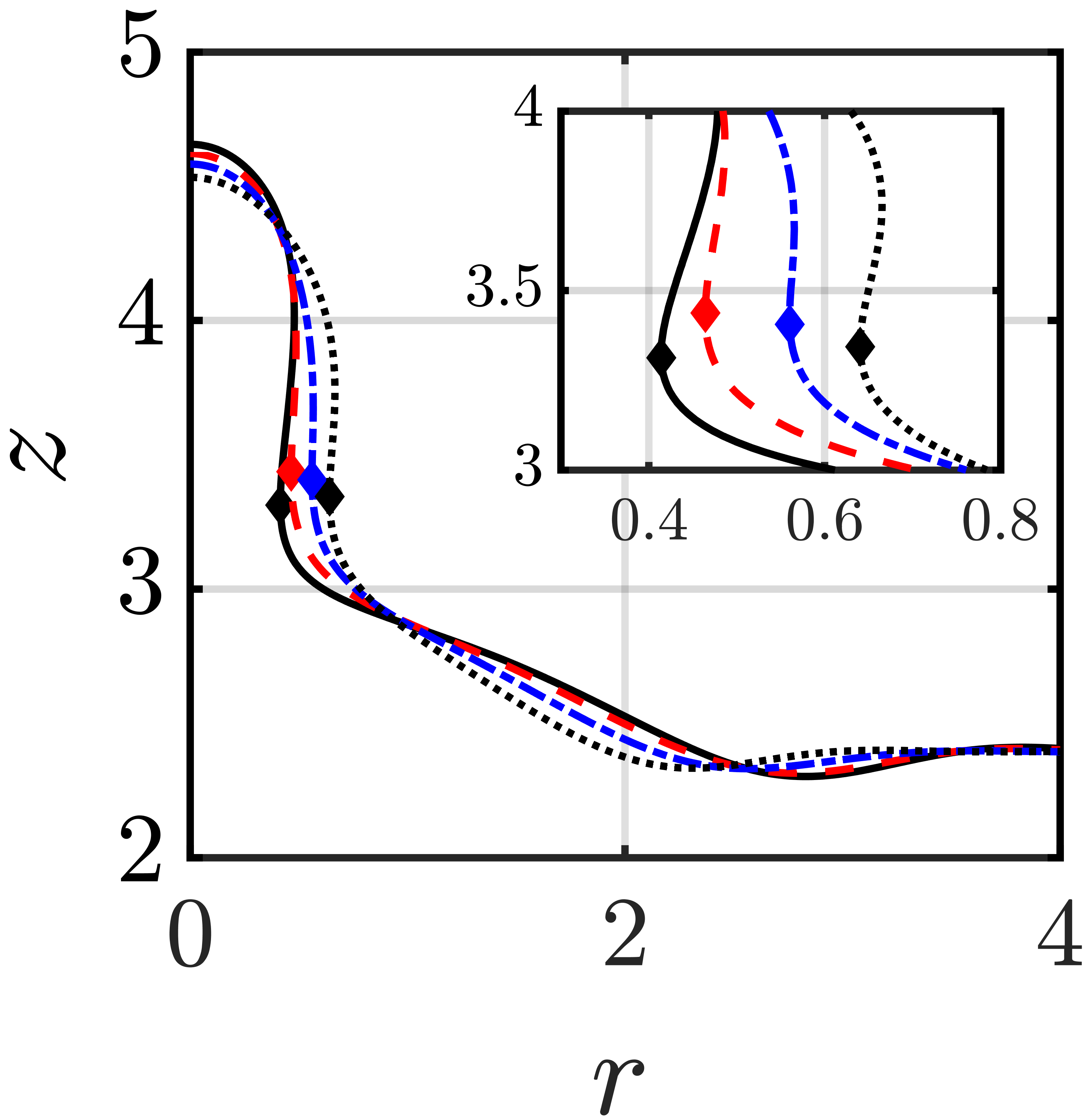}& 
\includegraphics[ width=0.25\linewidth]{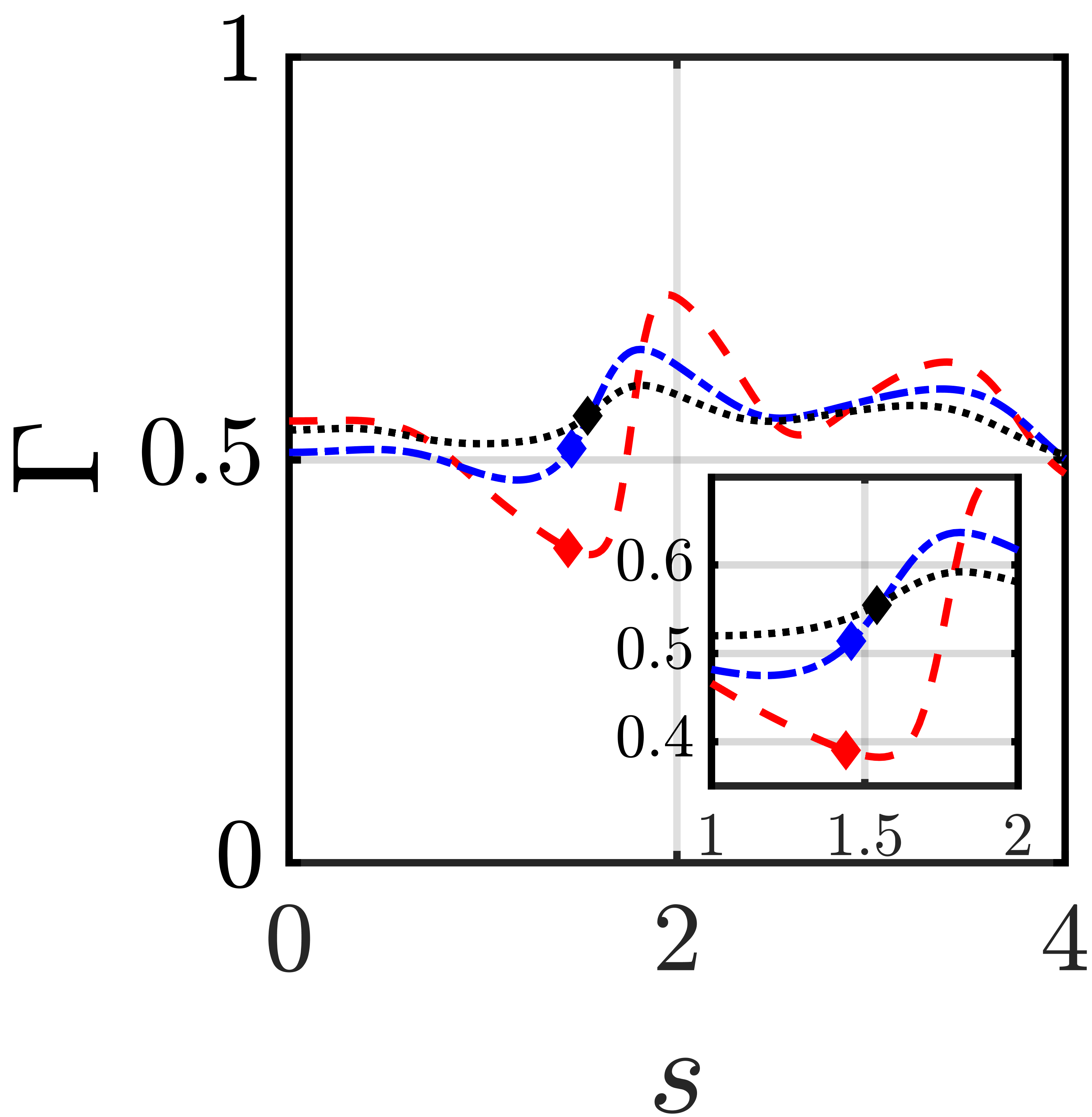}& 
\includegraphics[ width=0.25\linewidth]{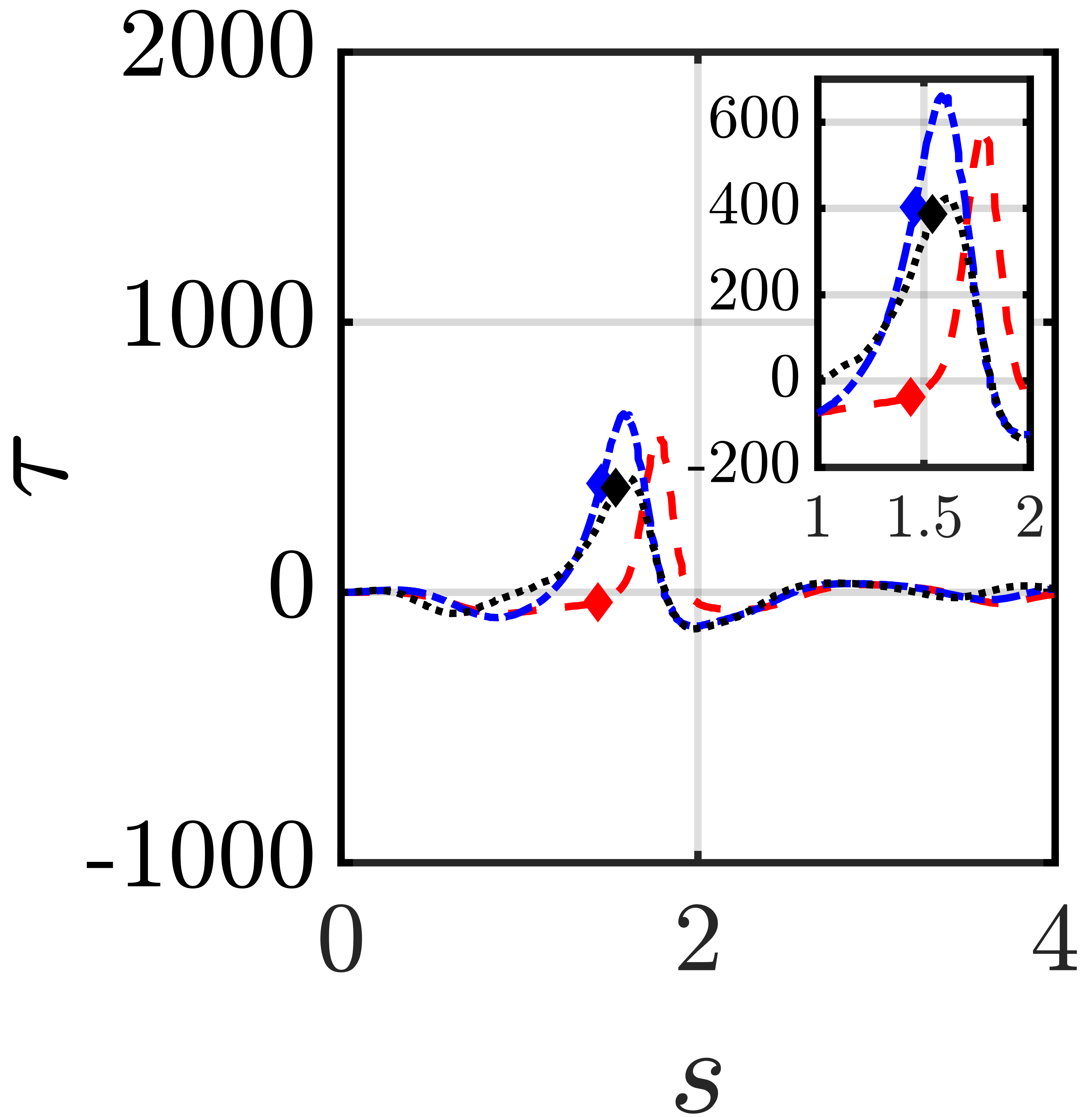}& 
\includegraphics[ width=0.25\linewidth]{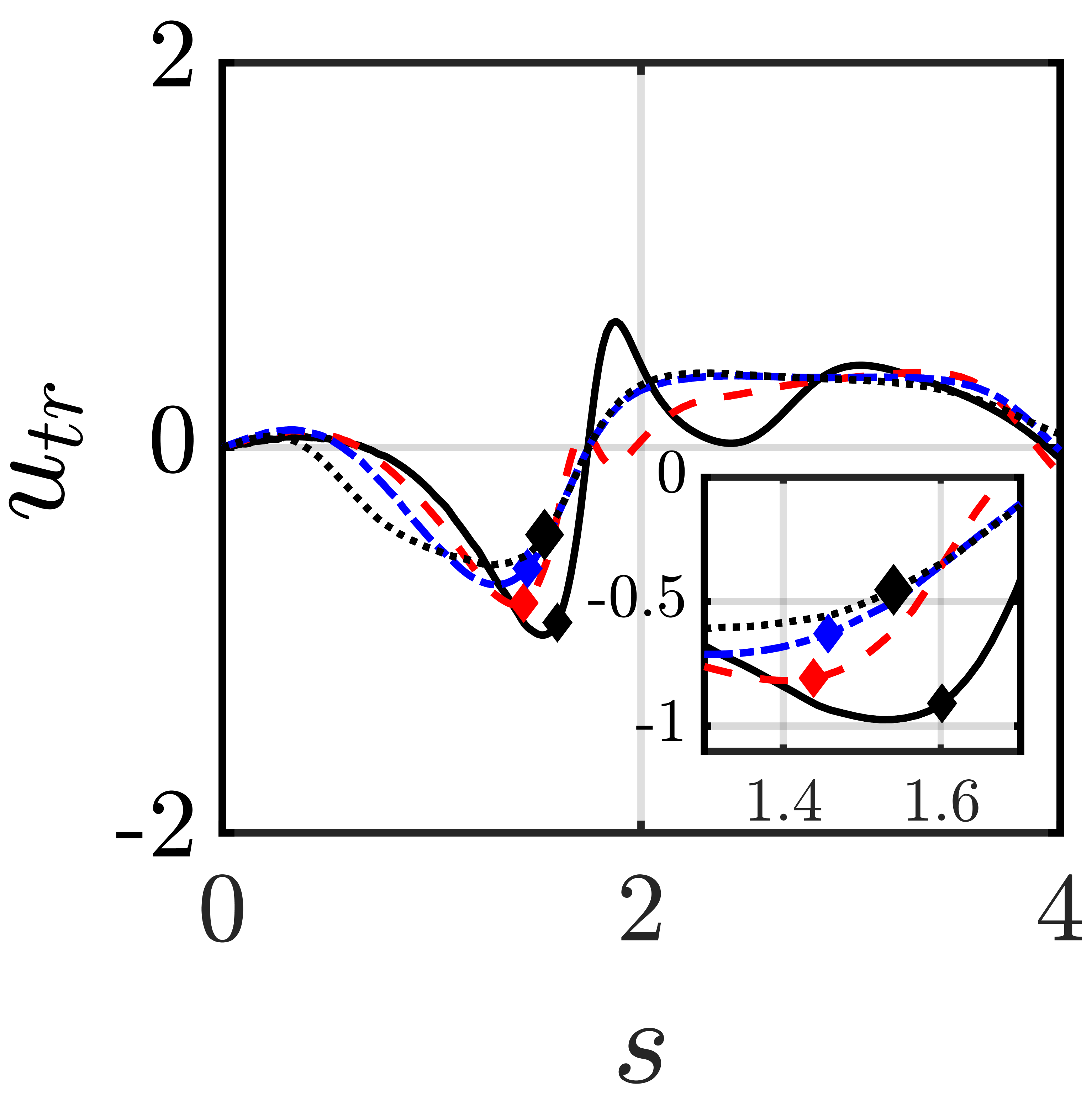}\\ 
(a) & (b) & (c) & (d)\\
\includegraphics[ width=0.25\linewidth]{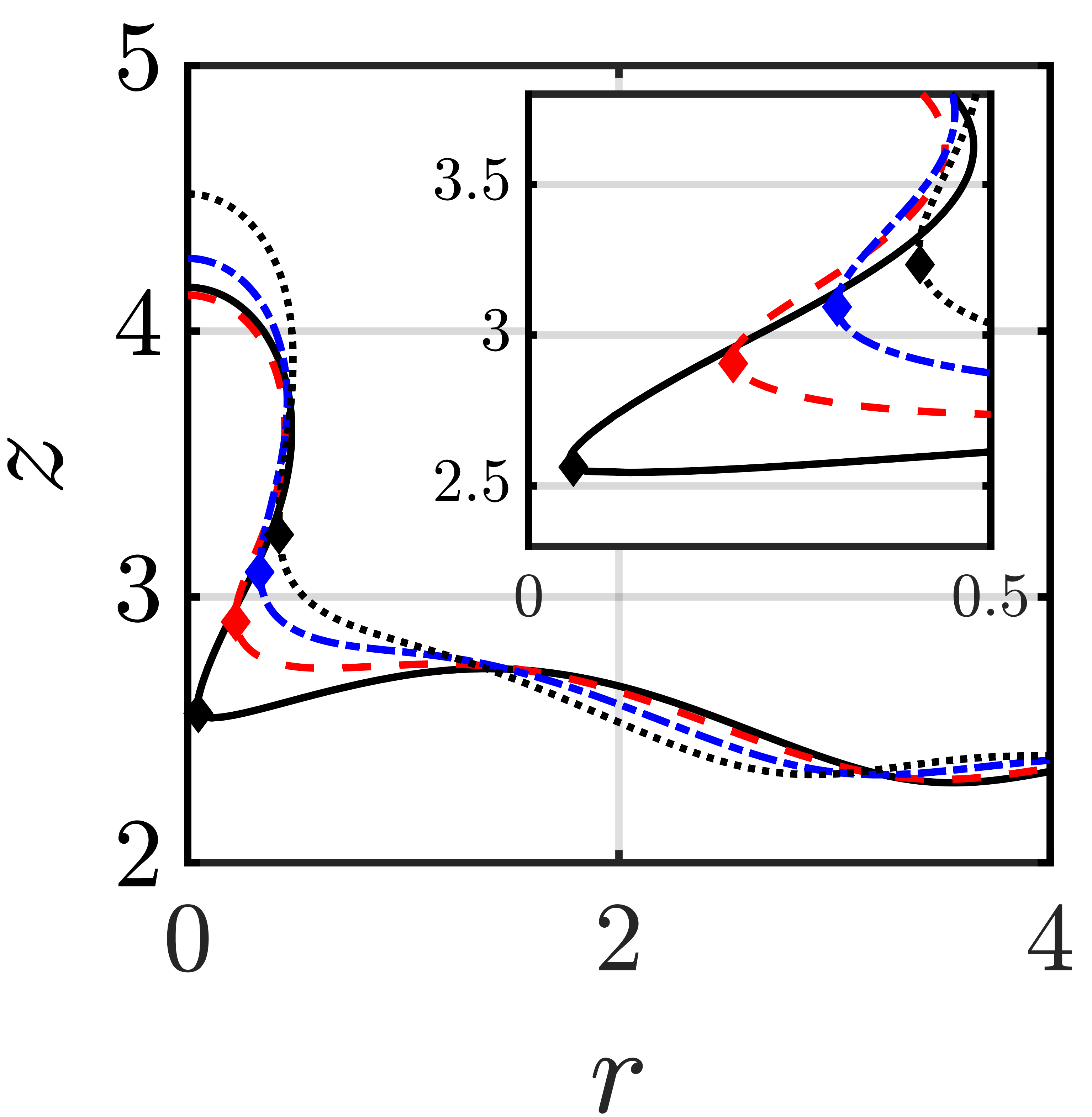}& 
\includegraphics[ width=0.25\linewidth]{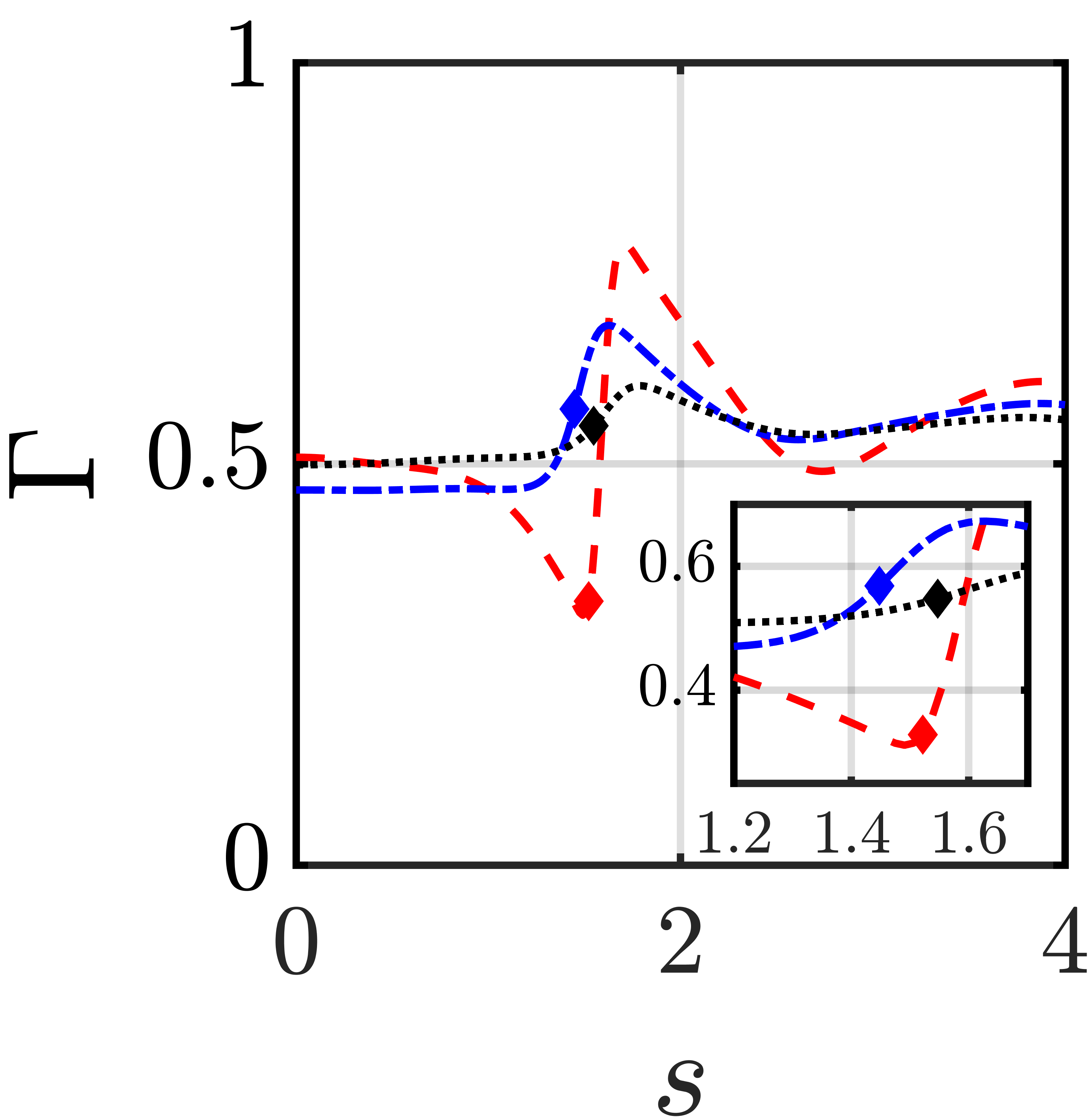}& 
\includegraphics[ width=0.25\linewidth]{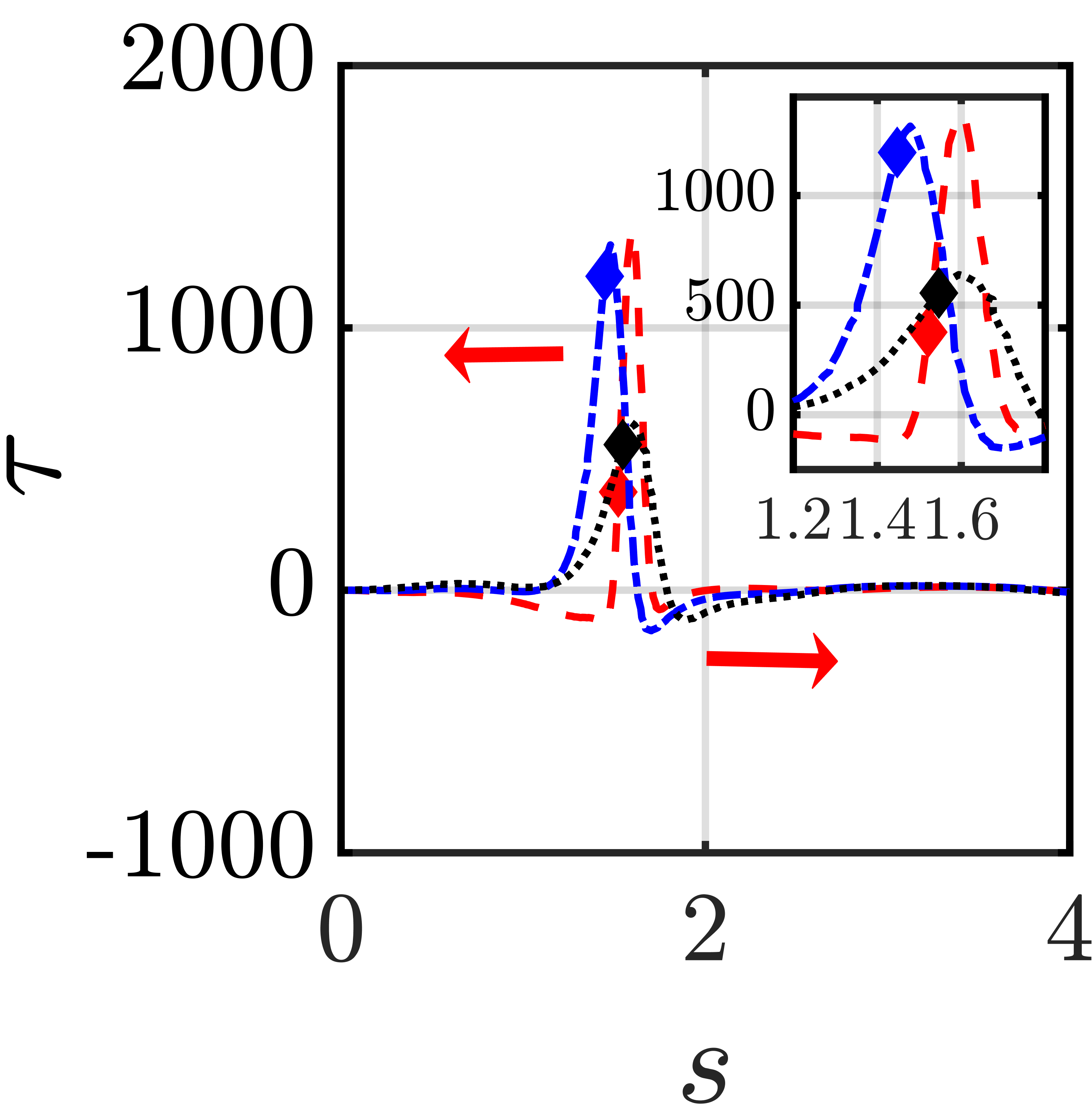}& 
\includegraphics[ width=0.25\linewidth]{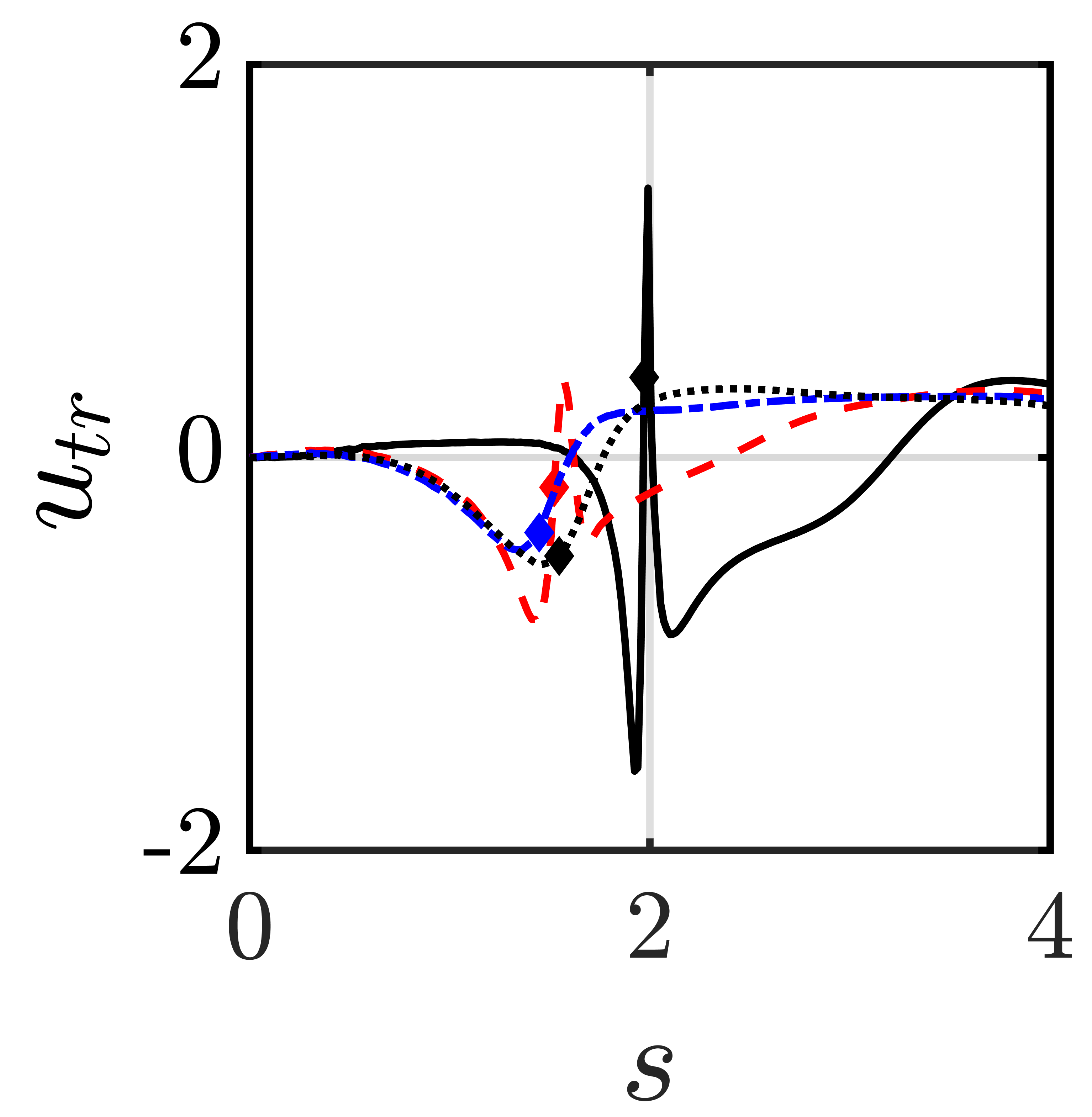}\\
(e) & (f) & (g) & (h)\\
\end{tabular}
\begin{tabular}{cc}
\includegraphics[ width=0.65\linewidth]{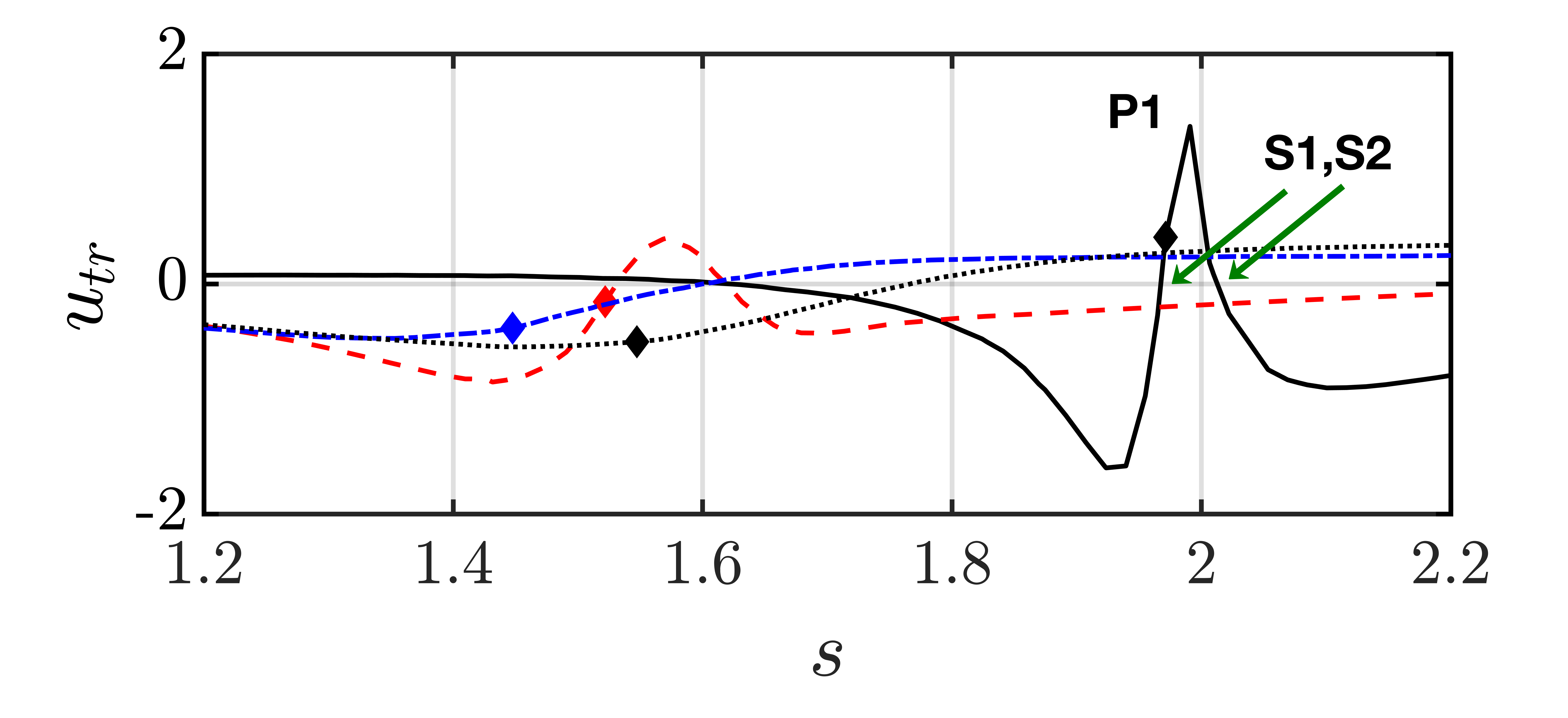}&
\includegraphics[ width=0.15\linewidth]{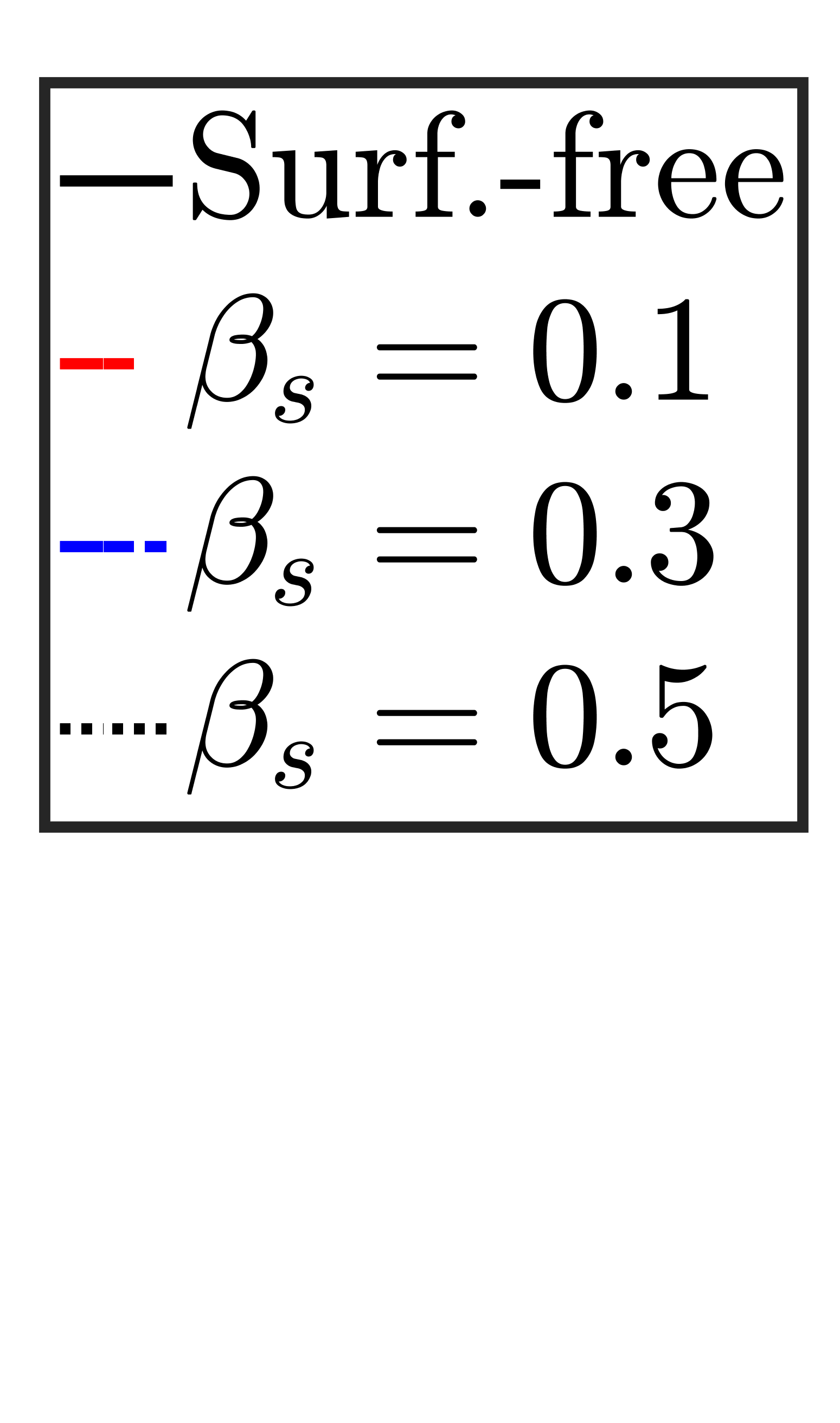}\\
(i)&\\
\end{tabular}
\begin{tabular}{cc}
Surfactant-free~~~~~~~~Surfactant-laden & Surfactant-free~~~~~~~~Surfactant-laden\\
\includegraphics[height=0.25\linewidth]{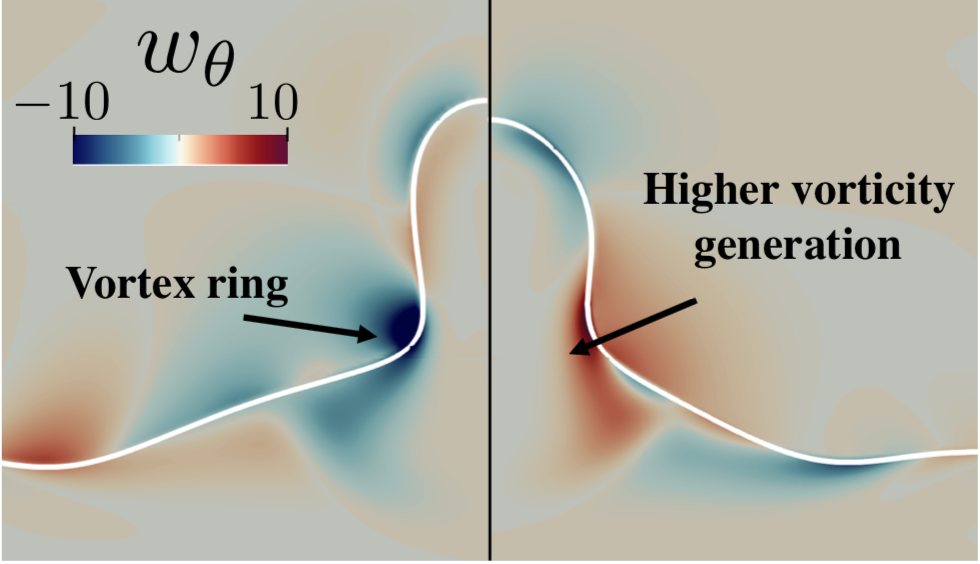}&
\includegraphics[height=0.25\linewidth]{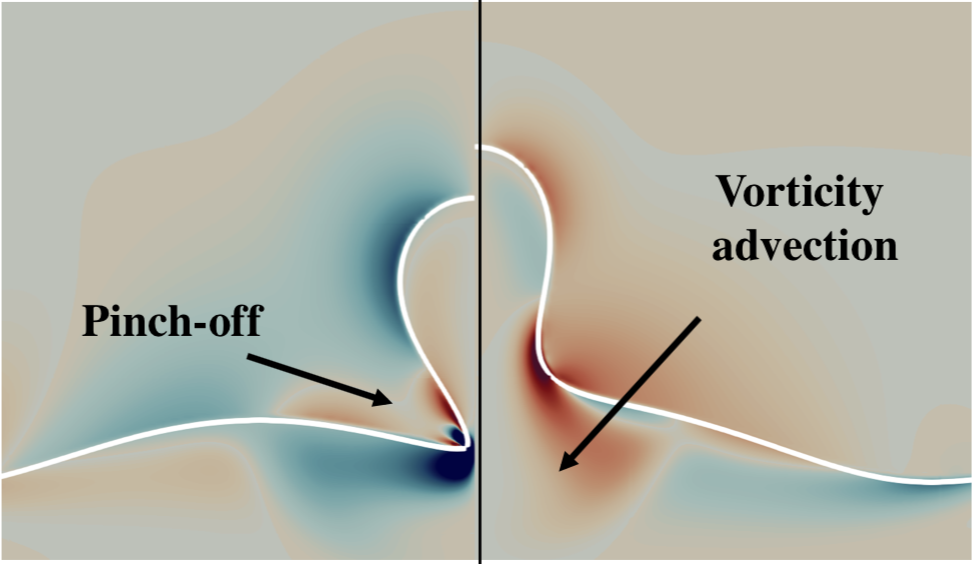}\\
%$t=1.2$ &   $t=1.68$ \\
(j) & (k)
\end{tabular}
\end{center} 
\caption{\label{beta_flow_fields} Panels (a-i): Effect of the elasticity parameter  $\beta_s$ on the flow and surfactant concentration fields associated with the drop-interface coalescence phenomenon.
Two-dimensional representation of the interface location, $\Gamma$, $\tau$, and the radial component of the interfacial velocity $u_{tr}$ are shown in (a)-(d) and (e)-(h) for $t=1.20$ and $t=1.68$, respectively. 
Note that the abscissa in (a) and (e) corresponds to the radial coordinate $r$, and in (b)-(d) and (f)-(h) to the arc length $s$.
The panel (i) represents a magnified view of (h). % columns one to four, respectively, for the  surfactant-free and surfactant-laden cases. In columns 1 and 2-4 the variation is with respect to the dimensionless radial coordinate and , $r$, and arc length, $s$, respectively. 
%The first, (a)-(d), and second, (e)-(h), rows  are given at $t=1.20$ and $t=1.68$, respectively.
The arrows in (g) indicate the directions of motion driven by the Marangoni stresses $\tau$; in (h) and (i), point P1, and S1 and S2 designate the peak in $u_{tr}$ and the stagnation points in the surfactant-free $u_{tr}$ profile, respectively. The diamond shapes in (i) show the location of the necks. The parameter values and the times for the radius are the same as in figure \ref{metrics_beta}.
Panels (j) and (k): Effect of surfactants on the azimuthal vorticity $\omega_{\theta}$ for the surfactant-free (left panels), and the surfactant-laden cases (right panels), for $\beta_s$=0.5, at $t=1.20$, $t=1.68$, respectively. All other parameters remain unchanged from figure \ref{metrics_beta}. The colour indicates the value of the azimuthal vorticity $\omega_{\theta}$.
} 
\end{figure}

The next part of the analyses focuses on the time evolution of a two-dimensional projection of the interfacial shape, $\tilde\Gamma$, $\tau$, and the radial component of the interfacial velocity, $u_{tr}$, presented in panels a-i of figure \ref{beta_flow_fields}. We also show the interplay  between the surface and the azimuthal component of the vorticity field (i.e., vorticity is defined as $\omega=\nabla \times \textbf{u}$), displayed in figures  \ref{beta_flow_fields}j and \ref{beta_flow_fields}k. In the surfactant-free case, it is seen from figure \ref{beta_flow_fields}d that $u_{tr}<0$ and $u_{tr}>0$ upstream and downstream of the developing neck, which drives flow away from this region. The narrowing of the neck induces capillary-driven flow that leads to further neck-thinning and the development of large peaks in $u_{tr}$, as shown in figure \ref{beta_flow_fields}e and h, which are typical of singularity formation. 
Close inspection of the $u_{tr}$ profile in figure \ref{beta_flow_fields}i for the surfactant-free case reveals that it is characterised by the presence of a large velocity peak ($P1$) and %of changing sign close to the droplet neck. 
%The spike implies the presence of 
two stagnation points (labelled $S1$ and $S2$)  with the neck sandwiched between them. Over time, the inertio-capillary-induced flow ultimately culminates in interfacial breakup %results in the formation of interfacial  singularity 
to form a daughter droplet. From the %We now turn our attention to the 
vorticity plots in figures \ref{beta_flow_fields}j and \ref{beta_flow_fields}k, it is seen that for the surfactant-free case, the vorticity generation is confined to the vicinity of the neck as the two stagnation points aid the fluid recirculation around the neck (so-called `vortex-ring', displayed in the left-panel of figure \ref{beta_flow_fields}j). As time evolves, the interfacial curvature of the neck increases, and a large vorticity generation can be observed on the side of the bulk accompanying %, and subsequently the normal motion of the fluid in the normal direction to the interface, leading to  
the eventual neck pinch-off, as depicted in the left-panel of figure \ref{beta_flow_fields}k. 
%
%\textcolor{red}{
More information regarding the mechanisms which induce the generation of vorticity at the liquid-gas interface is   provided below. %}

For the surfactant-laden cases, the accumulation of $ \Gamma$ near the nascent neck can be seen in figure \ref{beta_flow_fields}b thus giving rise to a local decrease of $\sigma$. The presence of $\Gamma$ gradients results in the generation of a large positive peak in the $\tau$ profile in the vicinity  of the neck region, which is largest for the intermediate value of $\beta_s=0.3$, as shown in figure \ref{beta_flow_fields}c (i.e., consistent with the non-monotonic response of dynamics observed in figure \ref{metrics_beta}). Upstream and downstream of the neck, $\tau > 0$ and $\tau <0$, respectively, which drives flow towards the drop summit and tail, reflected by $u_{tr}<0$ and $u_{tr}>0$, respectively. %; drop while  after the neck, $\tau > 0$, then $\tau$ acts towards the tail of the drop, and consequently, the Marangoni stresses counteract the direction of the inertial-capillary-induced flow in the neck. 
Although the overall shape of the $u_{tr}$ curve for the surfactant-free case is robust to the addition of insoluble surfactant, it is evident that the magnitude of $u_{tr}$ decreases with increasing $\beta_s$, particularly in the neck region; moreover, the oscillation in $u_{tr}$ in the surfactant-free case is damped out for $\beta_s>0$.

Additionally, by close inspection of the $u_{tr}$ plots for the surfactant-laden cases in figure \ref{beta_flow_fields}i, it becomes clear that only one stagnation point is present near the neck for $\beta_s=0.3$ and $0.5$; thus,  $u_{tr}>0$ towards its tail. The Marangoni-induced flow has therefore led to  %overcomes the inertio-capillary-induced flow in the vicinity of the neck, causing fluid to reverse its direction resulting in 
the suppression of one of the stagnation points. %, and eventually, to an %and to a change in the fate of the dynamics with the reopening of the neck (i.e., 
%escape from pinchoff. 
%
%
%We turn our attention to the 
Furthermore, by comparing the vorticity field pattern of the surfactant-free and the surfactant-laden cases, a change is observed as a result of the presence of surfactants, and the inhibition of a stagnation point. The generation of vorticity is also confined to the vicinity of the free-surface; however, the `vortex-ring'  no longer exists, as shown in the right panel of figure \ref{beta_flow_fields}j. 

Further in time, when the escape of capillary singularity commences, we observe that vorticity is separated from the vicinity of the interface, and advected towards the bulk of liquid reservoir, which is consistent with the findings of  \cite{Ananthakrishnan_wm_1994}, supporting  the reopening of the neck, as displayed in the right panel of figure \ref{beta_flow_fields}k. The reason behind the vorticity separation is %a result of the change of sign (i.e., direction) of $w_{\theta}$ close to the surface caused by 
the inhibition of one of the stagnation points. This behaviour is similar to the phenomenon explained  by \citet{Hoepffner_jfm_2013} in terms of capillary retraction of surfactant-free viscous ligaments where they  suggested  that the advection of $\omega_{\theta}$ plays a crucial role in their escape from breakup.
These observations also agree with the recent studies reported by \citet{Constante-Amores_prf_2020} and \citet{Constante-Amores_jfm_2021} in terms of the escape of capillary singularity during the capillary retraction of a liquid thread, and  the inhibition of jet-drops formation from bursting bubbles, respectively.

%\textcolor{red}{
Next, we turn our attention to the role of the Marangoni stresses in the generation of vorticity at the gas-liquid interface; this is consistent with the work of  \citet{batchelor_2000} who concluded that vorticity in a homogeneous fluid is generated at the boundaries only. %Therefore,  we aim to study the role of the Marangoni stresses owing to the presence  of surfactants in the generation of vorticity at the gas-liquid interface. %The vorticity generation mechanism can be obtained by taking the curl of the momentum equation (XX). 
%
%The theoretical work by \citet{brons_2014} demonstrated that the generation of vorticity  in a two phase system  can be understood by analysing the evolution of the circulation of a closed curve which crosses an interface separating phases. 
%
Furthermore, we can assume that the interface behaves as a viscous free-surface because of the vanishing viscosity and density ratios (similar assumptions have been made previously by \cite{Dooley1997} and \citet{Xia_2017}).  \citet{kamat_2020} demonstrated that for surface-tension-driven phenomena, vorticity generation  depends solely on the interfacial boundary conditions when $Oh<<1$. As a result, the tangential stress at the interface is balanced by the surface tension gradients, resulting in 
%}
%
\begin{equation}
\textbf{t} \cdot \textbf{D} \cdot \textbf{n} =\textbf{t} \cdot \nabla_s \sigma
\end{equation}
\noindent
%\textcolor{red}{
in which, $\textbf{D}$ represents  the rate of deformation tensor (the symmetric part of the velocity gradient tensor). By further mathematical manipulation \citep{lundgren_1999}, the generation of vorticity at the free-surface depends  entirely on the velocity field, interfacial geometry, and surface tension gradients:
%}
%
\begin{equation}
\label{eq:vorticity}
    \omega= \omega_n+ \omega_t+ \omega_{\tau}=-2 \frac{\partial \textbf{u}\cdot \textbf{n}}{\partial s}+ 2 \textbf{u} \cdot \textbf{t} \kappa  + \textbf{t} \cdot \nabla_s \sigma. \ 
\end{equation}
\noindent
%\textcolor{red}{
Similar results for $\omega_n$ and  $\omega_t$ have been reported by \citet{lundgren_1999} and \citet{brons_2014}. The first two terms on the right-hand-side  of equation (\ref{eq:vorticity}) correspond to the the normal and tangential velocity-driven vorticity generation, respectively, whereas the last term is representative of the Marangoni stress vorticity contribution.
%}
%

\begin{figure} 
\centering
\begin{tabular}{c}
\includegraphics[width=0.6\columnwidth]{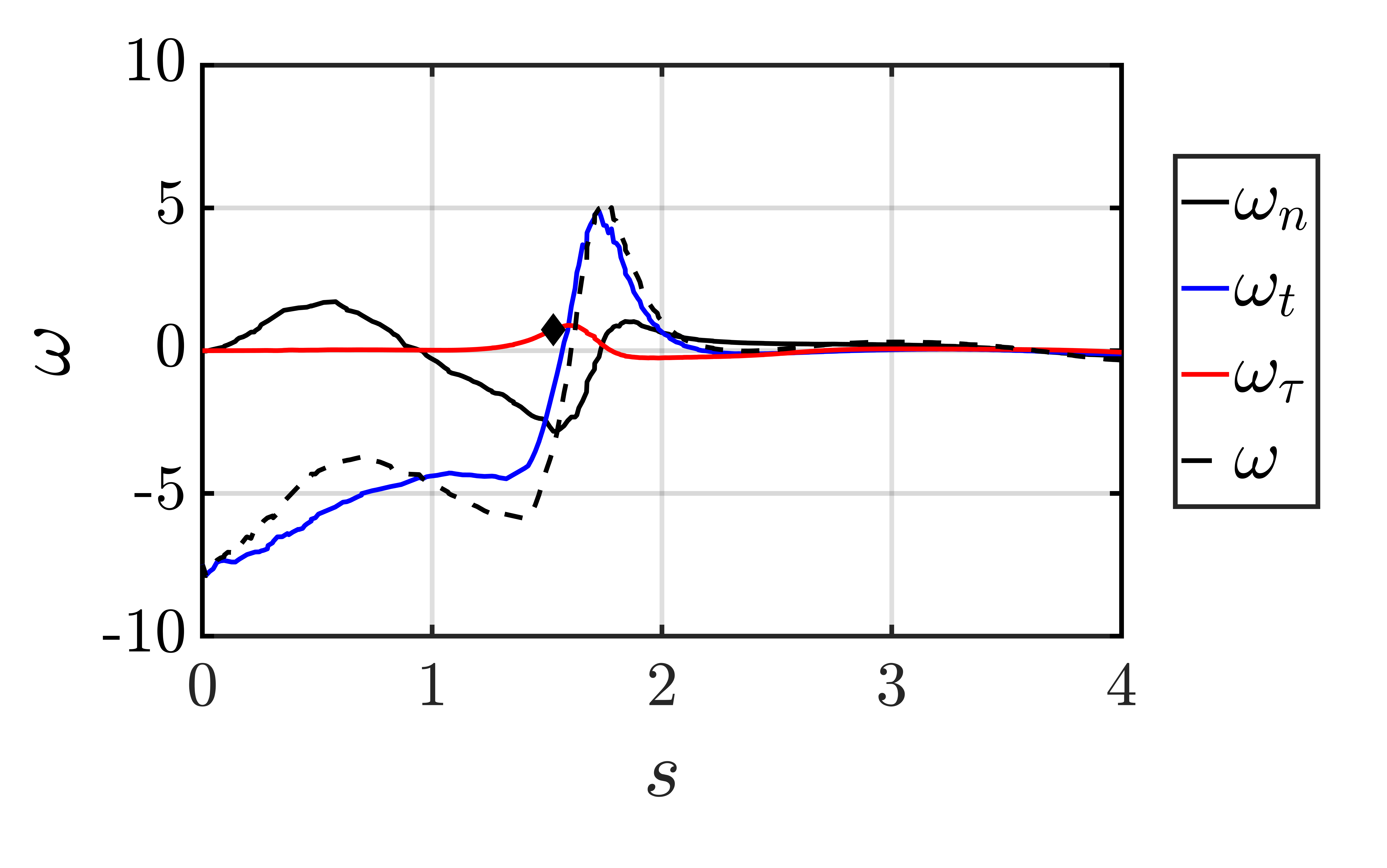}
\end{tabular}
\caption{\label{figure_w} 
%\textcolor{red}{ 
Vorticity production, $\omega$, along the gas-liquid interface expressed in terms of the local normal and tangential components from the velocity field and the Marangoni stresses, represented by $\omega_n$, $\omega_t$ and $\omega_{\tau}$, respectively. The surfactant-laden case is characterised by $\beta_s=0.5$ at $t = 1.68$. All other parameters remain unchanged from figure 3.  The diamond shape shows the location of the neck.
%}
}
 \end{figure}

%\textcolor{red}{
Figure \ref{figure_w} shows the vorticity distribution along the interface according to equation (\ref{eq:vorticity}).
%It is depicted that the highest peak of  positive vorticity values is located downstream the neck, which corresponds to the region in which there is the highest advection of vorticity towards the pool. Then, the vorticity values decay towards the tail of the drop. 
By close inspection of the  profiles,  we observe that the vorticity generation at the interface is highly dominated by the interfacial curvature term, $\omega_{t}$. Moreover, in the vicinity of the neck we discover the existence of a  positive  peak in the Marangoni stress-driven vorticity production (i.e., $\omega_{\tau}$). The peak of $\omega_{\tau}$ has a  different sign in comparison to $\omega_n$. The leading cause for this behaviour stems from the suppression of the stagnation points on both sides of the neck, as shown in figure \ref{beta_flow_fields}i. Ultimately, this analysis demonstrates the positive effect of surface tension-driven vorticity generation on the neck reopening process. % which is the result of the neck reopening of the neck due to the inhibition of stagnation points, and subsequently, counteracting the  capillary breakup of the neck.
% Downstream the neck,  $\omega_{\tau}$ and $\omega_{n}$ have values with similar order of magnitude. 
%
Additionally, this finding is in agreement with \citet{kamat_2020}, who concluded that the generation of vorticity arising from the presence of surfactants is generated over a time scale of similar magnitude to the capillary time scale. 
%}

%The horizontal collapse may, therefore, only induce pinch-off if the vertical collapse is sufficiently retarded. Such a delay is achieved by the convergence of capillary waves that are generated by the opening of the neck in the early stages of coalescence. These waves travel up the side of the drop, and carry enough momentum to significantly distort the drop as they converge on its summit. This convergence of capillary waves can then stretch the drop upwards by as much as 30 of its initial radius (Fig. 3). If the waves are sufficiently vigorous to counteract the vertical collapse, the horizontal collapse is then able to reach completion, producing a daughter droplet.

\begin{figure}
\begin{center} 
\begin{tabular}{ccc}
\includegraphics[ width=0.32\linewidth]{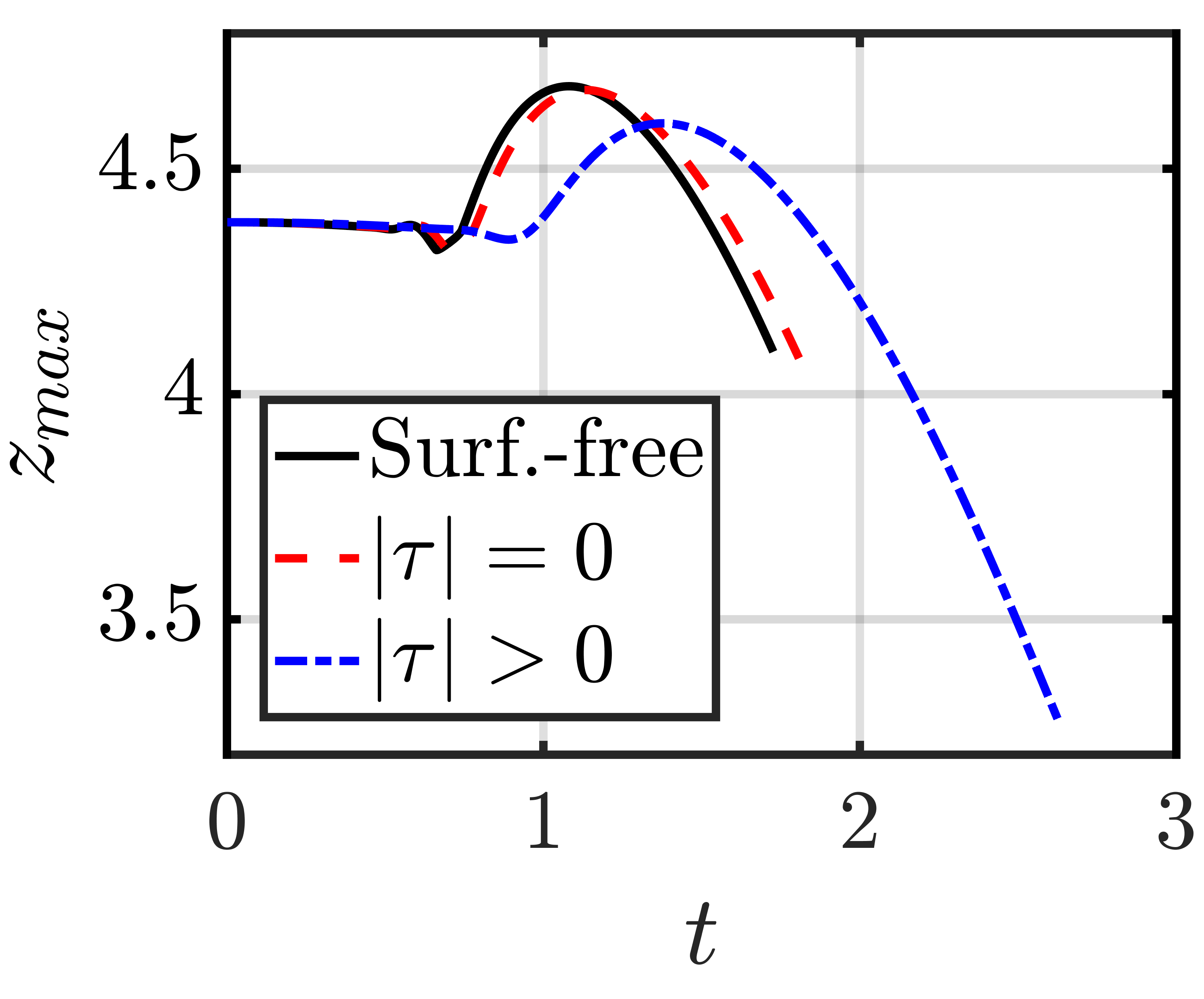}&
\includegraphics[ width=0.32\linewidth]{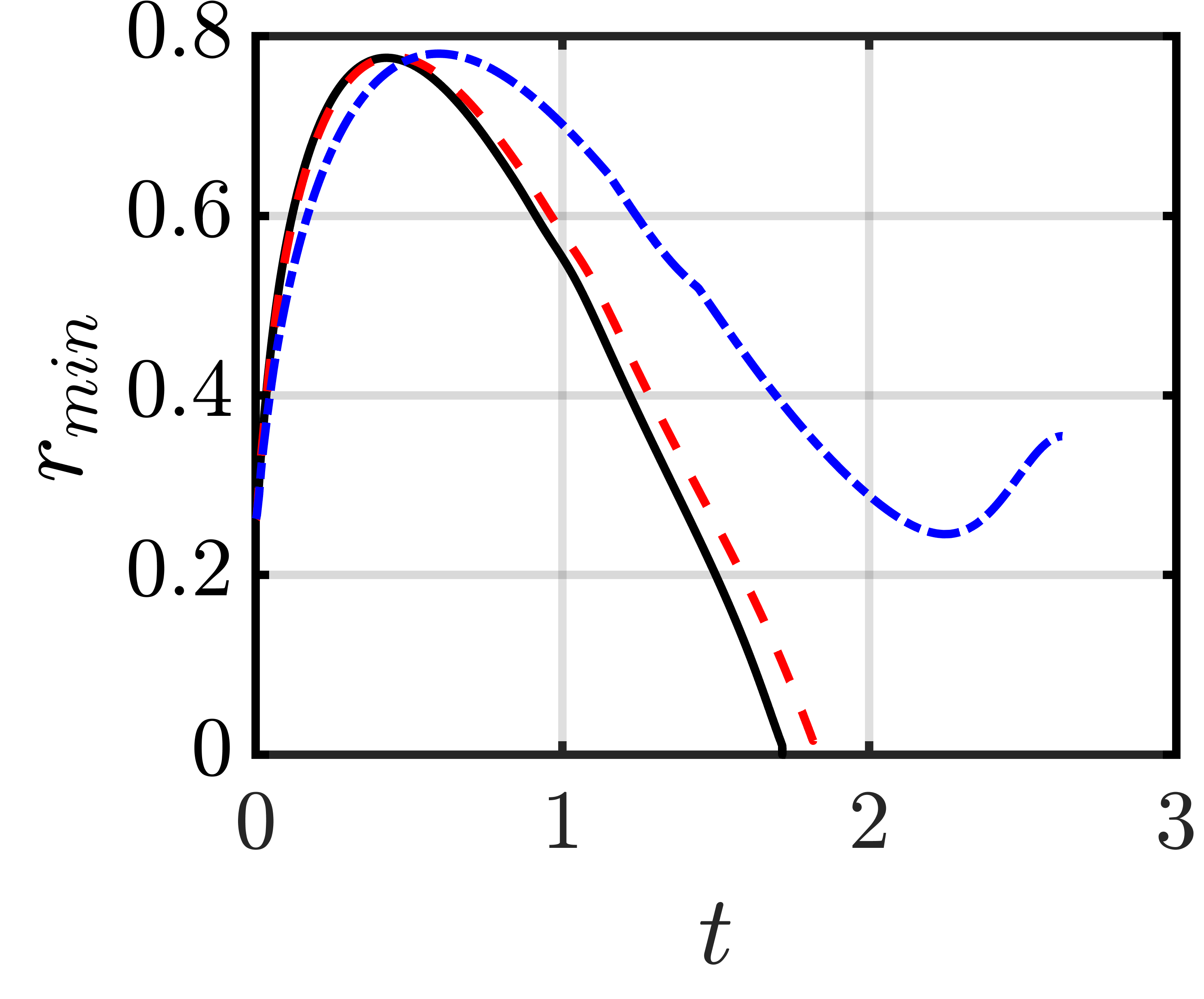}&
\includegraphics[ width=0.32\linewidth]{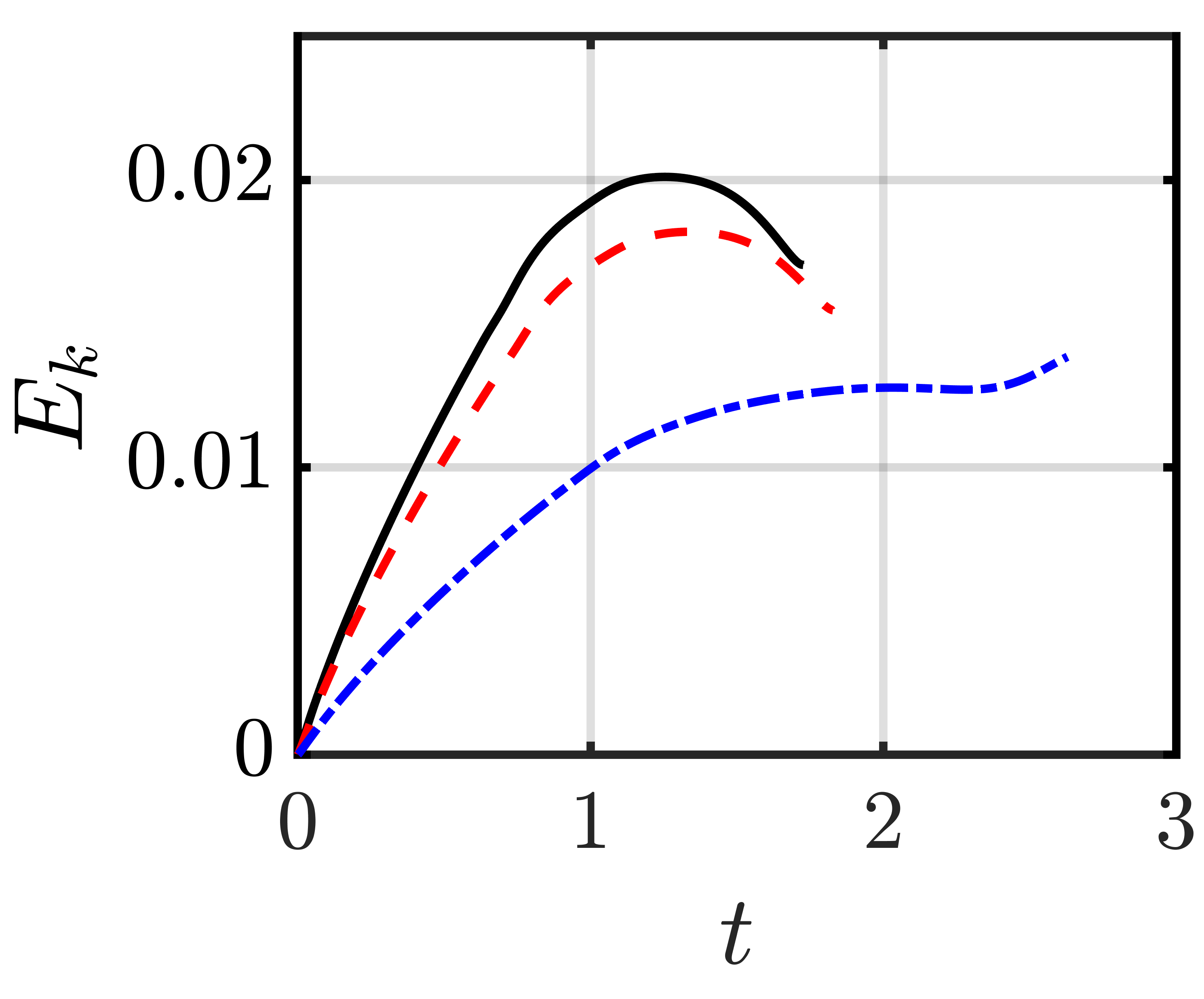}\\
(a) & (b) & (c)
\end{tabular}
\end{center} 
\caption{Demonstration that tangential Marangoni stresses are responsible for the inhibition of the interfacial singularity. Temporal evolution of the maximum vertical displacement of the interface, neck radius and kinetic energy; (a)-(c), respectively, for the surfactant-free, full-Marangoni $|\tau|>0$, and no-Marangoni cases $|\tau|=0$, for $Oh=0.02$, $Bo=10^{-3}$, $\beta_s=0.5$, $Pe_s=100$ and $\Gamma_o=0.5\Gamma_\infty$.
\label{noMaranogni}}
\end{figure}

Finally, we aim to provide  more conclusive evidence  that  the interfacial singularity inhibition is Marangoni-driven rather than a result of the reduction of the surface tension (i.e., capillary pressure reduction). For this reason,  we have performed an additional simulation in which, Marangoni stresses have been suppressed. Figure \ref{noMaranogni} reports the temporal evolution of the maximum axial position $z_{max}$, the neck radius $r_{min}$, and the kinetic energy $E_k$ for the surfactant-free and Marangoni-suppressed cases. Similar flow behaviour between the surfactant-free and Marangoni-suppressed cases are observed. The most remarkable finding is that for the Marangoni-suppressed-case, it is observed that the mean reduction of the surface tension does not prevent the horizontal collapse of the droplet (see figure \ref{noMaranogni}b). The inspection of the kinetic energy plot shows that the Marangoni-suppressed and surfactant-free cases have almost identical behaviour (see figure \ref{noMaranogni}c). Therefore, when  Marangoni stresses are enabled fully, a change of the fate of the coalescence is observed via the reopening of the neck.

%%%%%%%%%%%%%%%%%%%%%%%%%%%%%%%%%%%%%%%%%
%%%%%%%%%%%%%%%%%%%%%%%%%%%%%%%%%%%%%%%%%
%%%%%%%%%%%%%%%%%%%%%%%%%%%%%%%%%%%%%%%%%
\subsection{Soluble surfactants}
%%%%%%%%%%%%%%%%%%%%%%%%%%%%%%%%%%%%%%%%%
%%%%%%%%%%%%%%%%%%%%%%%%%%%%%%%%%%%%%%%%%
%%%%%%%%%%%%%%%%%%%%%%%%%%%%%%%%%%%%%%%%%

In this subsection we present a discussion of the results associated with the effects of surfactant solubility and sorption kinetics, parameterised by  $Bi$ and $k$, respectively. 
Unless stated otherwise, the parameters remain fixed to their `base' values: $Oh=0.02$, $Bo=10^{-3}$, $\beta_s=0.5$, and $Pe_s=100$; the interfacial surfactant concentration is initialised using its equilibrium surfactant concentration, thus $\Gamma_o=\chi=k/(1+k)$.
%$\Gamma_o=\Gamma_{eq} $. 
Once again, simulations are carried out until either neck pinch-off or reopening has been observed. %  rerun until there is a change of the neck-dynamics.

%%%%%%%%%%%%%%%%%%%%%%%%%%%%%%%%%%%%%%%%%
%%%%%%%%%%%%%%%%%%%%%%%%%%%%%%%%%%%%%%%%%
%%%%%%%%%%%%%%%%%%%%%%%%%%%%%%%%%%%%%%%%%
\subsubsection{Effect of the Biot number, $Bi$}
%%%%%%%%%%%%%%%%%%%%%%%%%%%%%%%%%%%%%%%%%
%%%%%%%%%%%%%%%%%%%%%%%%%%%%%%%%%%%%%%%%%
%%%%%%%%%%%%%%%%%%%%%%%%%%%%%%%%%%%%%%%%%
\begin{figure}
\begin{center}
\begin{tabular}{ccc}
\includegraphics[width=0.33\linewidth]{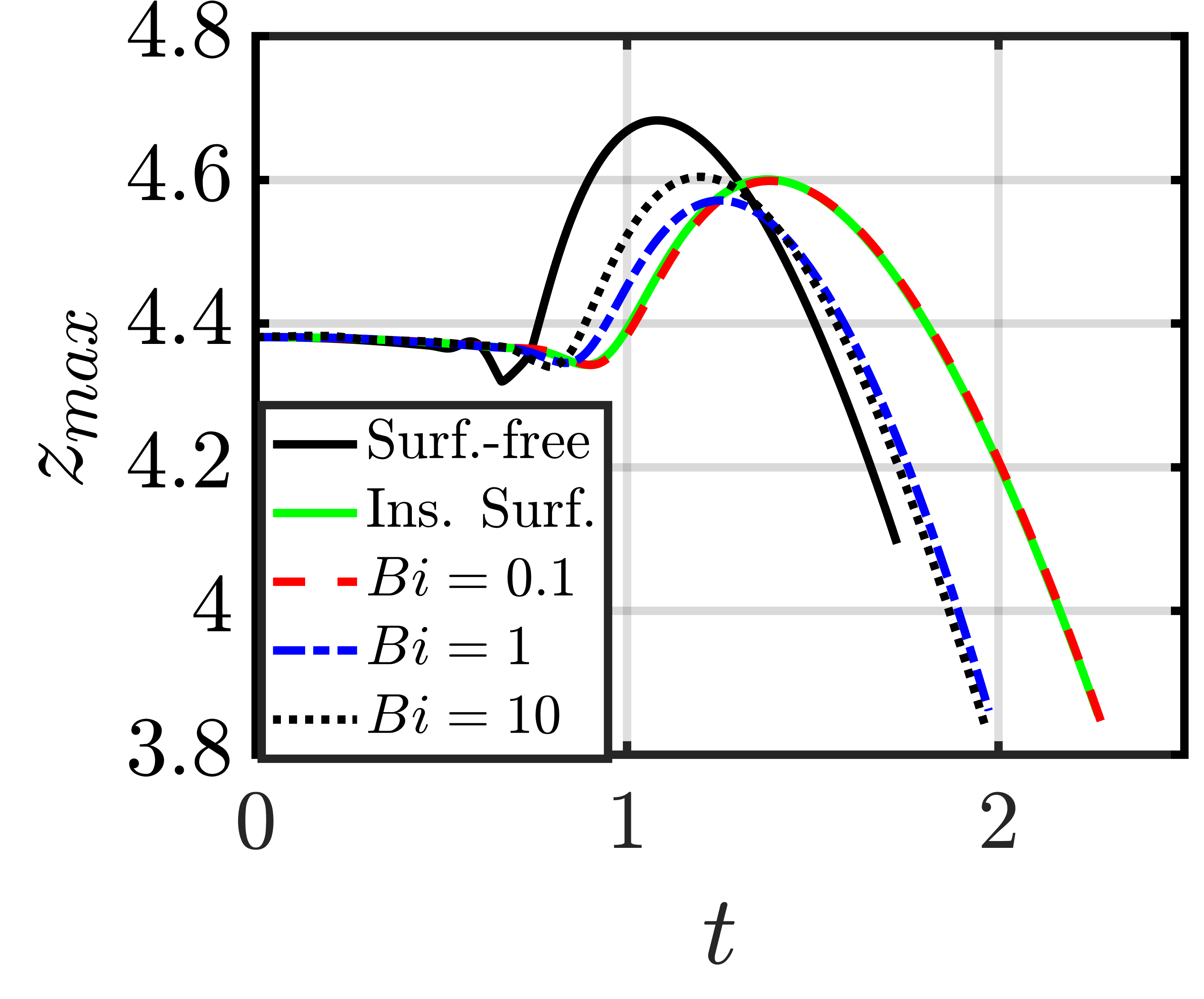}&
\includegraphics[width=0.33\linewidth]{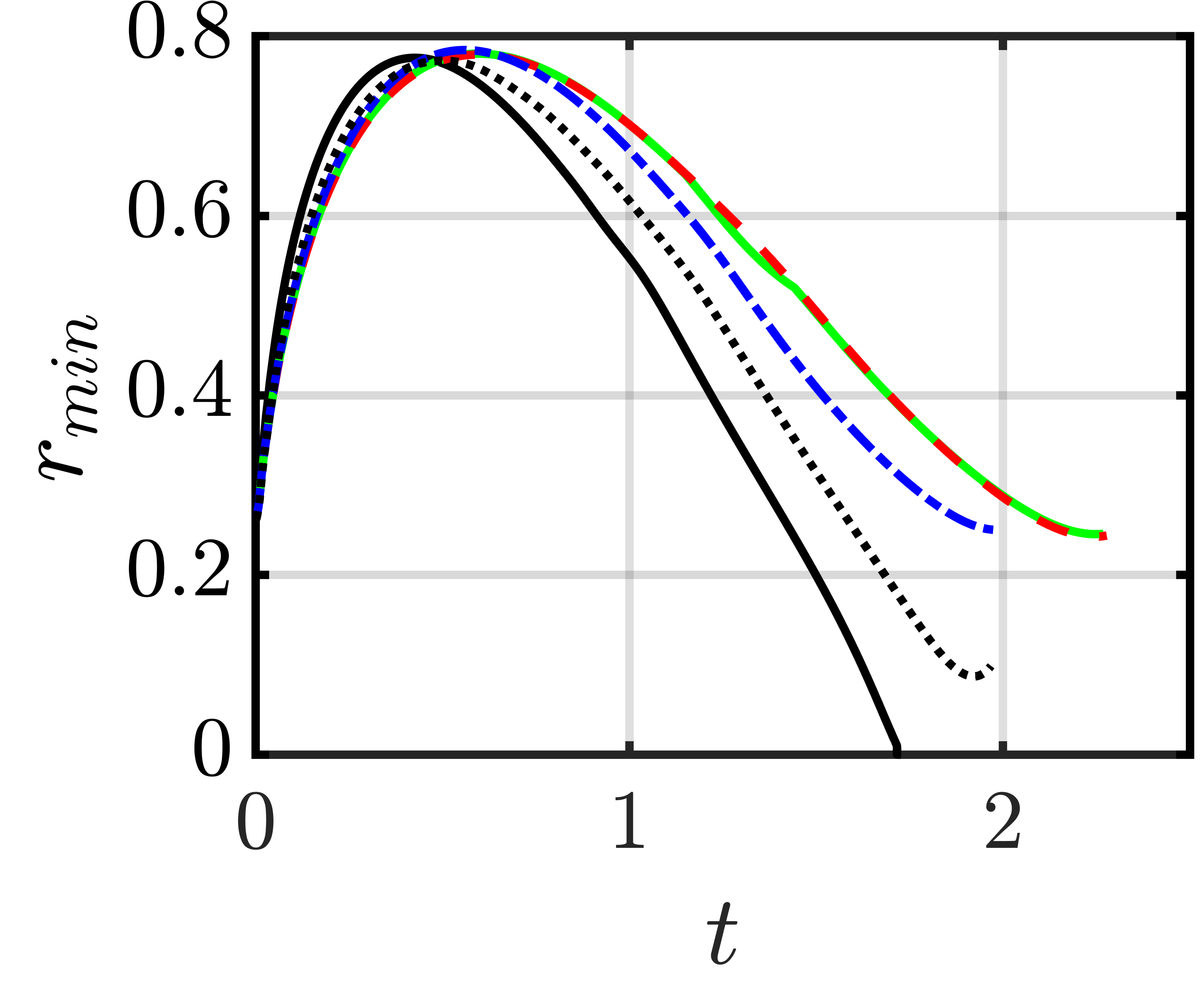}&
\includegraphics[width=0.33\linewidth]{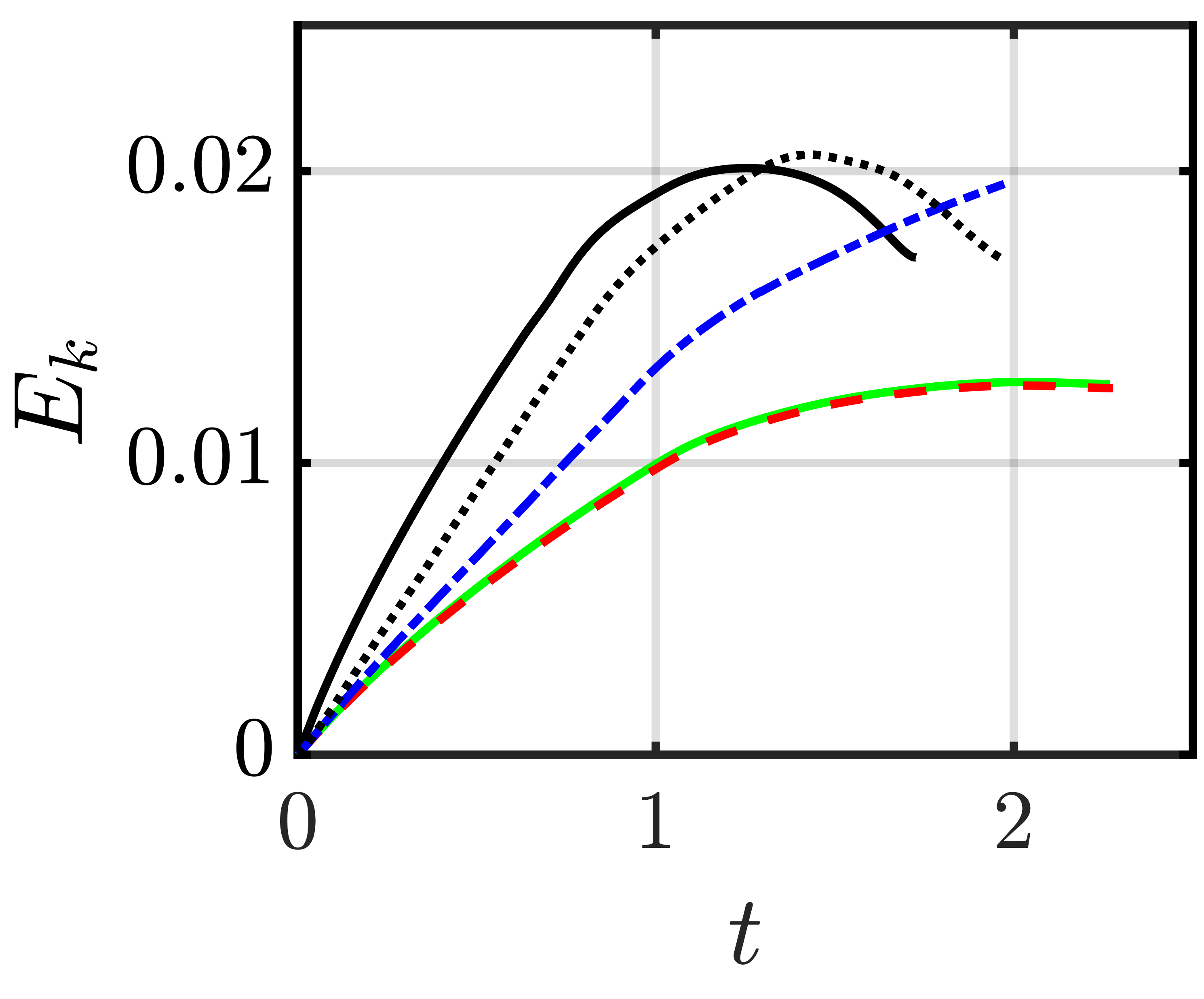}\\
     (a) & (b) & (c)
\end{tabular}
\end{center} 
\caption{
Effect of $Bi$ on the temporal dynamics of the vertical strength of the drop, (a),  minimum neck radius, (b), and kinetic energy (c),
when $Oh=0.02$, $Bo=10^{-3}$, $Pe_s=100$, $\beta_s=0.5$, $k=1$ and $\Gamma_o=\chi$.
\label{metrics_bi}} 
\end{figure}

\begin{figure}
\begin{center} 
\begin{tabular}{cccc}
\includegraphics[width=0.245\linewidth]{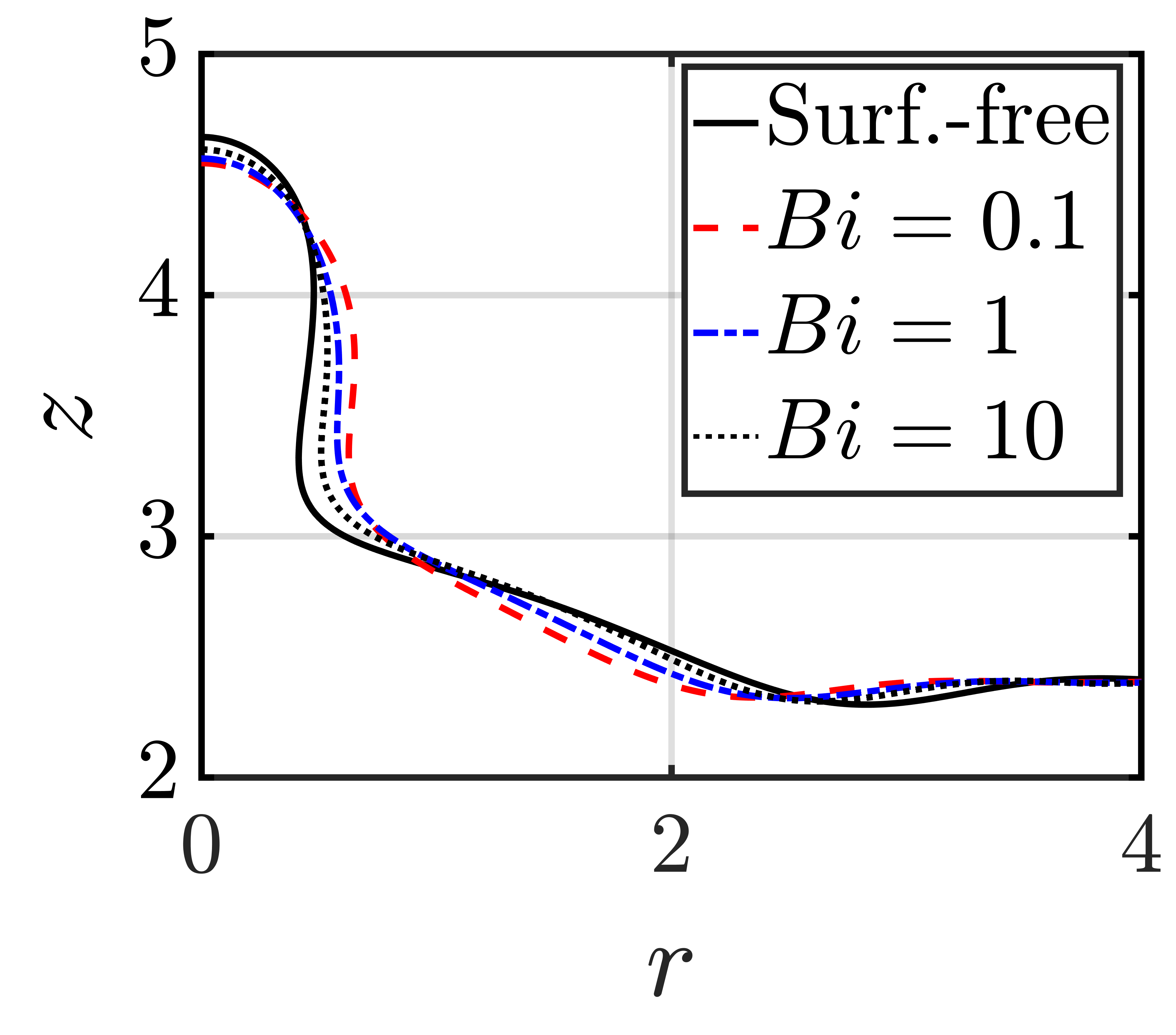}&
\includegraphics[width=0.245\linewidth]{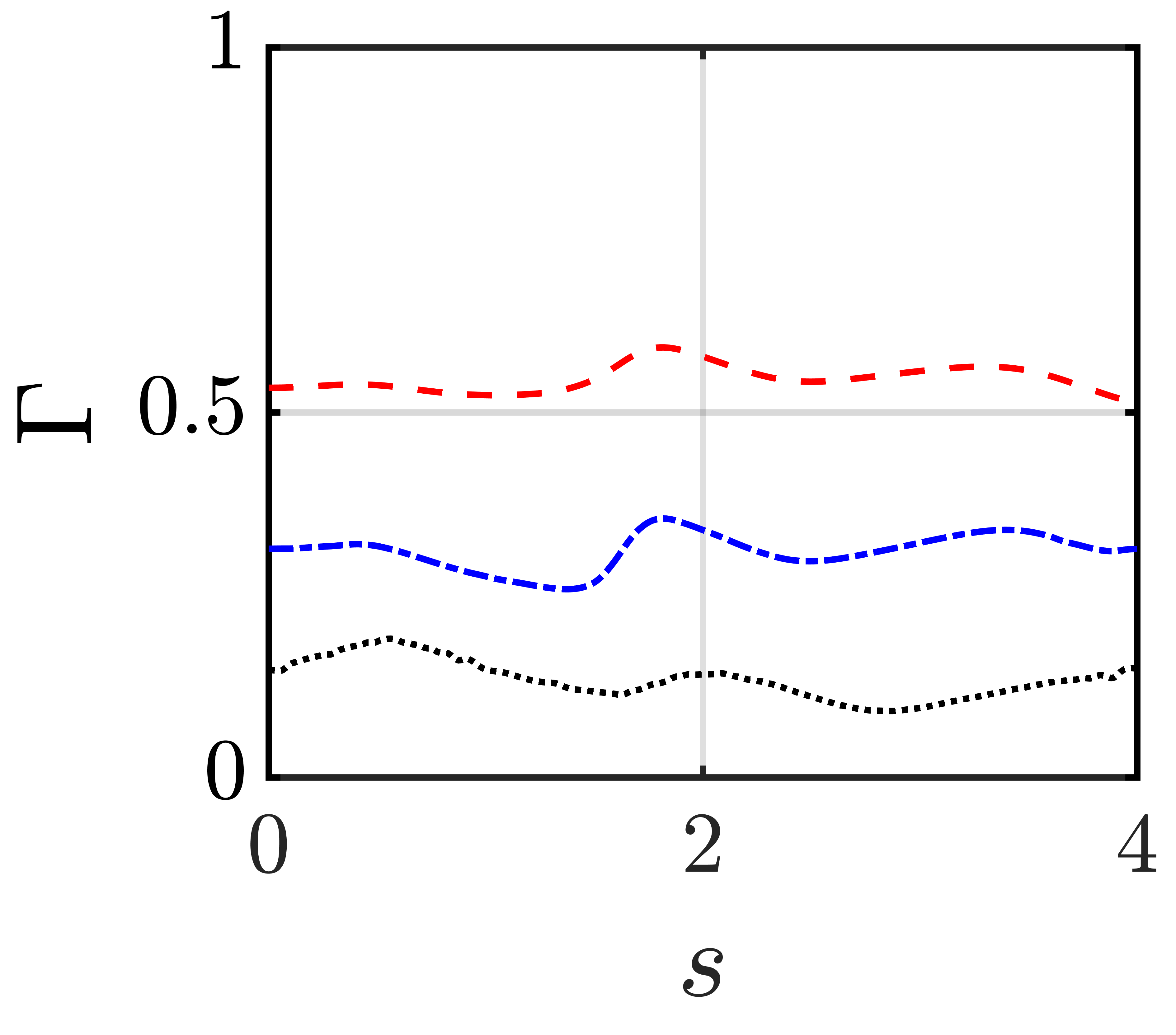}&
\includegraphics[width=0.245\linewidth]{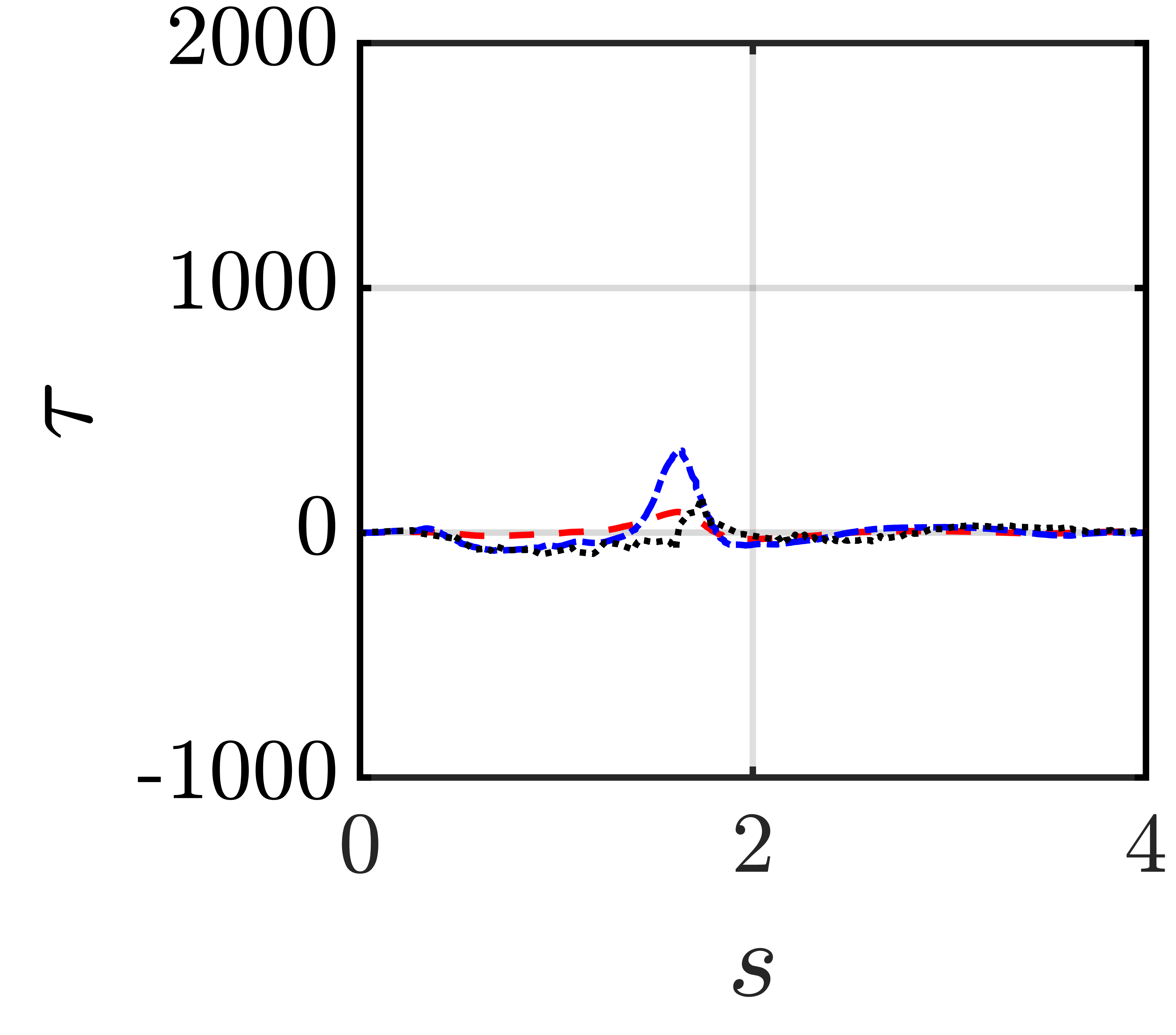}&
\includegraphics[width=0.245\linewidth]{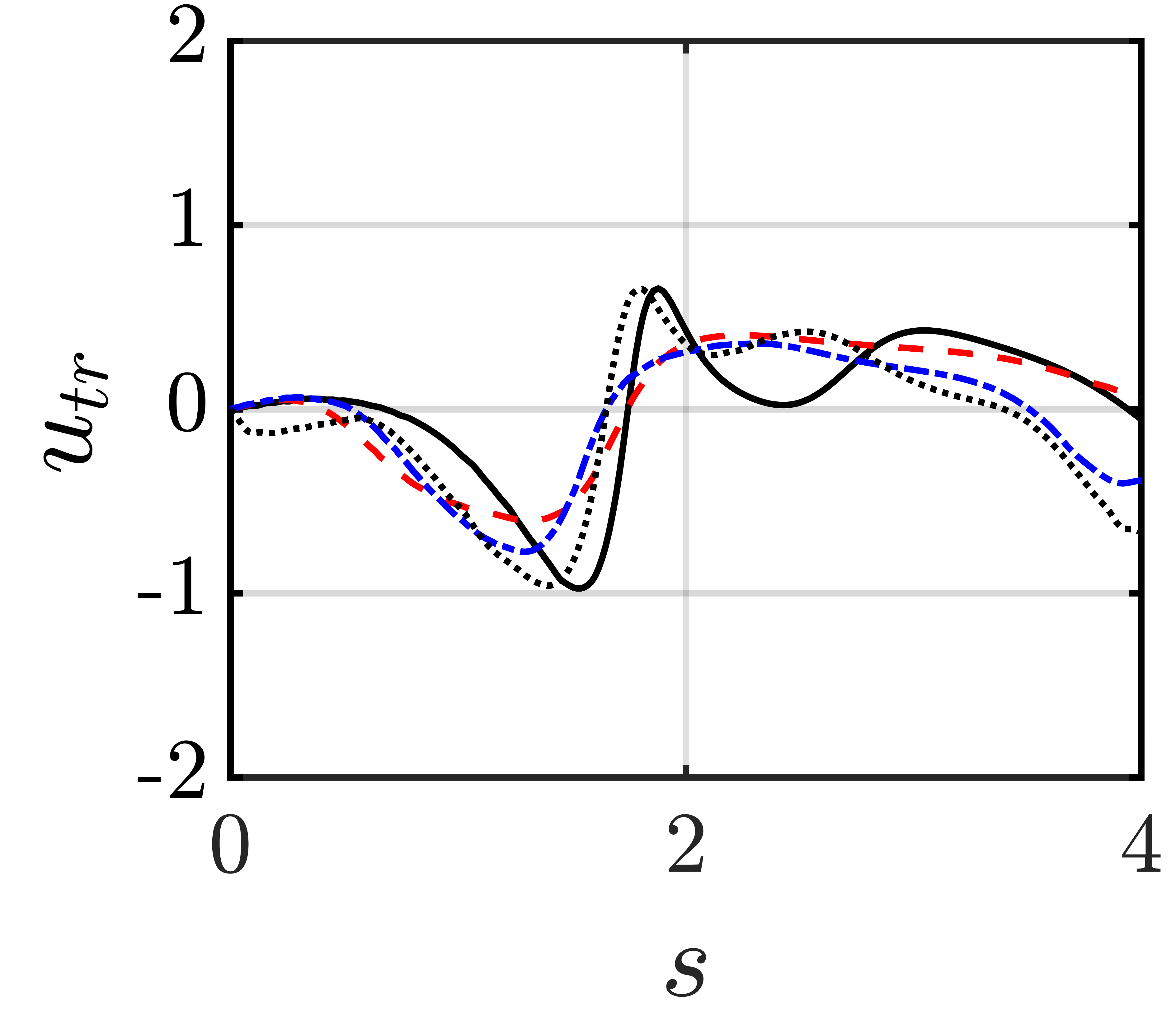}\\
(a) & (b)& (c) & (d)  \\
\includegraphics[width=0.245\linewidth]{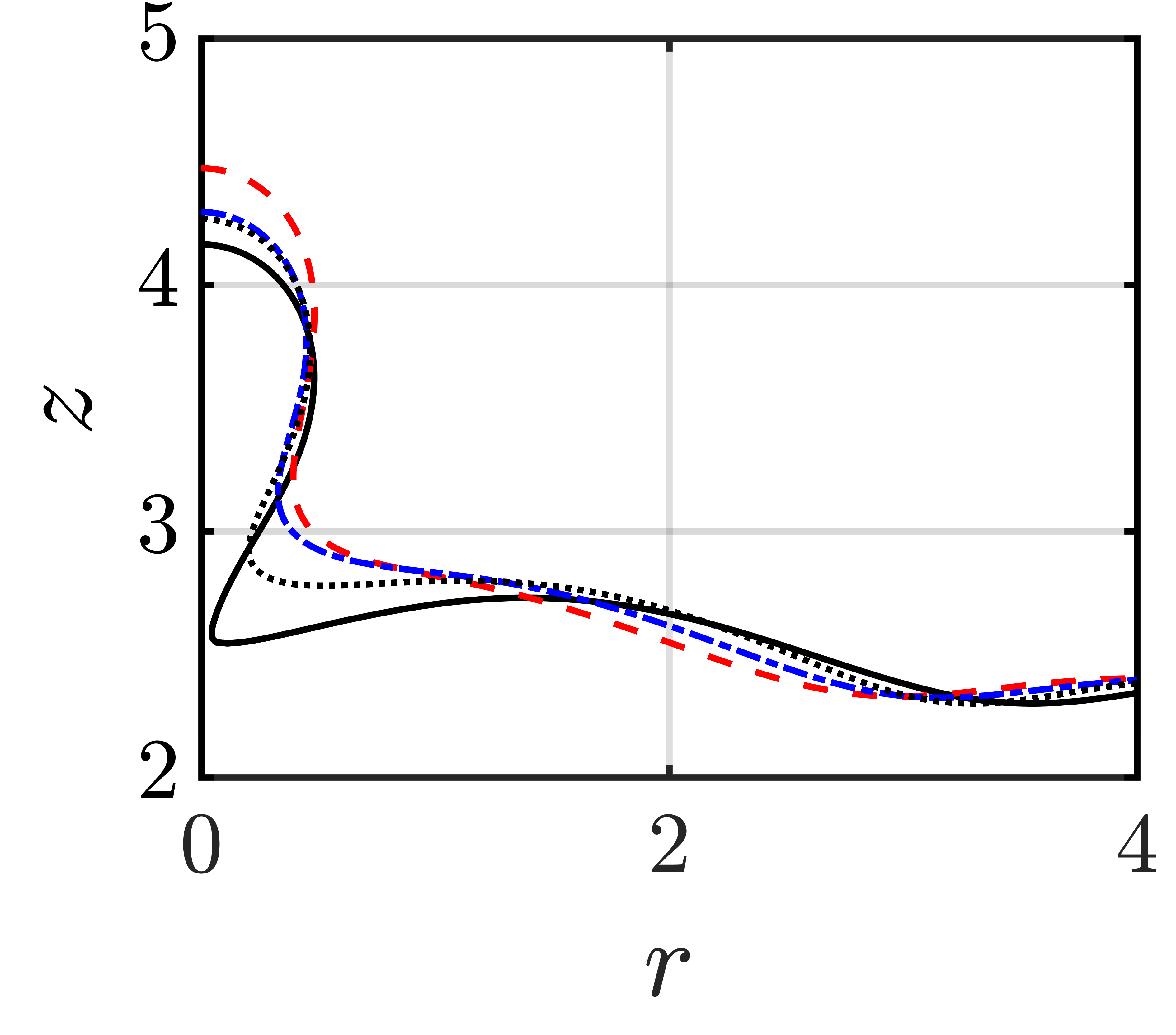}&
\includegraphics[width=0.245\linewidth]{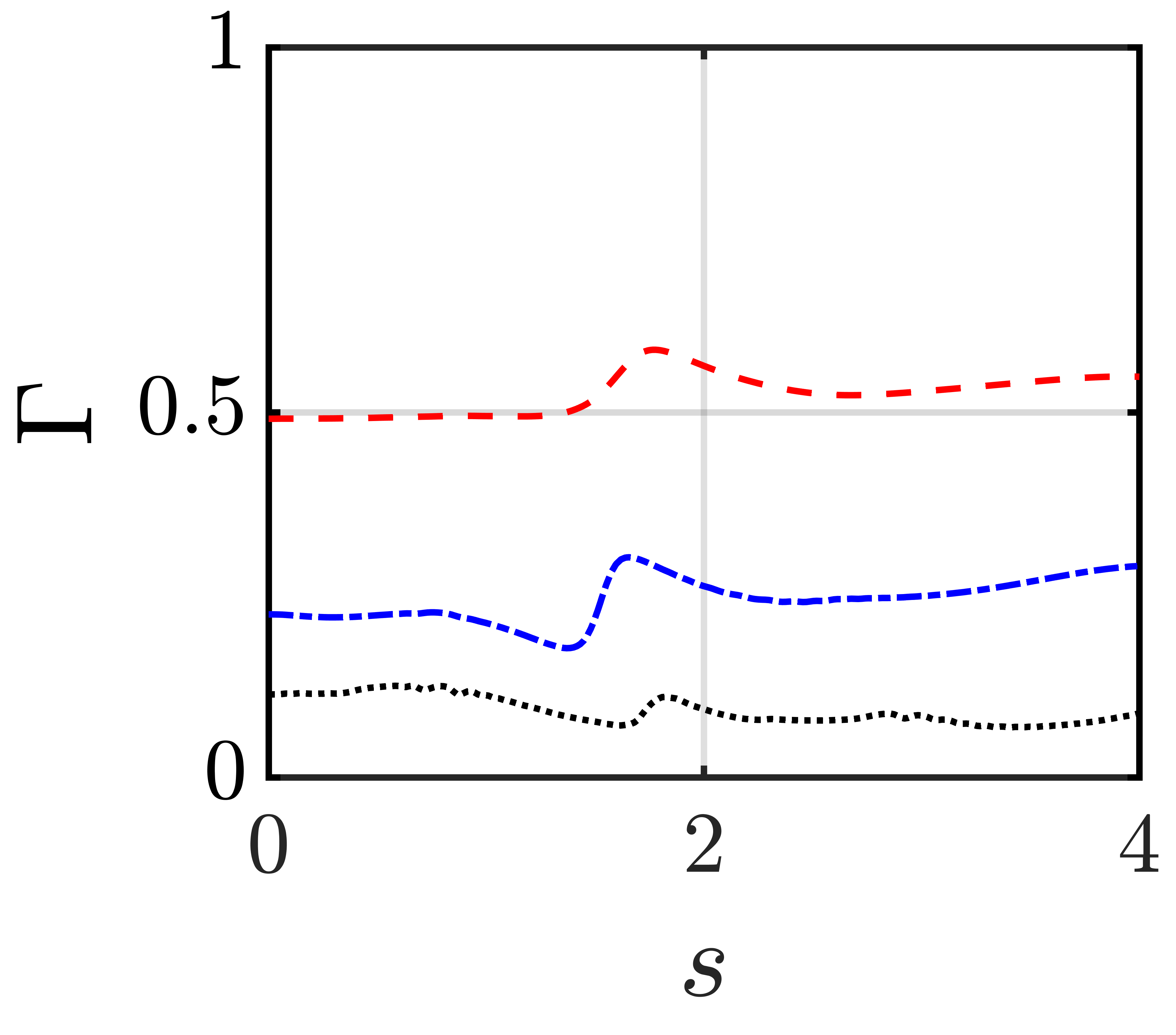}&
\includegraphics[width=0.245\linewidth]{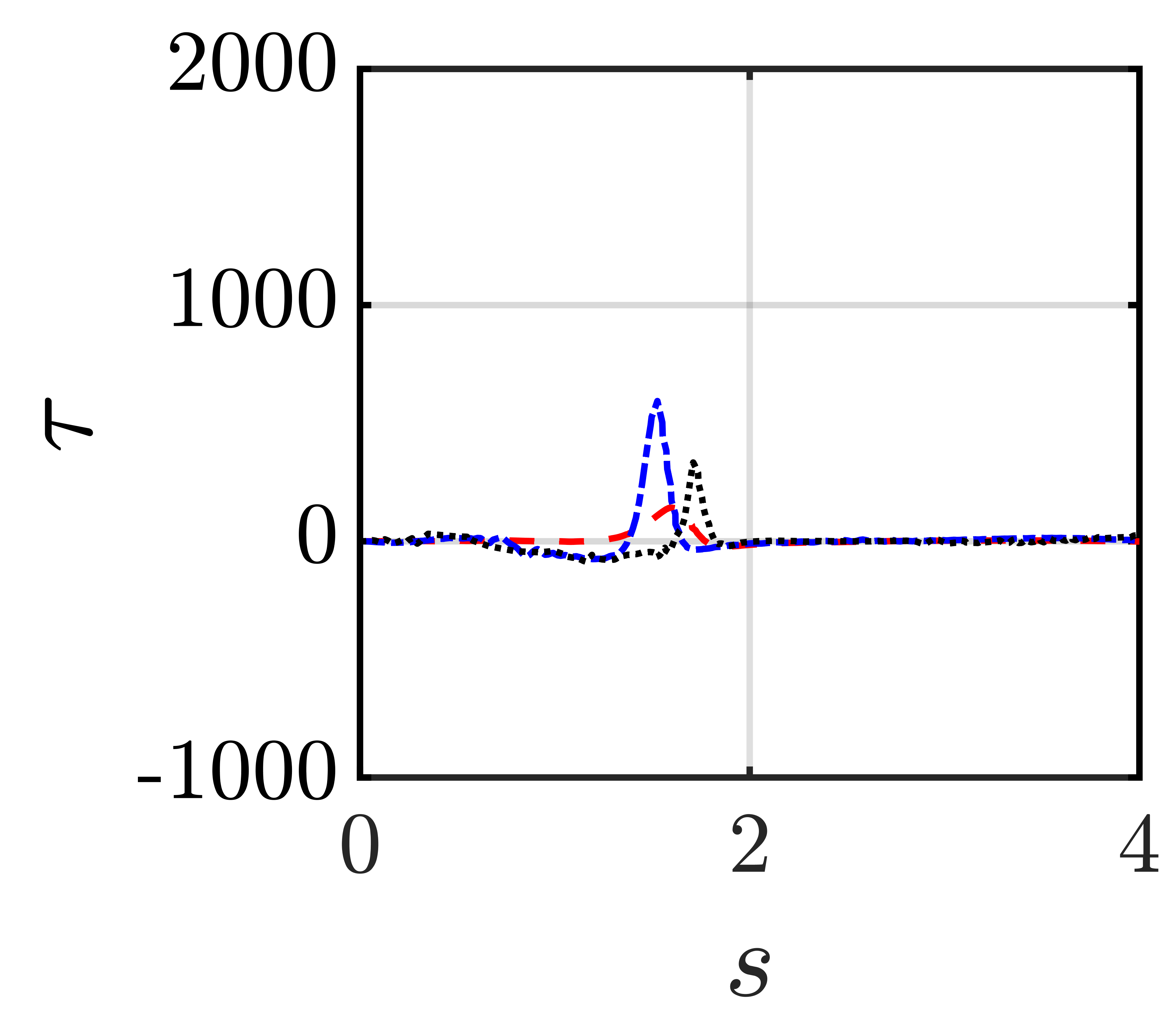}&
\includegraphics[width=0.245\linewidth]{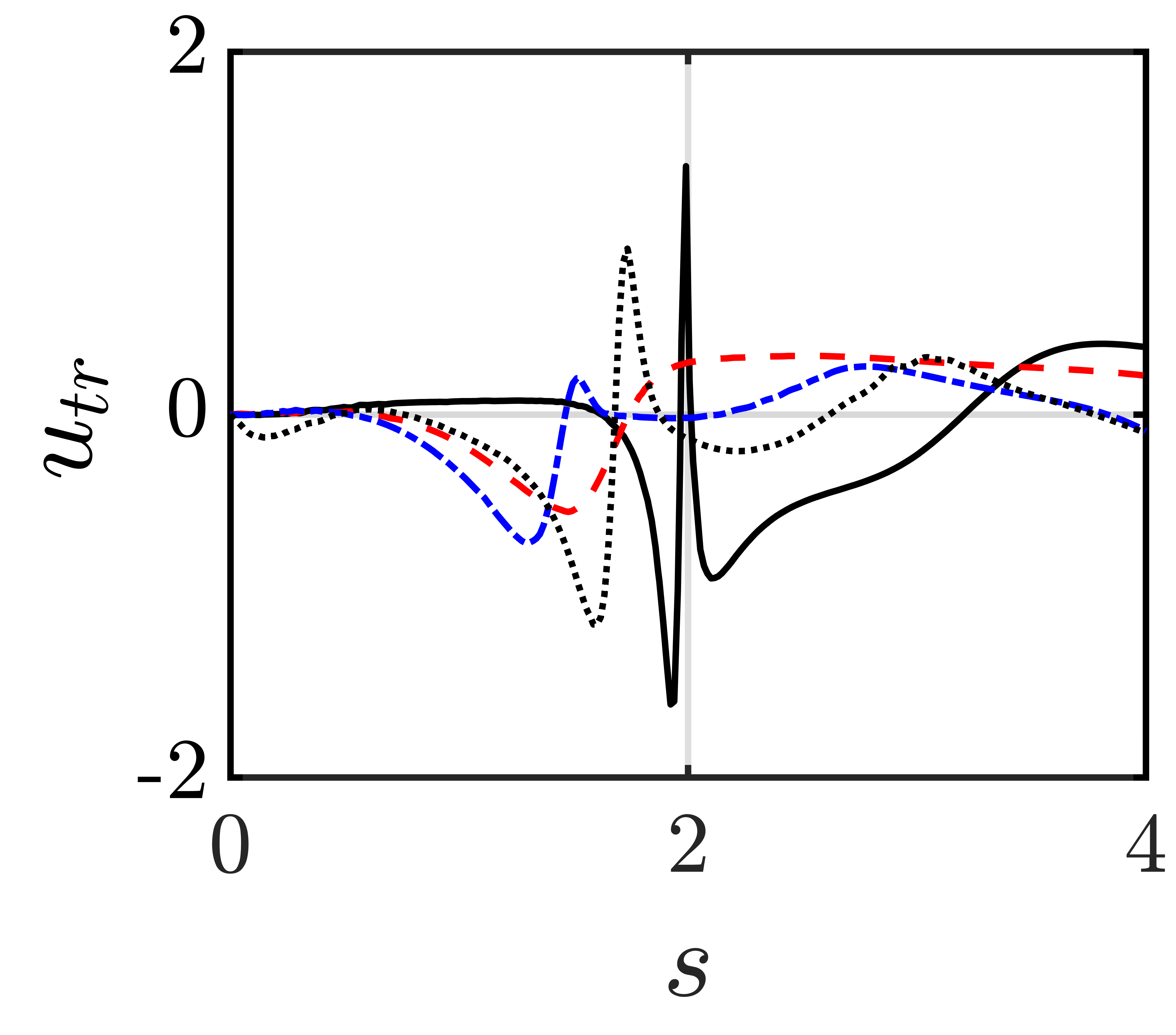}\\
(e) & (f)& (g) & (h)  
\end{tabular}
\end{center} 
\caption{\label{biot_flow_field} 
Effect of the solubility parameter  $Bi$ on the flow and surfactant concentration fields associated with the drop-interface coalescence phenomenon.
Two-dimensional representation of the interface location, $\Gamma$, $\tau$, and the radial component of the interfacial velocity $u_{tr}$ are shown in (a)-(d) and (e)-(h) for $t=1.20$ and $t=1.68$, respectively. 
Note that the abscissa in (a) and (e) corresponds to the radial coordinate $r$, and in (b)-(d) and (f)-(h) to the arc length $s$. Here, all other parameters remain unchanged from figure \ref{metrics_bi}.
%Effect of the solubility parameter  $Bi$ on the coalescence phenomenon. Two-dimensional representation of the interface location, $\Gamma$, $\tau$ and $u_{tr}$ are shown in columns one to four, respectively, for the  surfactant-free and surfactant-laden cases. In columns 1 and 2-4 the variation is with respect to the dimensionless radial coordinate and , $r$, and arc length, $s$, respectively. The parameter values and the times for the radius are the same as in figure \ref{metrics_bi}.The first, (a)-(d), and second, (e)-(h), rows  are given at $t=1.20$ and $t=1.68$, respectively. Here, all other parameters remain unchanged from figure \ref{metrics_bi}.
} 
\end{figure}

Figure \ref{metrics_bi} shows the effect of varying $Bi$ in the range $0.1-10$ on the drop maximal vertical extent, $z_{max}$, the neck radius $r_{min}$, and the kinetic energy $E_k$ with $k=1$; also shown are the curves associated with the insoluble surfactant and surfactant-free cases which respectively correspond to the $Bi \rightarrow 0$ and $Bi \rightarrow \infty$ (and/or $\beta_s \rightarrow 0$) limits.   %The ratio of adsorption to desorption time scales $k$ is fixed to $k=1$ and the Biot number varies in the range of  $Bi=0.1-10$. 
At the lower end of this range (i.e., $Bi =0.1$), the sorptive timescales are much larger than those associated with interfacial effects; therefore, the dynamics are  dominated by capillarity and Marangoni stresses, and are therefore expected to be similar to those observed for the insoluble surfactant case. This is confirmed upon inspection of figure \ref{metrics_bi} as the curves associated with the $Bi=0.1$ case practically overlap with those generated for the insoluble surfactant case. 
For large $Bi$, the monomers desorb rapidly from the interface, which represents the case of a highly-soluble surfactant characterised by dynamics that are similar to those that accompany the surfactant-free case. This is also confirmed by comparing the curves associated with the surfactant-free and $Bi=10$ cases, the latter corresponding to the largest $Bi$ value studied.

As depicted in figure \ref{metrics_bi}a, increasing the level of solubility leads to a decrease in the surfactant mass at the interface available to induce Marangoni stresses and, consequently, delays the retardation in the initialisation of the vertical stretch of the drop. %; this is as a consequence of the weakening of the Marangoni stresses due to the decrease in the surfactant mass present at the interface, as it is regulated by the mass transfer between the interface and the bulk. 
Interestingly, the lowest $z_{max}$ is associated with the intermediate $Bi=1$ case, thus $z_{max}$ exhibits a non-monotonic dependence on the surfactant solubility. %response of this parameter is predicted with the lowest drop-vertical stretch  for $Bi=1$. 
Turning attention towards the effect of solubility on the neck-size, $r_{min}$ displayed in figure \ref{metrics_bi}b, it is clearly seen that Marangoni-induced flow results in the inhibition of the capillary-driven singularity over the entire range of $Bi$ values studied. This effect becomes increasingly more pronounced with decreasing $Bi$ and the $r_{min}$ vs $t$ profiles for the surfactant-laden cases  are  bounded between the insoluble and the surfactant-free cases. %Interestingly, in the case of high solubility (i.e.,  $Bi=10$), the  neck size follows closely the surfactant-free case, but eventually, it escapes from its pinchoff  (e.g., $r=0.08$). 
Finally, the $E_k$ profiles depicted in \ref{metrics_bi}c show that decreasing $Bi$ leads to an overall reduction in the kinetic energy, which stems from the rigidifying effect of the Marangoni stresses; this is weakened by increasing the solubility %and the  once more, the overall reduction of  their values in the presence of surfactants (as depicted in figure \ref{metrics_bi}c). The higher the value of Biot, 
and the enhanced surfactant desorption from the interface which leads to a reduction in $\tau$. %, and  subsequently the reduction of the strength of the interfacial rigidification is observed.

%The next part of the analyses  focuses on the time evolution of a two-dimensional projection of the interfacial shape, $\tilde\Gamma$, $\tau$, and $u_{tr}$ presented in figure \ref{biot_flow_field}. 
Evidence of surfactant-induced immobilisation with decreasing $Bi$ is further provided in figure  \ref{biot_flow_field}a,e, and \ref{biot_flow_field}d,h, which depict the interface shape and $u_{tr}$, respectively; the rest of the parameters remain unaltered from figure \ref{metrics_bi}. %which shows that the interfacial rigidification  is larger, with a monotonic decrease, as $Bi \rightarrow 0$. 
It is also clearly seen from figure \ref{biot_flow_field}b,f % $\Gamma$ profiles, 
that higher surfactant desorption is observed as $Bi$ increases %and the $\Gamma$ field shows a more uniform distribution (displayed in figure \ref{biot_flow_field}b). The reduction of the interfacial surfactant concentration is 
driven by the mass transfer between the interface and bulk. Although the largest $\Gamma$ is associated with the smallest $Bi$ values, the largest gradients, and, therefore, Marangoni stresses, $\tau$, are found for the intermediate Biot number, $Bi=1$, as shown in %to reduce the strength of $\tau$ (see 
figure \ref{biot_flow_field}c,g; this is consistent with with the non-monotonic dependence of  the  vertical stretch of the drop on $Bi$, described in figure \ref{metrics_bi}a. Notably, a comparison of panels (b) and (f), and (c) and (g) of figure \ref{biot_flow_field} reveals that over time, the $\Gamma$ gradients become sharper, particularly for small and intermediate $Bi$ leading to an increase in $\tau$, as was also observed in the % grows in magnitude, and identically to the 
insoluble surfactant case. %prior to the neck $\tau > 0$ (e,g., acting towards the drop-summit) retarding its vertical collapse; after the neck radius, $\tau < 0$ (acting towards its  tail), then 
These Marangoni stresses counteract the direction of the inertio-capillary-induced flow, dampen the oscillations in $u_{tr}$ (see figure \ref{biot_flow_field}d,h), and act to 
% in the neck to 
prevent neck pinch-off %formation of interfacial singularity. Finally, the oscillatory motion of the radial tangential velocity component is also damped for all $Bi$ parameters, 
where the efficacy is once more dependent on the magnitude of $Bi$. Finally, the same analysis regarding the role of % the effect of the presence of 
surfactants in the inhibition of stagnation points, explained in subsection \ref{sec:insoluble}, can be extrapolated to the solubility parameter, $Bi$.

\subsubsection{Effect of the adsorption parameter, $k$}
%%%%%%%%%%%%%%%%%%%%%%%%%%%%%%%%%%%%%%%%%
%%%%%%%%%%%%%%%%%%%%%%%%%%%%%%%%%%%%%%%%%
%%%%%%%%%%%%%%%%%%%%%%%%%%%%%%%%%%%%%%%%%

\begin{figure}
\centering
\begin{tabular}{ccc}
\includegraphics[width=0.33\linewidth]{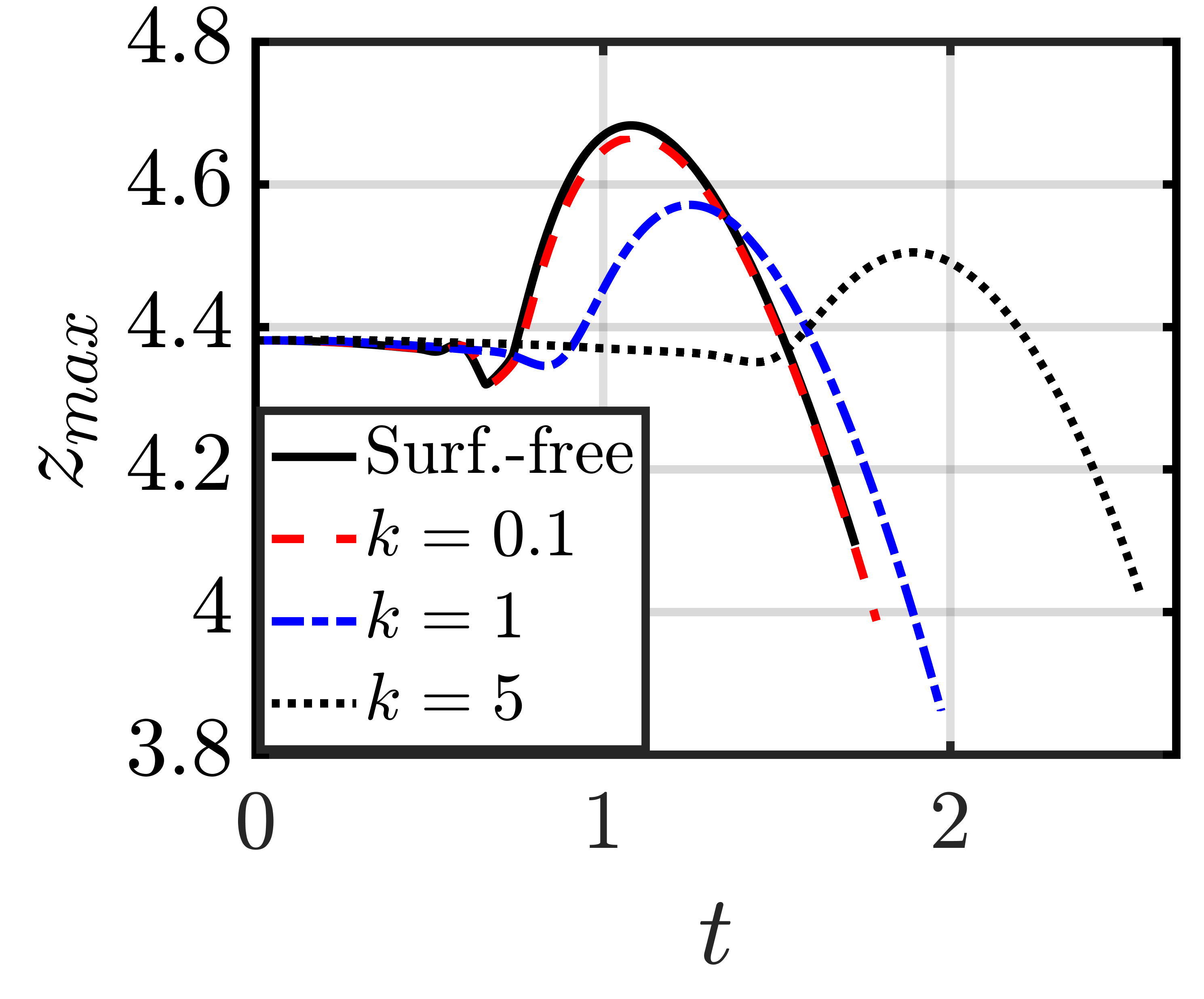} &
\includegraphics[width=0.33\linewidth]{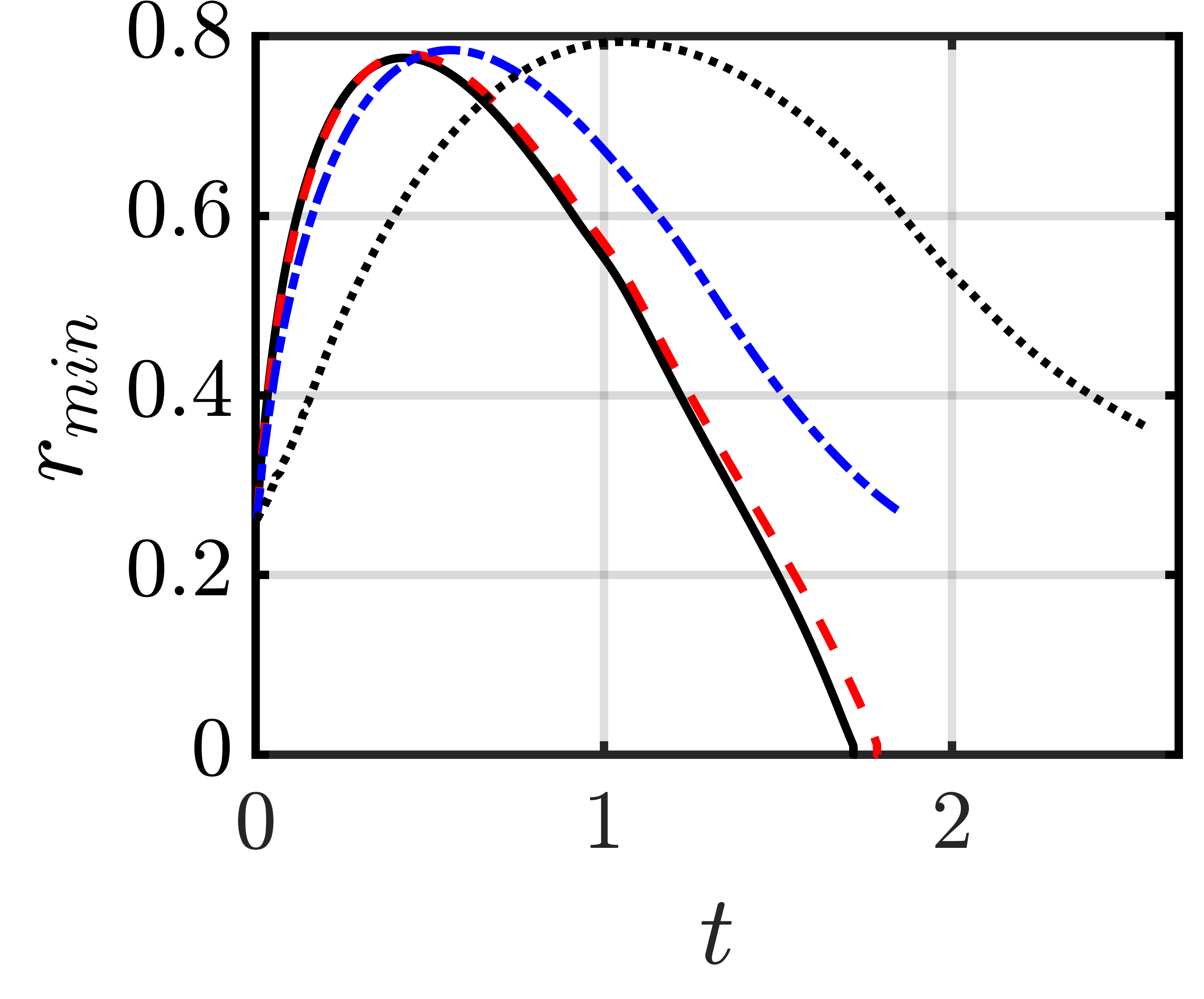} &
\includegraphics[width=0.33\linewidth]{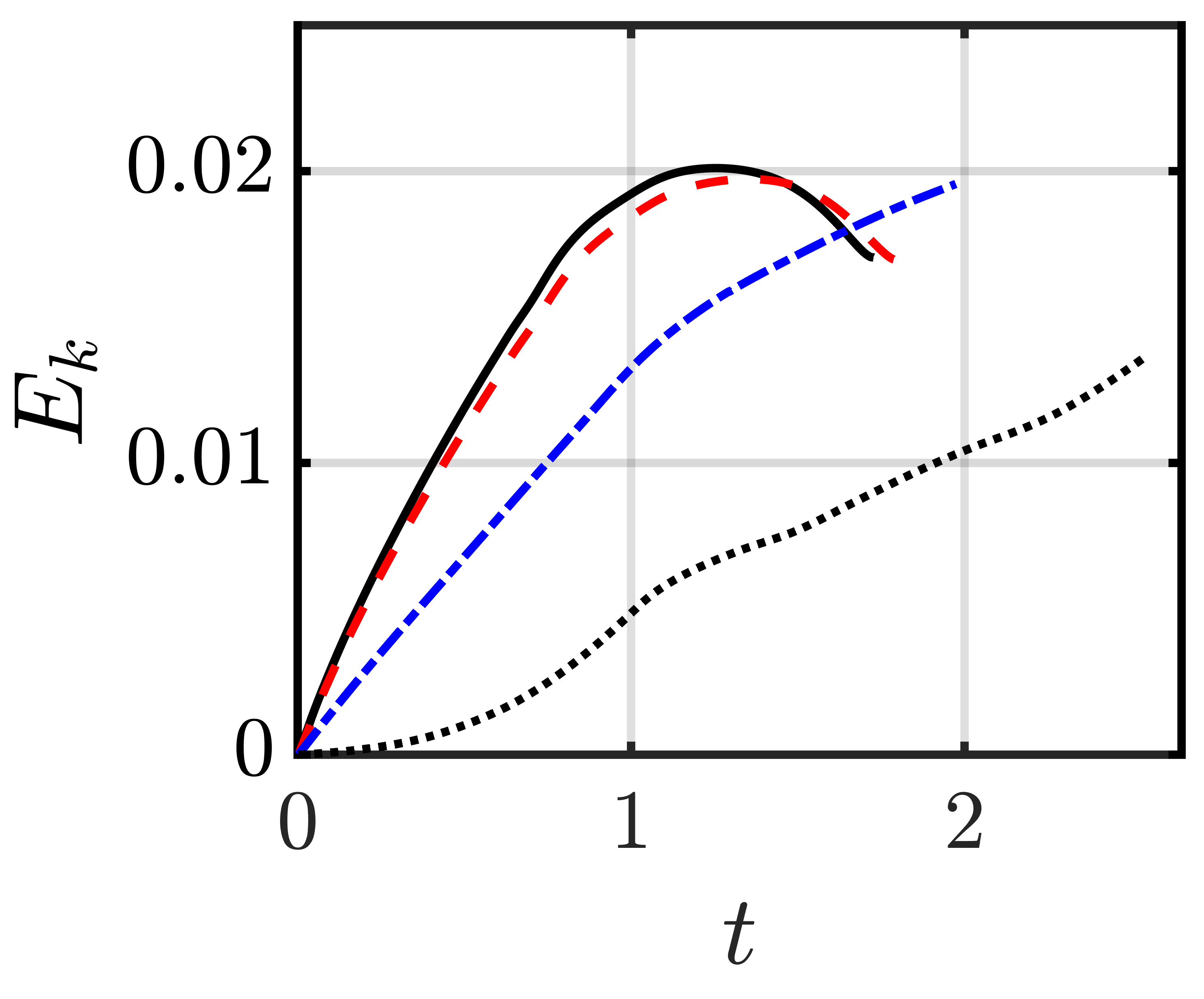} \\
(a) & (b) & (c) \\
\end{tabular}
\caption{\label{k_metrics}
Effect of the adsorption parameter $k$ on the temporal dynamics of the vertical strength of the drop, (a),  minimum neck radius, (b), and kinetic energy (c), when  $Oh=0.02$, $Bo=10^{-3}$, $\beta_s=0.5$, $Pe_s=Pe_b=100$, $Bi=1$ and $\Gamma_o=\chi$.}
\end{figure}

Figure \ref{k_metrics} shows the effect of the adsorption parameter, $k$, on the temporal dynamics of the drop-interface coalescence phenomenon through the analysis of the vertical stretch of the droplet, $z_{max}$, the neck radius, $r_{min}$, and the kinetic energy, $E_k$, for $Bi=1$, and $k=(0.01, 1, 5)$. 
We note that as $k \rightarrow 0$, $\chi \rightarrow 0$, and this corresponds to vanishingly small equilibrium interfacial concentrations, which were used to initialise the simulations. In addition, from eq. (2.5), $k \rightarrow 0$ implies that $\Gamma$ will remain small, and thus, in this limit, we expect the dynamics to be consistent with those associated with the surfactant-free case. For $k \gg 1$, on the other hand, the flow behaviour is similar to that observed in the insoluble surfactant case.
%At the lower end of this range (i.e., $k \ll 1$), the desorptive timescales are much larger than those associated with interfacial effects; therefore, the dynamics are expected to be similar to those observed for the surfactant-free case. Whereas at the upper end, a significant change is expected as surfactants tend to absorb more rapidly to the interface.

\begin{figure}
\begin{center} 
\begin{tabular}{cccc}
\includegraphics[width=0.245\linewidth]{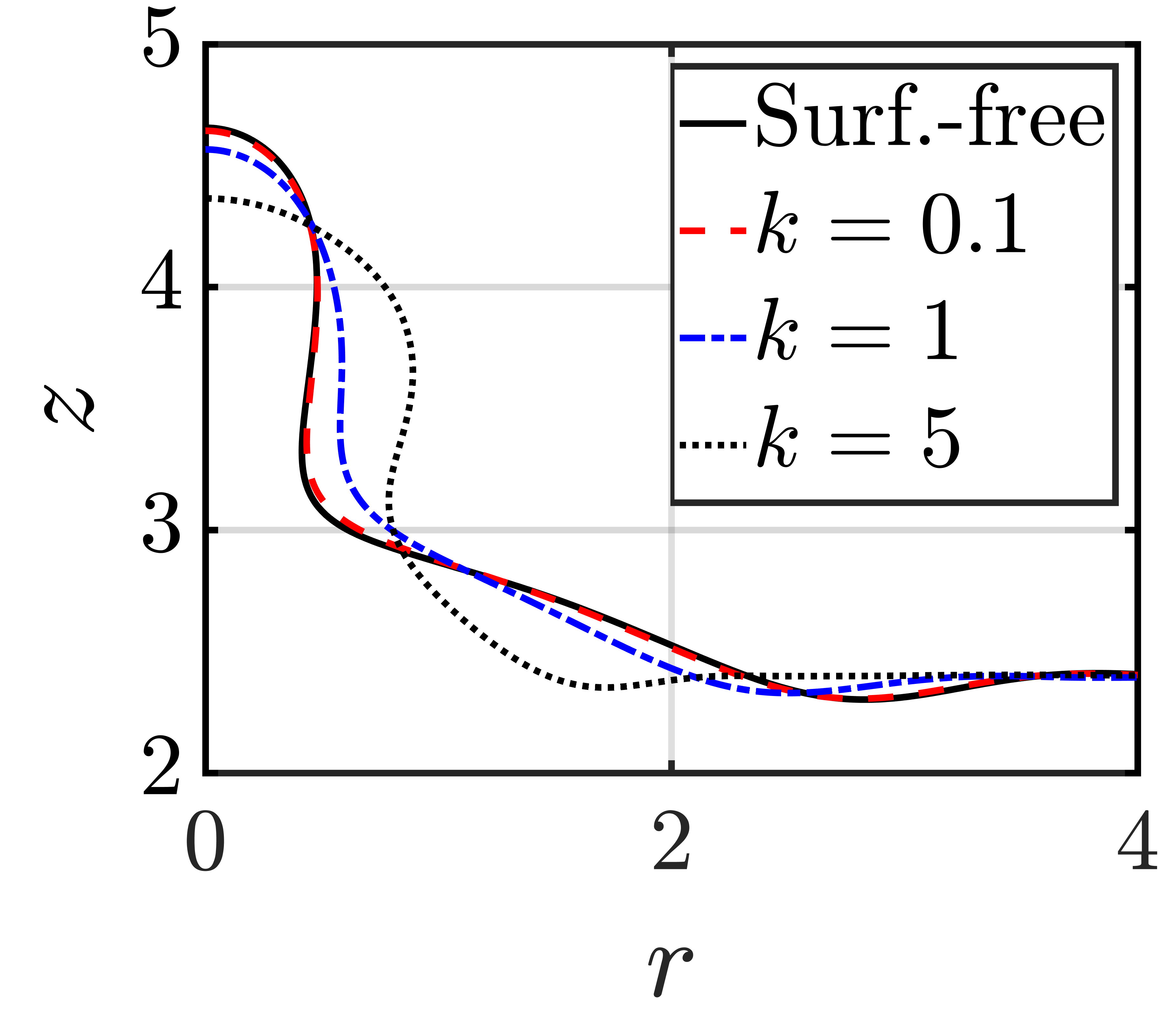}&
\includegraphics[width=0.245\linewidth]{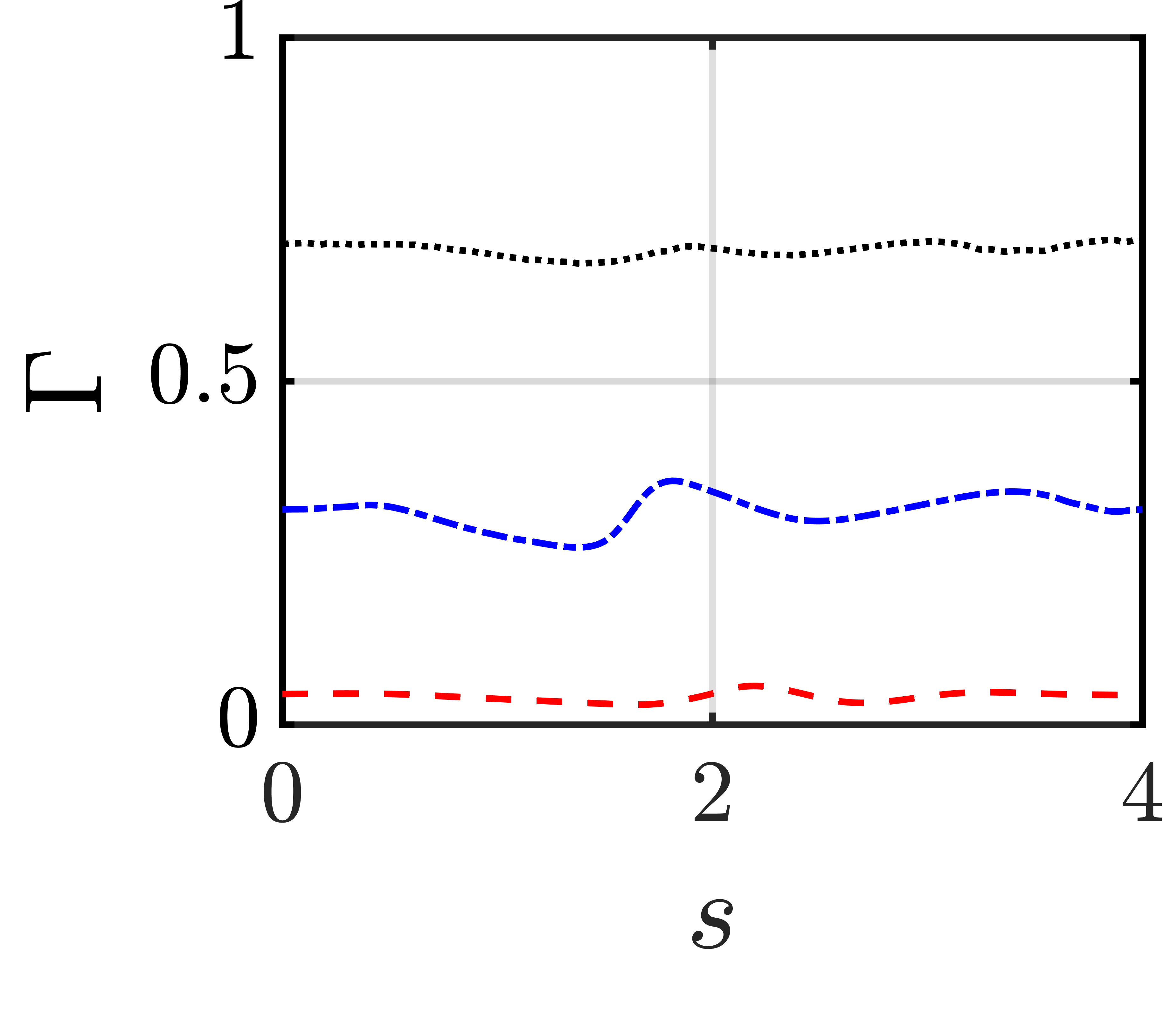}&
\includegraphics[width=0.245\linewidth]{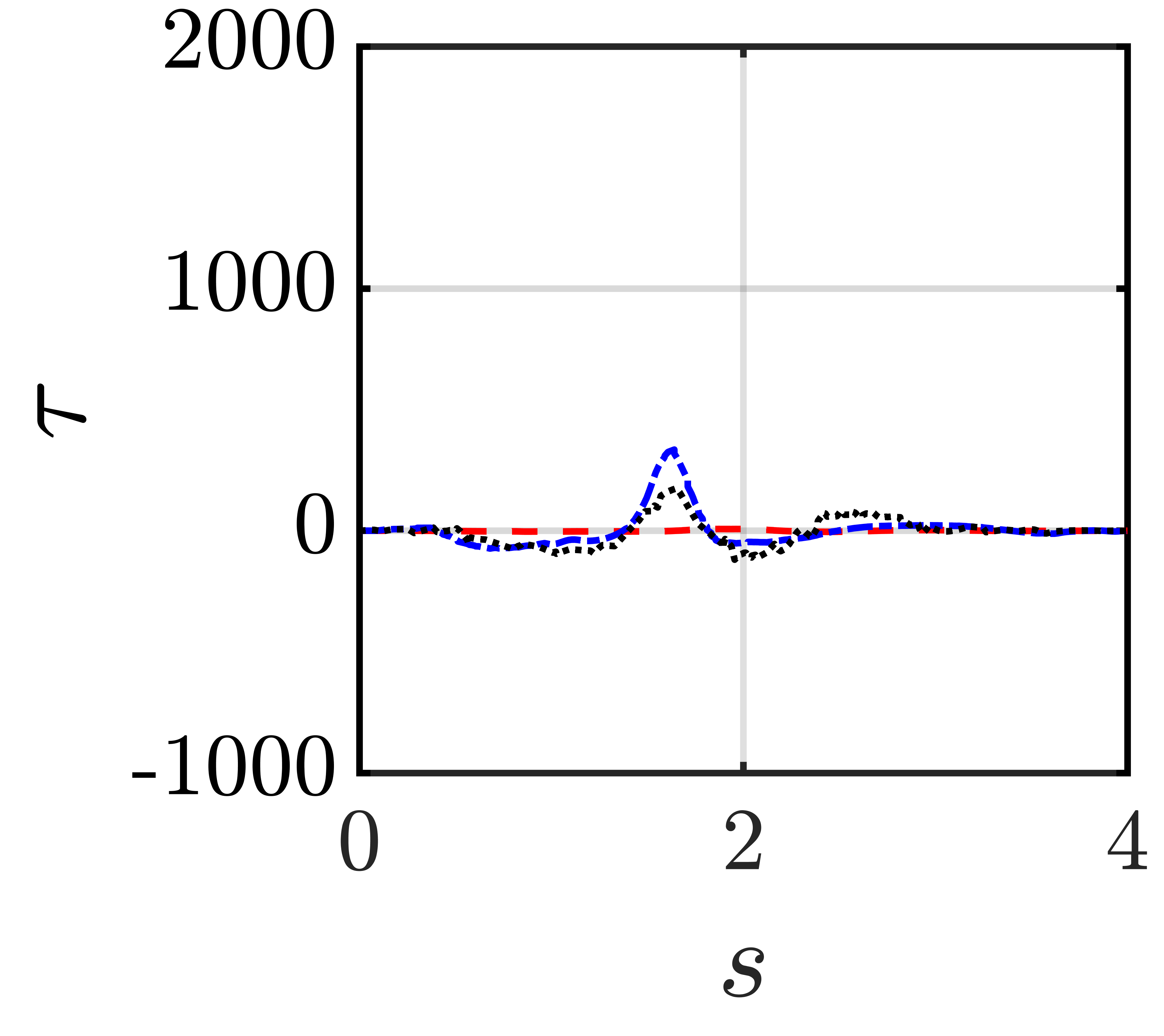}&
\includegraphics[width=0.245\linewidth]{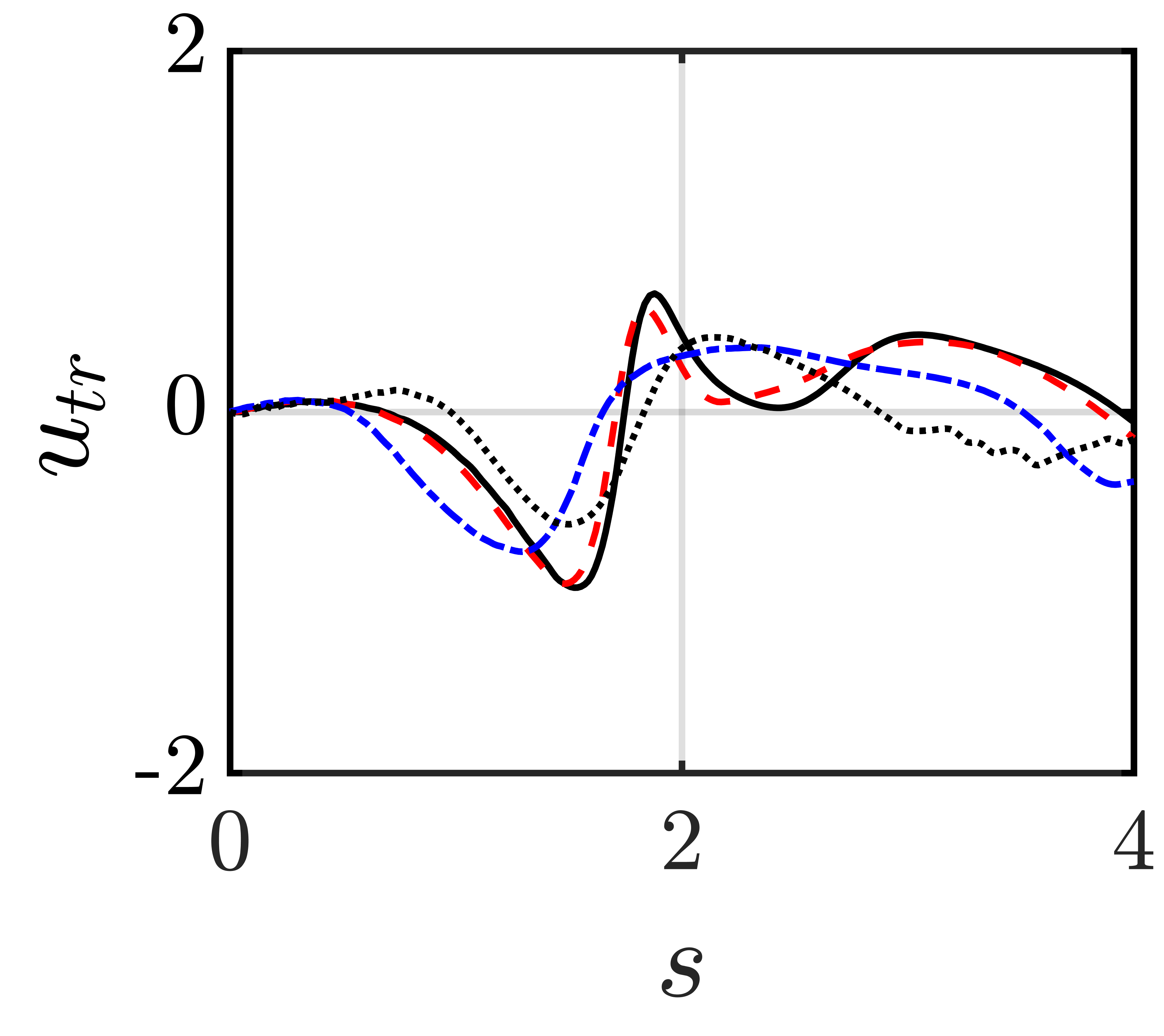}\\
(a) & (b)& (c) & (d)  \\
\includegraphics[width=0.245\linewidth]{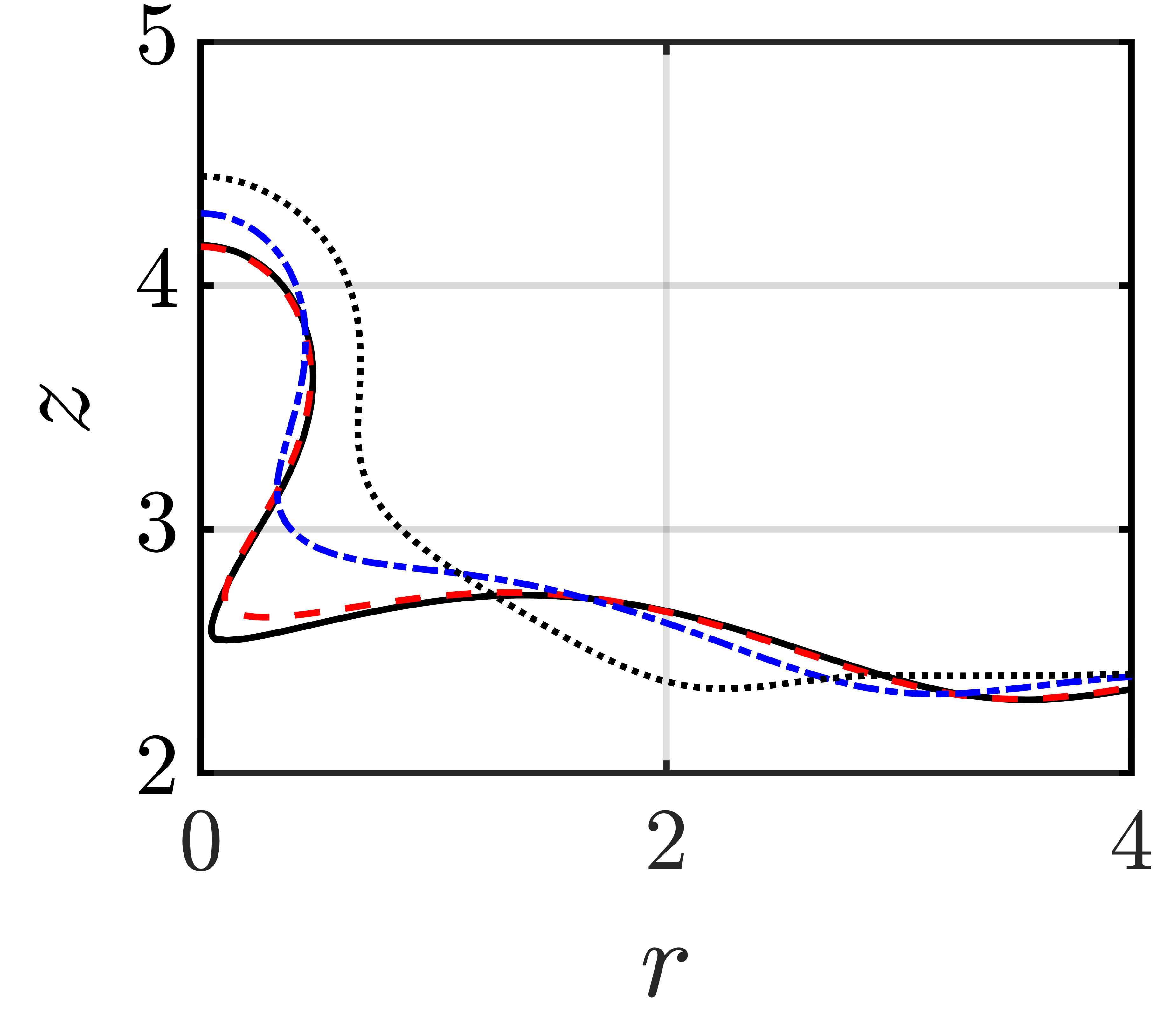}&
\includegraphics[width=0.245\linewidth]{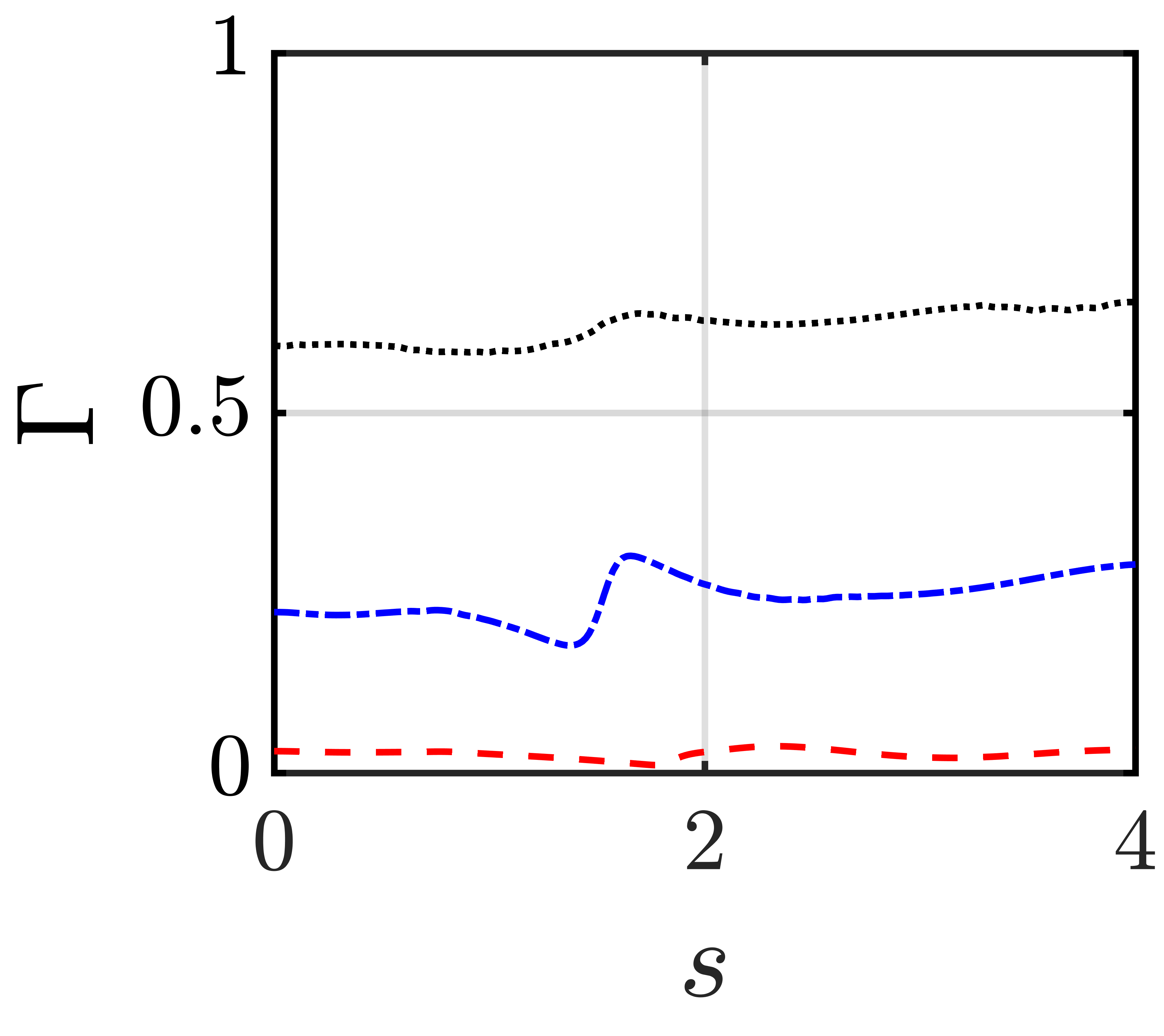}&
\includegraphics[width=0.245\linewidth]{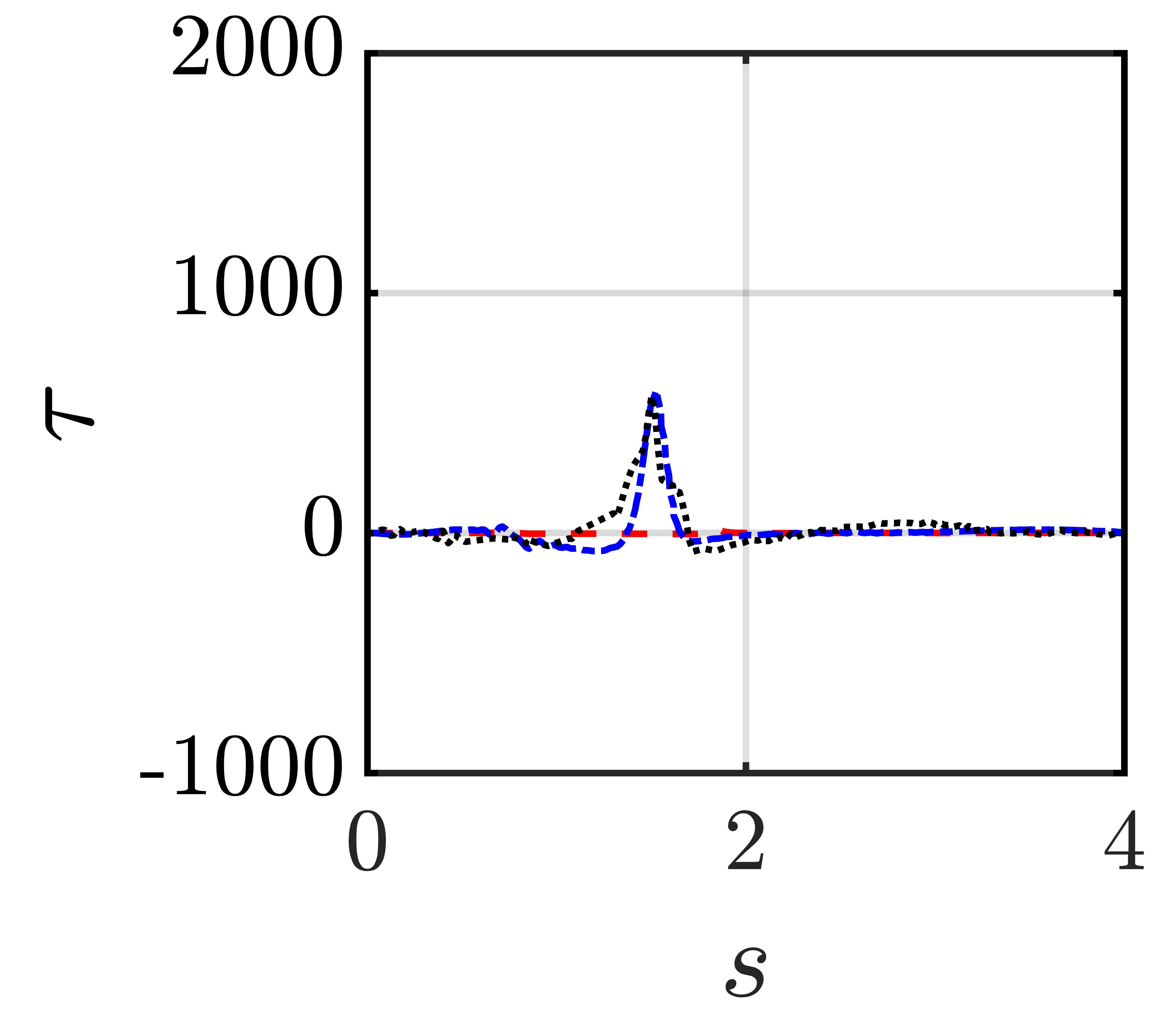}&
\includegraphics[width=0.245\linewidth]{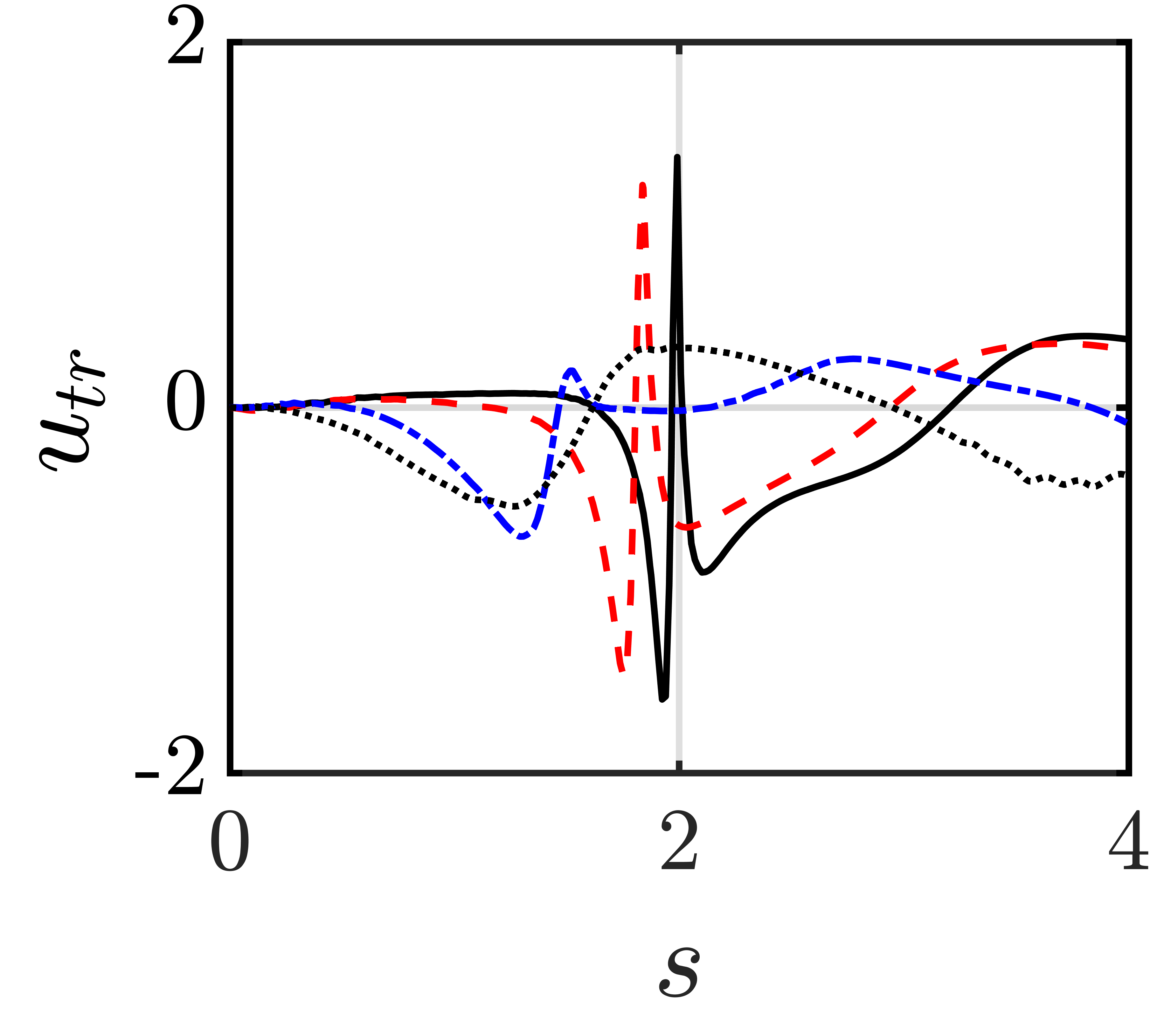}\\
(e) & (f)& (g) & (h)  \\
\end{tabular}
\end{center} 
\caption{\label{k_field_flows} 
Effect of the adsorption parameter $k$ on the flow and surfactant concentration fields associated with the drop-interface coalescence phenomenon.
Two-dimensional representation of the interface location, $\Gamma$, $\tau$, and the radial component of the interfacial velocity $u_{tr}$ are shown in (a)-(d) and (e)-(h) for $t=1.20$ and $t=1.68$, respectively. 
Note that the abscissa in (a) and (e) corresponds to the radial coordinate $r$, and in (b)-(d) and (f)-(h) to the arc length $s$. Here, all other parameters remain unchanged from figure \ref{k_metrics}.
%Effect of adsorption parameter $k$ on the interfacial dynamics. Interface location, $\Gamma$, $\tau$ and $u_{tr}$ are shown in rows one to four, respectively. In column 1 and 2, the dimensional times correspond to $t=1.20$ and $t=1.68$, respectively.  In rows 1 and 2-4 the variation is with respect to the dimensionless radial coordinate and, $r$, and arc length, $s$, respectively.  Here, all other parameters remain unchanged from figure \ref{k_metrics}.
} 
\end{figure}

We start the discussion of the effect of the $k$ parameter by analysing its effect on the vertical stretch of the droplet (shown in figure \ref{k_metrics}a). By inspection of the profiles, a  monotonic response of the vertical stretch is observed with decreasing $k$ values (e.g., $k=0.1$) % As $k \rightarrow 0$, the system tends to behave as a 
the dynamics are similar to that of the surfactant-free case characterised by neck formation and pinch-off, as shown in figure \ref{k_field_flows}a,e; this arises due to the increase in mass transfer from the interface to the bulk decreasing the interfacial concentration (see figure \ref{k_field_flows}b,f) and reducing % (i.e., decrease of surfactant adsoprtion) to reduce 
the magnitude of the Marangoni stress, which is maximal for $k=1$, as shown in figure \ref{k_field_flows}c,g. 
%The higher the value of $k$, the more adsorption of surfactant, and the higher the strength  of  interfacial  rigidification. 
The trends highlighted in figure \ref{k_metrics}a are mirrored in figure \ref{k_metrics}b that displays the temporal dynamics of the $r_{min}$ which also exhibits a monotonic dependence on $k$. %response depending on the $k$ parameter; 
Increasing $k$ alters the $u_{tr}$ profile in figure \ref{k_field_flows}d,h in a similar manner to that observed upon increasing $\beta_s$ and/or decreasing $Bi$ as was shown previously in figures \ref{beta_flow_fields}d,h and \ref{biot_flow_field}d,h, respectively. As a result, it is seen clearly  
%At low $k$ values, the lower end of the parameter, the presence of surfactants does not prevent the capillary singularity, and the system behaves similarly to the surfactant-free case; whereas at the upper end, 
that the presence of surfactants alters the fate of the coalescence phenomenon as Marangoni-driven flow induces neck reopening. Finally, the $E_k$ plots shown in figure \ref{k_metrics}c support, once more, the high interfacial rigidification brought about by the presence of surfactants.

\section{Conclusions \label{conclusion}}
%%%%%%%%%%%%%%%%%%%%%%%%%%%%%%%%%%%%%%%%%
%%%%%%%%%%%%%%%%%%%%%%%%%%%%%%%%%%%%%%%%%
%%%%%%%%%%%%%%%%%%%%%%%%%%%%%%%%%%%%%%%%%
A study of the effect of Marangoni-induced flow as a result of the presence of surfactants on the drop-interface coalescence was presented using a hybrid front-tracking/level-set method (\cite{Shin_ijnmf_2009}, \cite{Shin_jmst_2017}, \cite{Shin_jcp_2018}). The  surfactant transport equations were fully-coupled to the Navier-Stokes equations in which the surface tension  depends on the interfacial surfactant concentration through a nonlinear Langmuir equation of state.  The numerical framework has been validated against the  experimental work presented by \citet{Blanchette_np_2006} for the surfactant-free coalescence dynamics, and the inertio-viscous  scaling-lows regarding  the temporal evolution of the neck towards its capillary singularity presented by \citet{Eggers_prl_1993}. We have selected a surfactant-free base case characterised by the dimensionless quantities of $Oh=0.02$ and $Bo=10^{-3}$. The former parameter ensures a rich dynamics in the inertia-viscous-capillary flow regime, whereas the latter parameter ensures that  gravity forces do not affect the dynamics of the system, which could mask effects related to %any possible results owing to 
the presence of surfactants.

For  %The results began by the presentation  of the effect of 
insoluble surfactants, we have % (via $\beta_s$) and 
demonstrated that Marangoni stresses drive motion from regions of high-surfactant concentration (low surface tension) to low concentration (high tension) regions, resulting in retardation of the interfacial dynamics. This immobilising effect of the surfactants as a result of the Marangoni stresses is also observed via the strong reduction in the maximum stretching (by dampening the strength of the capillary waves which converge in the drop summit) of the drop and  kinetic energy.  We have also shown, that the condition for the capillary singularity is the existence of two stagnation points close to the drop neck, which leads to the generation of vorticity in this area (`vortex-ring'). In the presence of surfactants, Marangoni-induced flow suppresses one of the stagnation points resulting in the advection of vorticity towards the liquid bulk and the reopening of the neck. This effect is strongest for insoluble surfactants and, for soluble surfactants, is maximal for an intermediate range of solubility and sorption kinetic parameter values. 

%Additionally, we have considered  the role of soluble surfactants on the interfacial dynamics by analysing its sorptive kinetics (i.e., $Bi$ and $k$ parameters). For the studied range of $Bi$ number (e.g., $0.1<Bi<10$), the presence of surfactants is capable of inhibiting the capillary break-up, and subsequently changing the outcome of the surfactant-free system. 
%By altering the $k$ parameter, we observe a competition between the adsorption-desorption and hydrodynamics time scales which implies the change of the rate of  mass transfer from/to the interface  to weaken the effect of the Marangoni stresses, and consequently changing the fate of the drop-interface coalescence. 

Future directions are related to the performance of numerical simulations featuring three-dimensional behaviours occurring for large Bond numbers. Recently, the experimental work performed by \citet{Dong_pof_2019} suggested that the presence of surfactants induces the rupture of the interface (i.e., hole formation) in an off-axis location  at high Bond numbers. Thus, a fully three-dimensional retracting capillary wave will certainly affect the behaviour of the system rising a more complex coalescence dynamics, and constitute a fruitful area of future research. 

%%%%%%%%%%%%%%%%%%%%%%%%%%%%%%%%%%%%%%%%%
%%%%%%%%%%%%%%%%%%%%%%%%%%%%%%%%%%%%%%%%%
%%%%%%%%%%%%%%%%%%%%%%%%%%%%%%%%%%%%%%%%%
\subsection*{Acknowledgements}
%%%%%%%%%%%%%%%%%%%%%%%%%%%%%%%%%%%%%%%%%
%%%%%%%%%%%%%%%%%%%%%%%%%%%%%%%%%%%%%%%%%
%%%%%%%%%%%%%%%%%%%%%%%%%%%%%%%%%%%%%%%%%

This work is supported by the Engineering \& Physical Sciences Research Council, United Kingdom, through a studentship for CRCA in the Centre for Doctoral Training on Theory and Simulation of Materials at Imperial College London funded by the EPSRC (EP/L015579/1), and through the EPSRC MEMPHIS (EP/K003976/1) and PREMIERE (EP/T000414/1) Programme Grants. CRCA also acknowledges the funding and technical support from BP through the BP International Centre for Advanced Materials (BP-ICAM), which made this research possible. OKM also acknowledges funding from PETRONAS and the Royal Academy of Engineering for a Research Chair in Multiphase Fluid Dynamics. We also acknowledge HPC facilities provided by the Research Computing Service (RCS) of Imperial College London for the computing time. DJ and JC acknowledge support through computing time at the Institut du Developpement et des Ressources en Informatique Scientifique (IDRIS) of the Centre National de la Recherche Scientifique (CNRS), coordinated by GENCI (Grand Equipement National de Calcul Intensif) Grant 2020 A0082B06721. The numerical simulations were performed with code BLUE (\citet{Shin_jmst_2017}) and the visualisations have been generated using ParaView. \\

{\bf Declaration of interests:} The authors report no conflict of interest.
%%%%%%%%%%%%%%%%%%%%%%%%%%%%%%%%%%%%%%%%%
%%%%%%%%%%%%%%%%%%%%%%%%%%%%%%%%%%%%%%%%%
%%%%%%%%%%%%%%%%%%%%%%%%%%%%%%%%%%%%%%%%%
 \subsection*{APPENDIX: mesh study}

\begin{figure}
\begin{center} 
\includegraphics[ width=0.6\linewidth]{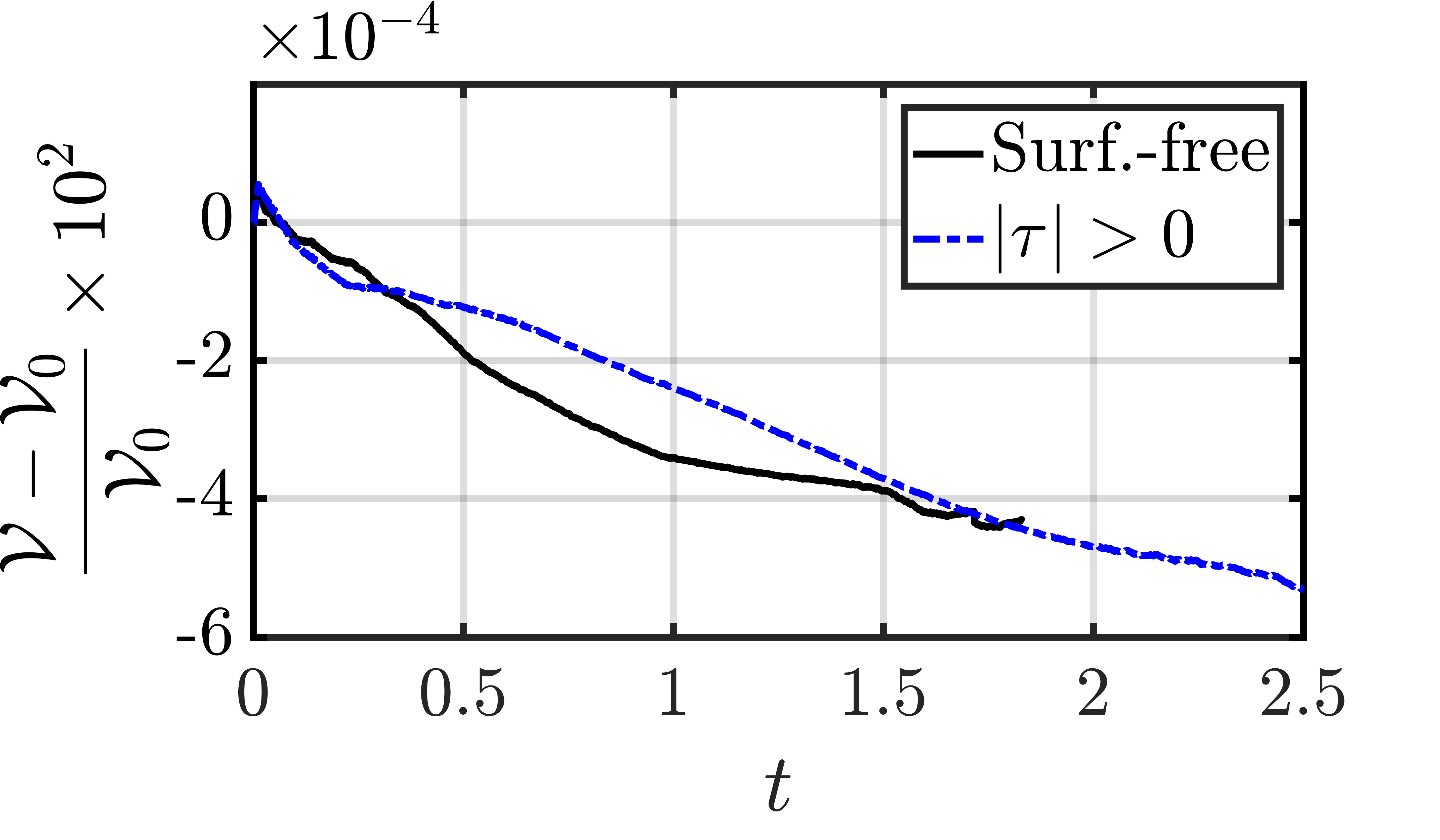}
\end{center} 
\caption{Relative variation of the liquid volume for the surfactant-free and full-Marangoni case $|\tau|>0$, for $Oh=0.02$, $Bo=10^{-3}$, $\beta_s=0.5$, $Pe_s=100$ and $\Gamma_o=0.5\Gamma_\infty$.
\label{mesh}}
\end{figure}

On account of showing mesh independent results, we have monitored the temporal variation of the liquid-volume of the system  for a resolution of $(386)^3$, which has been used throughout the entire study. Figure \ref{mesh} shows the plot profiles for  the surfactant-free and the surfactant-laden cases. It is evident that the numerical method is capable of capturing the rich interfacial dynamics with a conservation of volume under  $10 ^{-3}\%$. With respect to the accuracy of the surfactant equations, we refer to \cite{Shin_jcp_2018}, who carefully benchmarked  the formulation and numerical implementation of the surface gradients of surfactant concentration and surface tension. For the studied phenomenon, we observed the conservation of surfactant mass under  $10 ^{-2}\%$ for all the surfactant-laden cases. Additionally, extensive mesh studies for capillary phenomena, using the same numerical method, had been previously reported \citep{Bachvarov_prf_2020, Constante-Amores_prf_2020, 
Constante-Amores_jfm_2021}. 

% \bibliographystyle{jfm}
% \bibliography{references.bib}

\end{document}